\documentclass{article}

\usepackage[dandb, final]{neurips_2025}

\usepackage[utf8]{inputenc} % allow utf-8 input
\usepackage[T1]{fontenc}    % use 8-bit T1 fonts
\usepackage[hidelinks]{hyperref}
\usepackage{url}            % simple URL typesetting
\usepackage{booktabs}       % professional-quality tables
\usepackage{amsfonts}       % blackboard math symbols
\usepackage{nicefrac}       % compact symbols for 1/2, etc.
\usepackage{microtype}      % microtypography
\usepackage{xcolor}         % colors
\usepackage{graphicx}
\usepackage{multirow}
\usepackage{multicol}
\usepackage{amsmath}
\usepackage{subfigure}
\usepackage{subcaption}

\title{ArchPower: Dataset for Architecture-Level Power Modeling of Modern CPU Design}

\author{%
  Qijun Zhang, Yao Lu, Mengming Li, Shang Liu, Zhiyao Xie$^*$ \\
  Hong Kong University of Science and Technology \\
  \texttt{\{qzhangcs, yludf, mengming.li, sliudx\}@connect.ust.hk, eezhiyao@ust.hk}
}

\begin{document}

\maketitle

\begingroup\renewcommand\thefootnote{*}
\footnotetext{Corresponding Author}
\endgroup

\begin{abstract}

%Power is the primary design objective in IC design, especially for modern processor (i.e., CPU) design. To optimize the power, an efficient power estimation technique is critical. 
%The standard power estimation flow requires a significant amount of manpower and is time-consuming. 
%Therefore, architects require a fast yet accurate architecture-level power model to evaluate the power consumption before RTL implementation and the whole VLSI flow. However, the traditional analytical power models require significant human efforts and are inaccurate. Although the ML-based architecture-level power model is promising by being automatic and more accurate, there is a lack of an open-sourced dataset that should be high-quality and comprehensive. 

%Architects require a fast yet accurate architecture-level power model to evaluate the power consumption before RTL implementation and the whole VLSI flow.
%The ML-based architecture-level power model is promising because of its high accuracy compared with analytical models. 
%However, generating a dataset for ML-based architecture-level power modeling is substantially challenging, 

Power is the primary design objective of large-scale integrated circuits (ICs), especially for complex modern processors (i.e., CPUs). Accurate CPU power evaluation requires designers to go through the whole time-consuming IC implementation process, easily taking months. At the early design stage (e.g., architecture-level), classical power models are notoriously inaccurate. Recently, ML-based architecture-level power models have been proposed to boost accuracy, but the data availability is a severe challenge. Currently, there is no open-source dataset for this important ML application. A typical dataset generation process involves correct CPU design implementation and repetitive execution of power simulation flows, requiring significant design expertise, engineering effort, and execution time. Even private in-house datasets often fail to reflect realistic CPU design scenarios. In this work, we propose ArchPower, the first open-source dataset for architecture-level processor power modeling. We go through complex and realistic design flows to collect the CPU architectural information as features and the ground-truth simulated power as labels. Our dataset includes 200 CPU data samples, collected from 25 different CPU configurations when executing 8 different workloads. There are more than 100 architectural features in each data sample, including both hardware and event parameters. The label of each sample provides fine-grained power information, including the total design power and the power for each of the 11 components. Each power value is further decomposed into four fine-grained power groups: combinational logic power, sequential logic power, memory power, and clock power. ArchPower is available at https://github.com/hkust-zhiyao/ArchPower.

%here each ground-truth power has not only the total circuit power but also the fine-grained power values of four power groups. 

%ArchPower provides an open-sourced, high-quality, and comprehensive dataset. 
%The power label of the dataset is generated with commercial EDA tools and the commercial technology library. The SRAM implementation and clock-gating technique are integrated. The dataset is based on two different real CPU designs, including BOOM CPU and XiangShan CPU. ArchPower also provides a training-testing framework with an easily adopted user interface to evaluate ML-based architecture-level power models. 
%ArchPower is available at xxxxxxxxxxxxxxxxxxxxxxxxxxxxxxxxxx.
\end{abstract}

\section{Introduction}

\begin{figure}[!t]
\centering
%\vspace{-.05in}
\includegraphics[width=1\textwidth]{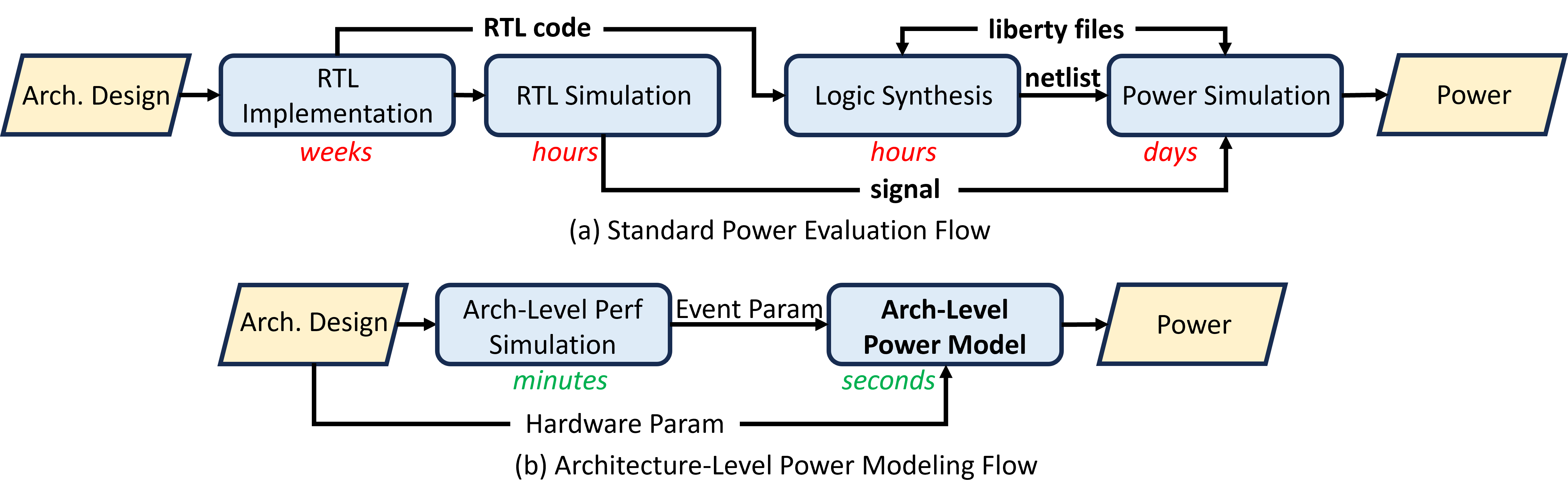}
\vspace{-.2in}
\caption{Comparison between (a) standard power evaluation flow and (b) architecture-level power evaluation flow. The architecture-level power modeling flow is significantly efficient compared with the standard power evaluation flow. ArchPower provides labeled data for ML-based architecture-level power modeling. }
\vspace{-.1in}
\label{overview}
\end{figure}

The rapid advancements of AI rely on the support of very large-scale integrated (VLSI) circuits. \emph{Power} is the primary design objective of integrated circuits (ICs), especially for complex modern processors (i.e., CPUs), which play a central role in various computing systems. 
Accurate yet efficient power estimation techniques are the premise of and key challenge of power optimization. However, as Fig.~\ref{overview}(a) shows, accurate CPU power evaluation requires designers to go through the whole time-consuming IC implementation process, easily taking months. As the complexity of CPU designs keeps increasing, the standard power estimation flow becomes increasingly costly.

To facilitate power estimation at early stages, designers will evaluate power consumption at the architecture level, before designing the RTL (e.g., in Verilog or VHDL) and going through the downstream implementation flow (i.e., circuit synthesis and layout). 
Fig.~\ref{overview}(b) illustrates the workflow of the architecture-level power modeling, using classical tools such as McPAT~\cite{li2009mcpat} and Wattch~\cite{brooks2000wattch}. 
Such a fast power estimation approach takes only tens of seconds, which is about 100$\times$ faster than the standard VLSI power estimation flow. 
%Analytical architecture-level power models such as McPAT~\cite{li2009mcpat} and Wattch~\cite{brooks2000wattch} have been widely adopted as fast alternatives for power estimation.
However, these classical analytical architecture-level power models are notoriously inaccuracy, as indicated in multiple existing studies~\cite{xi2015quantifying,zhai2022mcpat,lee2015powertrain,tang2014mcpat,guler2020mcpat}.

In recent years, machine learning (ML)-based architecture-level power model~\cite{zhai2022mcpat,zhang2023panda} has been explored for better power evaluation accuracy. The ML-based architecture-level power model takes both \emph{hardware parameters} and \emph{event parameters} as features to predict the CPU power consumption as its output. Hardware parameters are parameters to determine CPU configurations, such as \emph{FetchWidth} and \emph{DecodeWidth}. Event parameters are event statistics when a CPU executes a workload, collected from existing architecture-level performance simulators, such as the number of branch mispredictions and DCache misses. 
Based on a few training data collected on the target CPU architecture,  ML-based power models can mitigate the modeling error or bias incurred from analytical models that are built for outdated processors.

% The ML-based power model has two advantages compared with the analytical power models, including high accuracy and automatic modeling. First, 
% Second, the ML algorithm can automatically learn the correlation between the features and the ground truth power based on the training data, not requiring the engineering effort to analyze and calculate the power for each new architecture design. 

% As a machine learning task, ML-based architecture-level power models require an open-sourced high-quality dataset.

\begin{table*}[!t]
\centering
      \resizebox{1\textwidth}{!}{
\begin{tabular}{|c|cccc|}
\hline
Work & Commercial Tech Lib & Clock Gating & SRAM Implementation & Diverse Architectures \\
\hline
\hline
McPAT-Calib~\cite{zhai2022mcpat} & & & & \\
ASPDAC'23~\cite{zhai2023microarchitecture} & & & & \\
PANDA~\cite{zhang2023panda} & $\checkmark$ & & $\checkmark$ & \\
FirePower~\cite{zhang2025firepower} & $\checkmark$ & & $\checkmark$ & $\checkmark$ \\
AutoPower~\cite{zhang2025autopower} & $\checkmark$ & $\checkmark$ & $\checkmark$ & \\
\textbf{ArchPower (This Work)} & \textbf{$\checkmark$} & \textbf{$\checkmark$} & \textbf{$\checkmark$} & \textbf{$\checkmark$} \\
\hline
\end{tabular}
}
        \vspace{-.05in}
        \caption{Comparison between datasets in existing works and our proposed ArchPower dataset.}
        \vspace{-.2in}
        \label{limitation}
\end{table*}

However, despite emerging works in ML-based architecture-level power models~\cite{zhai2022mcpat,zhai2023microarchitecture,zhang2023panda,zhang2025firepower}, they all built their solutions on private datasets. There is no open-source dataset for such an important application, preventing the AI community from making its contribution. 
We find that some of them open-source their model implementation~\cite{zhai2022mcpat,zhang2023panda,zhang2025firepower}, however, none of them open-source their dataset for training and testing. Other related works, such as architecture-level design space exploration~\cite{bai2021boom,bai2024towards,yu2023dse,zhai2023vlsi,bai2023archexplorer}, also do not share their data. 
It is because the dataset generation for the ML-based architecture-level power modeling is challenging, requiring significant IC knowledge and engineering effort for the correct CPU design implementation and power simulation flow. % 1) To generate the architectural features, it requires setting up and configuring the architecture-level performance simulator. 2) Generating the power labels requires generating the RTL code that costs significant manpower, setting up scripts to invoke multiple EDA tools utilized in the VLSI flow that also require expensive licenses and substantial time overhead and computing power, and implementing the SRAM macros specific to the commercial technology library. 

% However, the dataset generation for the ML-based architecture-level power modeling is challenging, requiring significant IC knowledge and engineering effort for the correct CPU design implementation and power simulation flow. 1) To generate the architectural features, it requires setting up and configuring the architecture-level performance simulator. 2) Generating the power labels requires setting up the RTL generation flow based on the complex open-source CPU projects, writing correct and efficient scripts for multiple EDA tools utilized in the VLSI flow, and implementing the SRAM macros specific to the commercial technology library. 
% Therefore, even though there are many existing works targeting ML-based architecture-level power models~\cite{zhai2022mcpat,zhai2023microarchitecture,zhang2023panda,zhang2025firepower}, there is no open-source dataset for such an important application, 
% %there lack of an open-source dataset for the ML-based architecture-level power model evaluation, 
% hindering algorithm experts from making their contribution. 
% We find that some of them open-source their model implementation~\cite{zhai2022mcpat,zhang2023panda,zhang2025firepower} and example data~\cite{zhai2022mcpat}, however, none of them open-source their full dataset for training and testing. Other related works, such as architecture-level design space exploration~\cite{bai2021boom}, also do not share their data. 

Besides the unavailability, these in-house datasets also have other limitations, as shown in Table~\ref{limitation}. 1) Some datasets~\cite{zhai2022mcpat,zhai2023microarchitecture} do not include SRAM in their implementation. It is because of difficulties in implementing SRAM in RTL and the lack of SRAM support in some technologies. However, the SRAM is essential to build many important components of the CPU, such as the cache and the branch predictor, and consumes over 50\% power of the whole CPU. Therefore, the absence of SRAM leads to an unreliable evaluation.
2) Some other datasets~\cite{zhang2023panda,zhang2025firepower} do not adopt the clock-gating technique for logic synthesis, making the clock power far from the real processors. 
3) Most of these datasets~\cite{zhai2022mcpat,zhai2023microarchitecture,zhang2023panda} are only collected based on a single CPU architecture, unable to validate whether the evaluated models can also work for other architectures.

% where all existing works are evaluated on their own in-house datasets. The lack of such a dataset leads to three main problems.
% First, without an open-sourced dataset, algorithm experts can not easily contribute to such an important task. 
% Developing an in-house dataset is difficult, requiring significant engineering efforts, heavy expert knowledge in the integrated circuit design field, and multiple commercial tools. These difficulties hinder experts in the ML algorithm field from developing and improving the ML-based architecture-level power model. 
% Second, these in-house datasets adopted in existing works are not faithful. Some datasets~\cite{zhai2022mcpat,zhai2023microarchitecture} are based on unrealistic academic technology libraries where SRAM is not supported. However, the SRAM is essential to build many important components of the CPU, such as the cache and the branch predictor, and consumes over 50\% power of the whole CPU. Therefore, the absence of SRAM leads to an unreliable evaluation.
% Some other datasets~\cite{zhang2023panda,zhang2025firepower} do not adopt the clock-gating technique for logic synthesis, making the clock power far from the real processors. 
% Third, these in-house datasets are not comprehensive. Most of these datasets~\cite{zhai2022mcpat,zhai2023microarchitecture,zhang2023panda} are only collected based on a single CPU design, unable to validate whether the evaluated models can also work for other designs.  

To address the problems above, in this work, we propose ArchPower, the first open-source dataset for ML-based architecture-level power modeling of modern processor design. Our dataset includes 200 samples collected from 25 CPU configurations and 8 workloads. 
The architectural feature of each sample is a vector with 101 elements, including hardware parameters and event parameters. They can also be extracted as per-component features with our provided indexes. 
The power label of each sample includes the whole CPU ground-truth power and 11 per-component ground-truth powers. Each ground-truth power has not only the total circuit power but also the fine-grained power values of four power groups, including combinational logic power, sequential logic power, memory power, and clock power.

To build the dataset, we invest substantial engineering effort. We set up the frameworks~\cite{amid2020chipyard,xu2022towards} for the RTL code generation process of two widely adopted CPU architectures, including BOOM~\cite{zhao2020sonicboom} and XiangShan CPUs~\cite{xu2022towards}. 
We also integrate realistic SRAM macros for each memory block in the RTL designs. 
Based on the RTL implementation, we go through the complex VLSI flow and standard power evaluation flow with commercial EDA tools~\cite{vcs,ptpx,design-compilier}, with clock-gating considered. 
%To perform the VLSI flow, The 
The VLSI flow consumes a long runtime and significant computing power. 
% We also integrate realistic SRAM macros for each memory block in the RTL designs.

% with SRAM implementation and clock-gating considered. 

% First, ArchPower provides an \emph{open-sourced}, \emph{high-quality}, \emph{comprehensive} dataset for the ML-based architecture-level power model. 1) ArchPower is \emph{open-sourced} with easily adopted interfaces. 2) ArchPower is \emph{high-quality}. ArchPower is generated with commercial EDA tools and the commercial technology library. The clock-gating technique is adopted for logic synthesis, and the SRAM is integrated. 3) ArchPower is \emph{comprehensive}, providing data collected from two real Out-of-Order CPU architectures, including BOOM and XiangShan. 
% Second, besides the dataset, ArchPower also provides a ready-to-use training-testing framework for ML-based architecture-level power model evaluation. 

% Our contributions are summarized below.
% \begin{itemize} 
% \item We assemble an open-source high-quality dataset for the ML-based architecture-level power model. It is the \emph{first} open-source dataset for the architecture-level processor power model.
% \item We also provide a ready-for-use training-testing framework for the ML-based architecture-level power model. Users can design and evaluate different ML models with provided clear interfaces in our framework.
% \item We train and evaluate different ML models based on ArchPower, providing a comprehensive evaluation of them. 
% \end{itemize}

Our contributions are summarized below.
\begin{itemize} 
\item We release ArchPower, the \emph{first} open-source dataset for the ML-based architecture-level power model. ArchPower includes 200 data samples collected from 25 CPU configurations and 8 workloads. ArchPower reflects realistic design power since it considers the clock-gating and integrates realistic SRAM macros for power label collection. 
\item We also provide a training-testing framework for the ML-based architecture-level power model. It includes different setups of training and testing data that reflect the realistic CPU development scenarios, which can evaluate the generalization of ML-based power models.
\item We evaluate six different power models, including two analytical models and four ML-based models, based on ArchPower. The evaluation provides both the total power and per-component power modeling accuracy. 
\end{itemize}

% First, all existing works are evaluated on their in-house datasets. Developing an in-house dataset is difficult because of tool time-consuming  

\section{Preliminary}
In this section, we first describe the principle of the standard power evaluation flow in Sec.~\ref{sec:principle}, briefly introduce both classical analytical architecture-level power model and ML-based architecture-level power model in Sec.~\ref{sec:analytical}, and then introduce the modern CPU architecture in Sec.~\ref{sec:o3cpu}.

\subsection{Principle of VLSI Power Evaluation Flow}
\label{sec:principle}

The dynamic power dominates the power consumption, and the leakage power is small compared with the dynamic power. Therefore, we focus on the dynamic power evaluation. The dynamic power of the processor is the sum of the dynamic power of all cells. Denoting the dynamic power as $P$, the power calculation is shown in Eq.(\ref{eq:cellpower}), 
\begin{equation}
P = \,\sum_{c \in netlist} \alpha_{c} C_{c} V^2 f\label{eq:cellpower}
\end{equation}
where $c$ is the cell in the netlist, $\alpha_{c}$ is the switch activity of cell $c$, $C_c$ is the capacitance of cell $c$ and the load capacitance of its wire, $V$ is the voltage of the chip, and $f$ is the frequency of the chip. 

The standard power evaluation flow calculates the power by collecting all related values from inputs and conducting the calculation. The cell $c$ is from the netlist generated by logic synthesis, switch activity $\alpha_{c}$ is extracted from the signal information generated by RTL simulation, capacitance $C_c$ is from the liberty file, and the voltage $V$ and frequency $f$ are set at the chip level. Such a standard power calculation is complex and slow because both generating input files and calculating the final power are time-consuming.

\subsection{Architecture-Level Power Model}
\label{sec:analytical}

To avoid the time-consuming standard power evaluation, the architecture-level processor power model can provide fast power estimation at the early design stage. The architecture-level power model takes the hardware parameters $H$, which determine the CPU configurations, and event parameters $E$, which are the event statistics generated by the performance simulators like gem5~\cite{binkert2011gem5}, as input and outputs the power consumption $P$. 
% \begin{equation}
% P = M(H,E) \label{eq:archpower}
% \end{equation}

\textbf{Analytical Architecture-Level Power Model:} The analytical architecture-level power model estimates the processor power consumption following two steps, denoted as Eq.(\ref{eq:analyticalpower}). 
\begin{equation}
P = F_{event}(F_{op}(H),E) \label{eq:analyticalpower}
\end{equation}
In the first step, based on the hardware parameter $H$, the model instantiates each component of the processor based on empirical models to collect the per-operation energy. This step is denoted as $F_{op}$. In the second step, the model transforms the event parameters $E$ into the count of basic operations and calculates the power consumption. This step is denoted as $F_{event}$. However, because of the discrepancy between the real processor and the modeled one, the analytical $F_{op}$ and $F_{event}$ are usually inaccurate, leading to the low accuracy of analytical power models.

%\subsection{ML-based Architecture-Level Power Model}
%\label{sec:mlbased}

\textbf{ML-based Architecture-Level Power Model:} In recent years, to address the low accuracy of analytical power models, ML-based architecture-level power models~\cite{zhai2022mcpat,zhai2023microarchitecture,zhang2023panda,zhang2025firepower} have been proposed. The ML-based power model adopts a machine learning model $F_{ml}$ to directly learn the correlation between the input feature and the final power consumption, denoted as Eq.(\ref{eq:mlpower}). 
\begin{equation}
P = F_{ml}(H,E) \label{eq:mlpower}
\end{equation}
Among the ML-based architecture-level power models, some existing ML-based power models~\cite{zhai2022mcpat,zhai2023microarchitecture} adopt a purely black-box machine learning algorithm, and some others~\cite{zhang2023panda,zhang2025firepower} utilize a hybrid gray-box model, with some analytical information provided.
\subsection{Modern CPU Architecture}
\label{sec:o3cpu}

\begin{figure}[!t]
\centering
%\vspace{-.05in}
\includegraphics[width=1\textwidth]{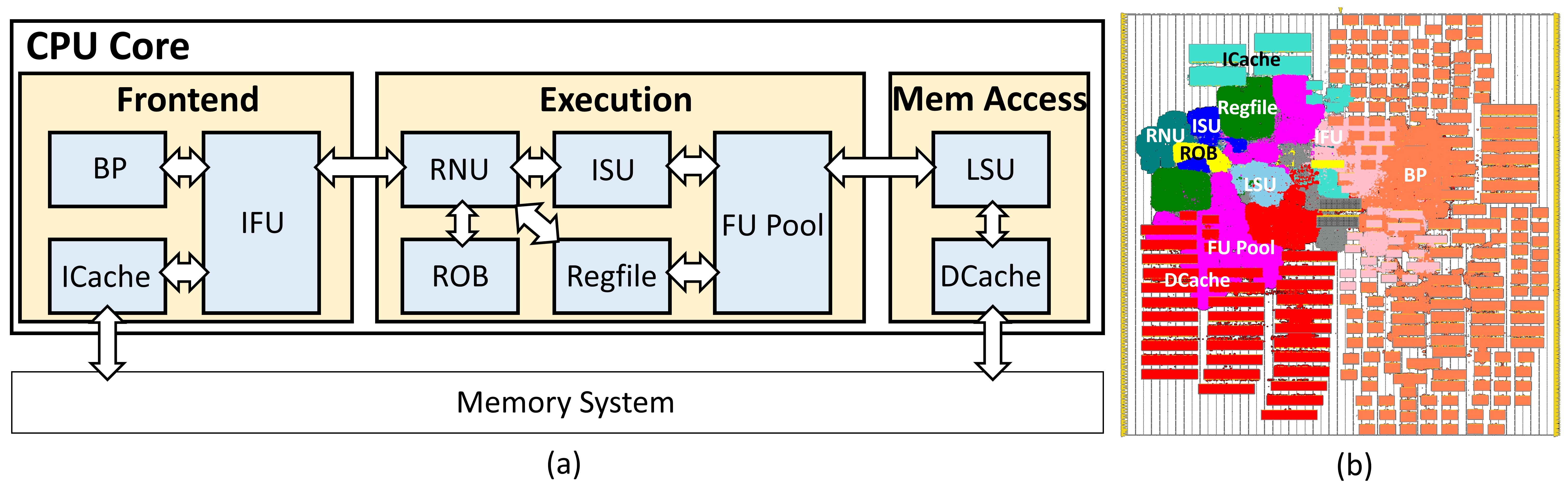}
\vspace{-.2in}
\caption{(a) The architecture of the modern high-performance CPU core. Blue blocks are major components. The yellow block represents the Other Logic. (b) A layout example of one BOOM CPU.}
\vspace{-.2in}
\label{cpu}
\end{figure}

Fig.~\ref{cpu} shows the basic architecture of modern high-performance CPUs that our dataset targets. Modern CPUs usually adopt out-of-order execution to improve instruction-level parallelism, which can significantly boost CPU performance. 
%The Out-of-Order CPU is the mainstream of modern high-performance CPUs, adopting out-of-order execution to improve instruction-level parallelism that can significantly boost the CPU performance. Therefore, our dataset targets the Out-of-Order CPUs. Fig.~\ref{cpu} shows the basic architecture of the Out-of-Order CPU core. 
The CPU has three major blocks, including Frontend, Execution, and Mem Access, with each block consisting of multiple components. 
1) The Frontend includes 3 components: branch predictor (BP), instruction cache (ICache), and instruction fetch unit (IFU). 
2) The Execution consists of 5 components: renaming unit (RNU), reorder buffer (ROB), issue unit (ISU), register file (Regfile), and function unit pool (FU Pool). 
3) The Mem Access has 2 components: load-store unit (LSU) and data cache (DCache).
4) The logic not covered by the major components above is referred to as Other Logic. ArchPower provides per-component power labels for each of the 11 components.

% \begin{itemize}
%     \item The Frontend includes 3 components: branch predictor (BP), instruction cache (ICache), and instruction fetch unit (IFU).
%     \item The Execution consists of 5 components: renaming unit (RNU), reorder buffer (ROB), issue unit (ISU), register file (Regfile), and function unit pool (FU Pool).
%     \item The Mem Access has 2 components: load-store unit (LSU) and data cache (DCache).
%     \item The logic not covered by the major components above is referred to as Other Logic.
% \end{itemize}

% BOOM~\cite{zhao2020sonicboom} and XiangShan~\cite{xu2022towards} are two representative, widely adopted, and open-source modern CPU architectures. Therefore, our dataset is built on these two architectures.

\section{Related Work}
% \subsection{Architecture-Level Power Model Dataset}
Despite many existing works exploring architecture-level power modeling, there is \emph{no} existing open-source dataset for the architecture-level power model. Some existing works~\cite{zhai2022mcpat,zhai2023microarchitecture,zhang2023panda,zhang2025firepower} release their model implementation code, and some of them~\cite{zhai2022mcpat,zhai2023microarchitecture} also provide example data for demonstration with only one or two samples. However, none of them release their full dataset for training and testing.

Besides unavailability, these in-house datasets also have other problems. 
%Some works~\cite{zhai2022mcpat,zhai2023microarchitecture} utilize the academic technology library like ASAP7~\cite{clark2016asap7} for VLSI flow, which is far from realistic scenarios. These 
Some works~\cite{zhai2022mcpat,zhai2023microarchitecture} exclude the SRAM in the processors, while the SRAM is the basic building block of many important components and dominates the power consumption of modern CPUs. Some other works~\cite{zhang2023panda,zhang2025firepower} do not adopt the clock-gating technique when performing logic synthesis, which is an essential optimization for processor designs. Therefore, ignoring the clock-gating technique makes their CPU implementation and power labels far from the real processors. 
Therefore, an open-source high-quality dataset is in great need for the development of ML-based architecture-level power models. 
\section{Dataset Description}

\subsection{Dataset Overview}

\begin{figure}[!t]
\centering
%\vspace{-.2in}
\includegraphics[width=1\textwidth]{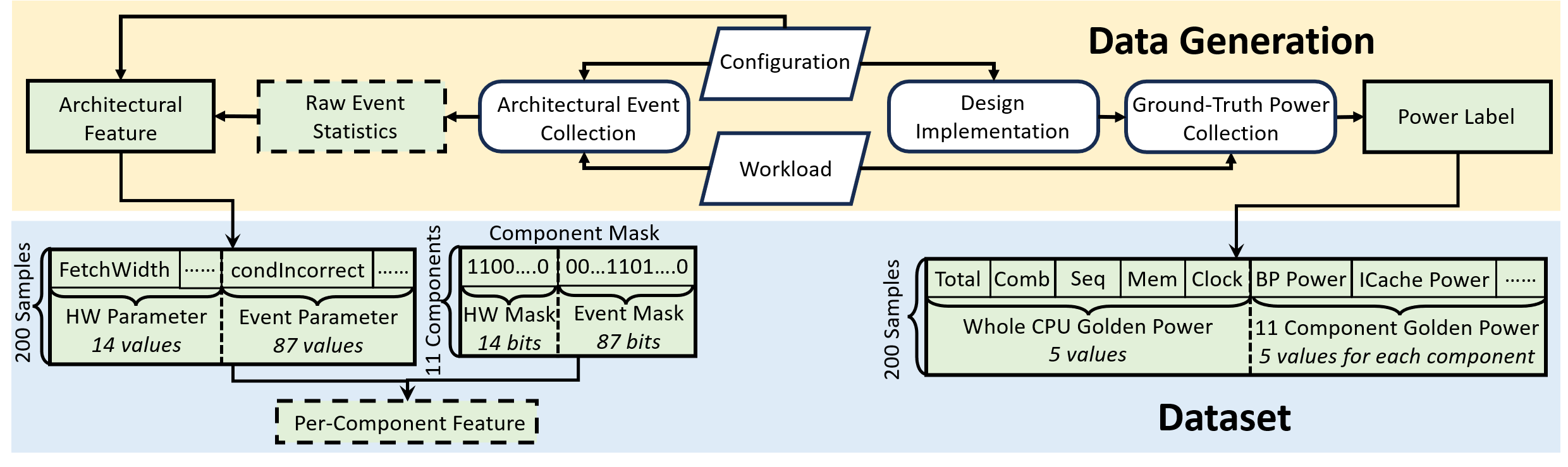}
\vspace{-.2in}
\caption{Dataset and data generation process of ArchPower. Our dataset mainly includes the architectural power modeling features and power labels. The features can further be masked with component masks to generate per-component features. ArchPower generates architectural power modeling features through the architectural event collection and generates power labels through design implementation and ground-truth power collection.}
\vspace{-.2in}
\label{dataset}
\end{figure}

Our ArchPower dataset consists of $25\times 8 = 200$ data samples collected from 25 CPU configurations when executing 8 different workloads. Each data sample describes a CPU configuration when executing a workload, providing both architecture-level \emph{features} and power \emph{labels}. Fig.~\ref{dataset} provides an overview of our ArchPower dataset. The architectural feature of each sample is a vector with 101 elements, including 14 hardware parameters $H$ and 87 event parameters $E$. The corresponding ground-truth power label of each sample is collected by going through the design implementation and simulation flow. 

%which can also be extracted as per-component features. 

% \yao{[Introduce the per-component feature, label, and power groups here: ``in addition to the feature and power of the whole design, ...'']} 
In addition to the feature and power label of the whole design, we provide component masks to indicate per-component features and include per-component power labels for 11 components. For both the whole design and per-component, we also provide fine-grained power labels for four power groups: combinational logic power, sequential logic power, memory power, and clock power. The power label of each sample has 60 values in total. 
% The power label of each sample has 60 power values, including the whole CPU ground-truth power and 11 per-component ground-truth powers, where each ground-truth power has not only the total circuit power but also the fine-grained power values of four power groups. 
The raw event statistics files of each sample are also included for potential use in future research.

\subsection{Detailed Dataset Description}
Given the hardware parameter $H$ of the target CPU configuration and event parameter $E$ when the CPU runs the target workload, the ML-based architecture-level power model predicts the power consumption $P$, as shown in Eq.(\ref{eq:mlpower}). Therefore, in ArchPower, a sample represents a CPU configuration running a workload. 
ArchPower provides features and labels for both the whole CPU and each component. In this subsection, we describe the feature and label in detail. 

\subsubsection{Architectural Power Modeling Feature}

\begin{table*}[!t]
      \centering
      %\vspace{-.2in}
      \renewcommand{\arraystretch}{1.05}
      \resizebox{1\textwidth}{!}{
        \begin{tabular}{ |c|c|c| } 
        \hline
        Component &  Hardware parameters of each component  &  Event parameters of each component   \\
        \hline
         \hline
BP &  FetchWidth, BranchCount  & BTBLookups, condPredicted, condIncorrect, commit.branches  \\
         \cline{1-3}
\multirow{4}{*}{IFU} &  \multirow{3}{*}{FetchWidth, DecodeWidth}   & fetch.\{insts, branches, cycles\}, numRefs, numStoreInsts, numInsts,     \\
         &           & decode.\{runCycles, blockedCycles, decodedInsts\}, numBranches,  \\
         &  FetchBufferEntry, ICacheFetchBytes  & intInstQueueReads, intInstQueueWrites, intInstQueueWakeupAccesses,  \\ 
         & & fpInstQueueReads, fpInstQueueWrites, fpInstQueueWakeupAccesses  \\
         \cline{1-3}
% I-TLB &  ICacheTLBEntry  & itb.accesses, itb.misses   \\
%          \cline{1-3}
%I-Cache &  ICacheWay, ICacheFetchBytes  & icache.overallAccesses, icache.overallMisses, icache.ReadReq.mshrHits, icache.ReadReq.mshrMisses, icache.tagAccesses & \\ 
%\hline
         %& & ReadReq.mshrMisses, tagAccesses   \\
\multirow{2}{*}{ICache} &  \multirow{2}{*}{ICacheWay, ICacheFetchBytes} & overallAccesses, overallMisses, ReadReq.mshrHits,  \\  
& & ReadReq.mshrMisses, tagAccesses  \\
         \hline

\multirow{2}{*}{RNU} & \multirow{2}{*}{DecodeWidth}  & intLookups, renamedOperands, fpLookups, renamedInsts,  \\
& & runCycles, blockCycles, committedMaps  \\

% RNU & DecodeWidth  & intLookups, renamedOperands, fpLookups, renamedInsts, runCycles, blockCycles, committedMaps  \\
         \cline{1-3}
ROB &  DecodeWidth, RobEntry  & reads, writes  \\
         \cline{1-3}
\multirow{3}{*}{ISU} &  DecodeWidth, Mem/FpIssueWidth,   & IssuedMemRead, IssuedMemWrite, IssuedFloatMemRead,    \\
             &  IntIssueWidth  & IssuedFloatMemWrite, IssuedIntAlu, IssuedIntMult,  \\
             & & IssuedIntDiv, IssuedFloatMult, IssuedFloatDiv \\
        \cline{1-3}
\multirow{3}{*}{Regfile} &  DecodeWidth, IntPhyRegister,  & intRegfileReads, fpRegfileReads, intRegfileWrites,     \\
&  FpPhyRegister  & fpRegfileWrites, functionCalls    \\
        % & & fpRegfileWrites, functionCalls\\
         \cline{1-3}
FU Pool &  Mem/FpIssueWidth, IntIssueWidth  & intAluAccesses, fpAluAccesses   \\
        \hline

LSU &  LDQ/STQEntry, MemIssueWidth  & MemRead, InstPrefetch, MemWrite  \\
         \cline{1-3} 
% D-TLB &  DCacheTLBEntry  & dtb.accesses, dtb.misses  \\
%          \cline{1-3}
% \multirow{2}{*}{DCache} &  DCacheWay, DCacheTLBEntry,  & dcache.ReadReq.accesses, dcache.WriteReq.accesses, dcache.ReadReq.misses, dcache.WriteReq.misses,   \\
%          & DCacheMSHR, MemIssueWidth  & dcache.overallMisses, dcache.MshrHits, dcache.MshrMisses, dcache.tagAccesses  \\
%         \hline

\multirow{2}{*}{DCache} &  DCacheWay, DCacheTLBEntry,  & ReadReq.accesses, WriteReq.accesses, ReadReq.misses, tagAccesses, \\
         & DCacheMSHR, MemIssueWidth  & WriteReq.misses, overallMisses, MshrHits, MshrMisses \\
        \hline \hline
         \multirow{7}{*}{CPU Level} &  \multirow{7}{*}{-}  & totalIpc, totalCpi, numCycles, idleCycles, numLoadInsts, \\
         & & numSquashedInsts, committedInsts, commit.\{numDist::mean, memRefs\},\\
         & & mmu.dtb.\{accesses, misses\}, iew.writebackCount, numIssuedDist::mean, \\
         & & statIssuedInstType\_0::total, fuBusy, mmu.itb.\{accesses, misses\}, \\
         & & conflictingLoads, conflictingStores, insertedLoads, insertedStores, \\
         & & mem\_ctrls.\{readReqs, writeReqs, bytesReadSys\}, \\
         & & icache.tags.totalRefs, dcache.\{overallAccesses::total, tags.totalRefs\} \\
        \hline
        \end{tabular}
        }
        \caption{Hardware parameters and event parameters of each component. The 14 hardware parameters and 87 event parameters in the architectural power modeling feature are the union of all components and the CPU-level parameters. Other Logic adopts all features and is not listed.}
        \label{tbl:config_event}
        %\vspace{-.25in}
\end{table*}

ArchPower has 200 data samples, and each sample has 101 features, including 14 hardware parameters and 87 event parameters. Therefore, we provide the architectural power modeling feature as a $200\times101$ matrix in our dataset. 
%20200 feature values in total, shaped as a $200\times101$ matrix in our dataset. 
1) The first 14 columns are the 14 hardware parameters that determine the CPU configuration, including \emph{FetchWidth}, \emph{DecodeWidth}, \emph{FetchBufferEntry}, \emph{RobEntry}, \emph{IntPhyRegister}, \emph{FpPhyRegister}, \emph{LDQ/STQEntry}, \emph{BranchCount}, \emph{Mem/FpIssueWidth}, \emph{IntIssueWidth}, \emph{DCache/ICacheWay}, \emph{DTLBEntry}, \emph{MSHREntry}, and \emph{ICacheFetchBytes}. 2) The last 87 columns are the 87 event parameters that are event statistics generated by the architecture-level performance simulator when a CPU configuration executes a workload, such as \emph{condIncorrect}, \emph{icache.overallMisses}, and \emph{dcache.ReadReq.access}. %, generated by the architecture-level performance simulator. 
Table~\ref{tbl:config_event} lists hardware parameters and event parameters of each component. The 14 hardware parameters and 87 event parameters in the architectural power modeling feature are the \emph{union} of all components and the CPU-level parameters.

% The 14 hardware parameters and 87 event parameters are the \emph{union} of events related to each component listed in Table.~\ref{tbl:config_event}.

% We also provide features for each component. 
% For each component, the hardware and event parameters only include those related to this component and are a subset of our provided 101 features. 
Our dataset provides a component mask to select the features for each component, as shown in Fig.~\ref{dataset}. The Other Logic adopts all features and is not listed. For each component, the component mask has a 14-bit hardware mask to select from 14 hardware parameters and an 87-bit event mask to select from 87 event parameters. With the component mask, the per-component features can be extracted from our provided architectural power modeling feature.

%Similar to the whole CPU power modeling, the feature is a vector including hardware parameters and event parameters related to this component based on Table.~\ref{tbl:config_event}. The component-level features are subsets of the whole CPU power modeling feature, therefore, these features are not provided as standalone values and can be extracted from the whole CPU features with our provided indexes. 

\begin{figure}[!t]
\centering
%\vspace{-.05in}
\includegraphics[width=1\textwidth]{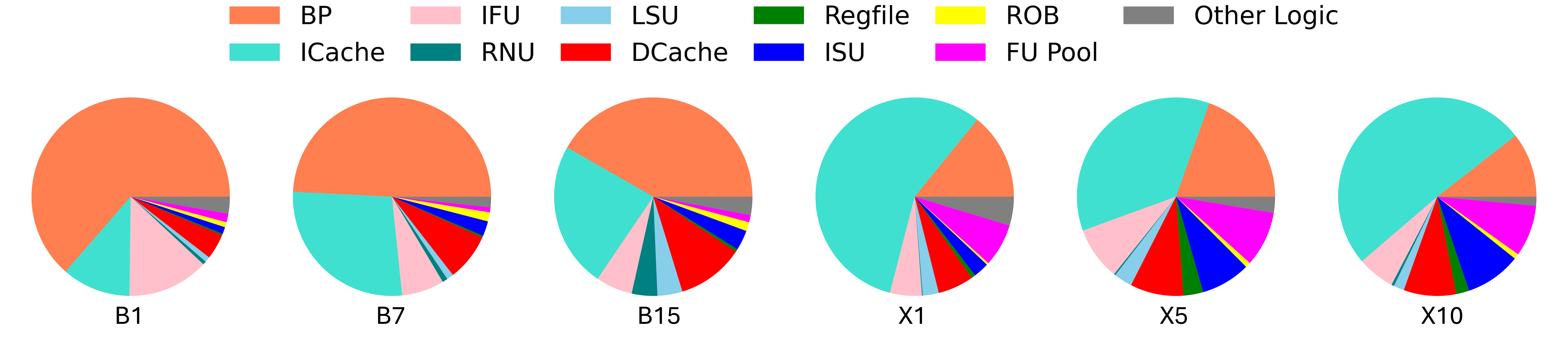}
\vspace{-.1in}
\caption{The power distributions across 11 components of 6 different CPU configurations (B1, B7, B15, X1, X5, X10) with different scales.}
\vspace{-.2in}
\label{compdist}
\end{figure}

% \begin{figure}[!t]
% \centering
% %\vspace{-.05in}
% \includegraphics[width=1\textwidth]{_fig/group_dist.png}
% %\vspace{-.05in}
% \caption{The power breakdown across different power groups.}
% %\vspace{-.15in}
% \label{groupdist}
% \end{figure}

\subsubsection{Power Label}
For the 200 data samples, each sample has $(1 + 11) \times (1+4) =\,60$ fine-grained power labels, for both the whole CPU and 11 components. We further decouple the power into four power groups, including combinational logic power, sequential logic power, memory logic power, and clock power. 
Therefore, we provide our power label as a $200 \times 60$ matrix in our dataset. 
%Therefore, we provide 12000 golden power values in total in the dataset, shaped as a $200 \times 60$ matrix. 
For each sample, the label includes the whole CPU ground-truth power collected from the standard VLSI power evaluation flow. Besides the whole CPU ground-truth power, we also provide the labels at the component level, which include ground-truth power for the 11 components, including BP, IFU, ICache, RNU, ROB, ISU, Regfile, FU Pool, LSU, DCache, and Other Logic, respectively. %Therefore, there are 12 ground-truth powers in total. 
Fig.~\ref{compdist} shows the power distribution across different components for configurations with different scales, where Bi is the $i^\text{th}$ configuration of BOOM architecture and Xi is the $i^\text{th}$ configuration of XiangShan architecture.

\subsection{Training-Testing Data Setup}
\label{sec:setup}

In addition to the dataset, we also provide a ready-for-use training-testing framework for the evaluation of ML-based architecture-level power models. In our framework, we set up training scenarios based on the three unique characteristics of processor developments. 
1) Training and testing samples are divided based on configurations, where all data collected from training configurations are training data, and data from testing configurations are testing data. 
2) Because of the significant manpower and time overhead, the training configurations are usually limited in real scenarios. Therefore, we set up few-shot scenarios with only three training configurations. 
3) In the industry, architects usually also work on configurations that have different scales from available training configurations. Therefore, we set up three training scenarios with different training data distributions. 
% \begin{itemize}
%     \item The CPU design implementation is more expensive compared with RTL simulation and power estimation. Therefore, \emph{training and testing samples are divided based on configurations}, where all data collected from training configurations are training data, and data from testing configurations are testing data.
%     \item Because of the significant manpower and time overhead, the training configurations are usually limited in real scenarios. Therefore, \emph{we set up few-shot scenarios with only three training configurations.}
%     \item In the industry, architects usually also work on configurations that have different scales from available training configurations. Therefore, \emph{we set up three training scenarios with different training data distributions.} 
% \end{itemize}
Therefore, we set up three training-testing scenarios named \emph{Balance}, \emph{Small}, and \emph{Large}:
1) \emph{Balance}. We evenly select the configurations as available training configurations based on the scale: B1, B8, and B15 for BOOM, X1, X6, and X10 for XiangShan. 
2) \emph{Small}. We select the smallest configurations as available training configurations: B1, B2, and B3 for BOOM, X1, X2, and X3 for XiangShan. 
3) \emph{Large}. We select the largest configurations as available training configurations: B13, B14, and B15 for BOOM, X8, X9, and X10 for XiangShan.

% \begin{itemize}
%     \item \emph{Balance}. We evenly select the configurations as available training configurations based on the scale: B1, B8, and B15 for BOOM, X1, X6, and X10 for XiangShan. 
%     \item \emph{Small}. We select the smallest configurations as available training configurations: B1, B2, and B3 for BOOM, X1, X2, and X3 for XiangShan. 
%     \item \emph{Large}. We select the largest configurations as available training configurations: B13, B14, and B15 for BOOM, X8, X9, and X10 for XiangShan. 
% \end{itemize}

Besides directly utilizing the training-testing data setup in our provided evaluation framework, users can also evaluate their own training-testing data setup based on the provided dataset of ArchPower.

%\subsection{Feature Description}

\section{Dataset Generation Process}

\subsection{Adopted CPU Configurations and Workloads}

We adopt two highly configurable CPU architectures, BOOM~\cite{zhao2020sonicboom} and XiangShan~\cite{xu2022towards}, to generate multiple CPUs with different configurations. 
% RISC-V~\cite{URL:riscv} is one of the most promising open-sourced instruction set architectures (ISA), and is widely adopted by high-performance open-sourced CPU design projects. 
The Berkeley Out-of-Order Machine (BOOM)~\cite{zhao2020sonicboom} implemented in Chisel~\cite{bachrach2012chisel} is a synthesizable and parameterizable open-source out-of-order core. It can be configured to cores with different scales, given a variety of hardware parameters $H$. XiangShan~\cite{xu2022towards} is a high-performance open-source CPU project. Similar to the BOOM, the XiangShan is also implemented with Chisel and is highly configurable. 
In our dataset, we adopt 15 configurations of the BOOM CPU named B1-B15. 
We adopt 10 configurations of the XiangShan CPU in our dataset named X1-X10. The CPU configurations that we adopted are carefully selected to be similar to real-world commercial CPUs. Different hardware parameters within each configuration also configure a CPU where components are balanced.
Some representative CPU configurations are listed in Table~\ref{configtable} because of the page limitation. All CPU configurations are provided in the appendix. 

%\subsubsection{Riscv-Tests}

For the workloads executed on the CPU, we utilize workloads from the riscv-tests~\cite{URL:riscvtests}. Riscv-tests is the official test benchmark for the RISC-V processors. We collect 8 widely adopted real-world workloads from riscv-tests, including dhrystone, median, multiply, qsort, rsort, towers, spmv, and vvadd. These workloads are across different lengths, from thousands of cycles to several hundred thousand cycles.

\begin{table*}[!t]
\centering
      %\vspace{-.35in}
    %  \renewcommand{\arraystretch}{1.05}
      \resizebox{1\textwidth}{!}{
        \begin{tabular}{ |c||c c c c c c c c c||c c c c c c| } 
\hline
Hardware Parameter  & B1 & B2 & B4 & B6 & B7 & B9  & B11 & B13 & B15 &      X1 & X3 & X5 & X7 & X8  & X10\\
\hline
\hline
FetchWidth          & 4  & 4  & 4  & 8  & 8  & 8   & 8   & 8   & 8 &        4  & 4  & 4  & 8  & 8   & 8\\
\hline
DecodeWidth         & 1  & 1  & 2  & 2  & 3  & 3   & 4   & 5   & 5 &        2  & 2  & 3  & 4  & 4   & 5\\
\hline
FetchBufferEntry    & 5  & 8  & 8  & 24 & 18 & 30  & 32  & 30  & 40 &       8  & 24 & 24 & 24 & 32  & 24\\
\hline
RobEntry            & 16 & 32 & 64 & 80 & 81 & 114 & 128 & 125 & 140 &      16 & 48 & 64 & 81 & 96  & 112\\
\hline
IntPhyRegister      & 36 & 53 & 64 & 88 & 88 & 112 & 128 & 108 & 140 &      36 & 68 & 80 & 88 & 110 & 108\\
\hline
FpPhyRegister       & 36 & 48 & 56 & 72 & 88 & 112 & 128 & 108 & 140 &      36 & 68 & 80 & 88 & 110 & 108\\
\hline
LDQ/STQEntry        & 4  & 8  & 12 & 20 & 16 & 32  & 32  & 24  & 36 &       16 & 24 & 24 & 24 & 32  & 32\\
\hline
BranchCount         & 6  & 8  & 10 & 14 & 14 & 16  & 20  & 18  & 20 &       7  & 7  & 7  & 7  & 7   & 7\\
\hline
Mem/FpIssueWidth    & 1  & 1  & 1  & 1  & 1  & 2   & 2   & 2   & 2 &        2  & 2  & 2  & 2  & 2   & 2\\
\hline 
IntIssueWidth       & 1  & 1  & 1  & 2  & 2  & 3   & 4   & 5   & 5 &        2  & 2  & 4  & 4  & 6   & 6\\
\hline
DCache/ICacheWay    & 2  & 4  & 4  & 8  & 8  & 8   & 8   & 8   & 8 &        4  & 8  & 4  & 8  & 8   & 8\\
\hline
DTLBEntry           & 8  & 8  & 8  & 16 & 16 & 32  & 32  & 32  & 32 &       8  & 16 & 8  & 16 & 16  & 32\\
\hline
MSHREntry           & 2  & 2  & 2  & 4  & 4  & 4   & 4   & 8   & 8 &        2  & 4  & 2  & 4  & 4   & 4\\
\hline
ICacheFetchBytes    & 2  & 2  & 2  & 4  & 4  & 4   & 4   & 4   & 4 &        2  & 2  & 2  & 2  & 2   & 2\\
         \hline
        \end{tabular}
        }
    %   \vspace{-.08in}
        %\vspace{-1mm}
        \caption{Some representative CPU configurations across different scales from our dataset. B1-B15 denote 15 configurations of BOOM, and X1-X10 denote 10 configurations of XiangShan. }
        \vspace{-.2in}
        \label{configtable}
\end{table*}

\subsection{Data Collection Flow}
%\subsection{Data Description}

\textbf{Architectural Event Collection:} For a CPU configuration when executing a workload, we adopt the gem5~\cite{binkert2011gem5} as our performance simulator to generate the event statistics. We configure the O3CPU in gem5 with the hardware parameters of the simulated configuration and execute the workload. All of our generated raw event statistic files are available in our dataset, and our script for automatic configuration and simulation is also open-sourced in ArchPower.

\textbf{Design Implementation:} To implement a CPU design with a configuration, we perform RTL code generation with Chipyard~\cite{amid2020chipyard} v1.8.1 and OpenXiangShan~\cite{xu2022towards} for BOOM and XiangShan, respectively. To get the netlist, we perform logic synthesis with Synopsis Design Compiler\textsuperscript{\textregistered}~\cite{design-compilier}, during which is clock-gating technique is turned on. The technology library utilized in our implementation is 40nm standard cell library. We also implement the SRAM in the processor using the Memory Compiler of the 40nm technology library.

\textbf{Ground-Truth Power Collection:} For a configuration executing a workload, we perform the standard power evaluation flow to collect the ground-truth power. We perform RTL simulation with Synopsys VCS\textsuperscript{\textregistered}~\cite{vcs}. 
%We perform logic synthesis with Synopsis Design Compiler\textsuperscript{\textregistered}~\cite{design-compilier}, during which is clock-gating technique is turned on. 
We perform post-synthesis power simulation with PrimePower~\cite{ptpx} and collect the power data as the label. 
% The technology library utilized in our VLSI flow is a commercial library, TSMC 40nm~\cite{URL:tmsc40nm} standard cell library. We also implement the SRAM in the processor by ourselves using the Memory Compiler of the TSMC 40nm technology library. 
% Through the VLSI power evaluation flow, we extract ground-truth power from the power report generated by the PrimePower. 
% Our scripts for RTL code generation and RTL simulation are available in ArchPower.

\textbf{Advanced Technology Library:} 
Besides the primary 40nm technology library on which our experiment is performed in this paper, we also provide an additional dataset collected with a 28nm technology library. This additional dataset has the same organization as our primary 40nm dataset, and can be adopted for benchmarking with the same benchmark framework. In the future, if we get access to any high-quality FinFET technology library in the future, we will update our dataset and provide the data with the FinFET technology library.
\section{Experiments}

\subsection{Benchmarked Models}

There are many ML-based architecture-level power modeling works~\cite{zhai2022mcpat,zhai2023microarchitecture,zhang2023panda,zhang2025firepower}. 
We benchmark two representative existing ML-based architecture-level power models, McPAT-Calib~\cite{zhai2022mcpat} and PANDA~\cite{zhang2023panda}, based on ArchPower. We also derive two ML-based power models, McPAT-Calib-Component and McPAT-Calib-CompGroup, based on McPAT-Calib, utilizing fine-grained component-level and power-group-level power labels. We also evaluate the classical analytical power models for comparison, including McPAT~\cite{li2009mcpat} and our enhanced version, McPAT-Plus.

We describe our 6 evaluated power models below. 
(a) McPAT~\cite{li2009mcpat}: A widely adopted analytical power model. Its input includes the hardware parameter and the raw event statistics. Therefore, it can also be evaluated based on our dataset.
(b) McPAT-Plus: An enhanced version of McPAT. It fits a scaling factor on training data, and then scales the output of McPAT when testing. 
(c) McPAT-Calib~\cite{zhai2022mcpat}: It utilizes an ML model, XGBoost~\cite{chen2016xgboost}, to learn the correlation between the input feature and the final total power label.
(d) McPAT-Calib-Component: An enhanced version of McPAT-Calib. It builds one ML model for each component based on per-component power labels. The per-component power predictions are summed up for total power. %When predicting the total power, it predicts per-component power respectively and sums them up.
(e) McPAT-Calib-CompGroup: It is derived from the McPAT-Calib-Component, building one ML model for each group of each component based on per-power-group power labels. When predicting the total power, it predicts per-component power respectively and sums them up.
(f) PANDA~\cite{zhang2023panda}: It adopts resource functions to capture the major correlation between hardware parameters and the power of each component, and multiplies it by the ML model for the final power prediction. 
We adopt the mean absolute percentage error (MAPE) and the correlation coefficient $R$ between label and prediction to evaluate the power modeling accuracy of the ML-based architecture-level power model.

\begin{table*}[!t]
\centering
      %\vspace{-.35in}
    %  \renewcommand{\arraystretch}{1.05}
      \resizebox{1\textwidth}{!}{
\begin{tabular}{|c|cccc|cccc|cccc|}
\hline
\multirow{3}{*}{Scenario}         & \multicolumn{4}{c|}{McPAT}                               & \multicolumn{4}{c|}{McPAT-Plus}                       & \multicolumn{4}{c|}{McPAT-Calib}                                                                   \\ \cline{2-13} 
         & \multicolumn{2}{c|}{BOOM}     & \multicolumn{2}{c|}{XiangShan} & \multicolumn{2}{c|}{BOOM}       & \multicolumn{2}{c|}{XiangShan} & \multicolumn{2}{c|}{BOOM}     & \multicolumn{2}{c|}{XiangShan}  \\
 & MAPE & \multicolumn{1}{c|}{R} & MAPE            & R            & MAPE  & \multicolumn{1}{c|}{R}  & MAPE            & R            & MAPE & \multicolumn{1}{c|}{R} & MAPE            & R                        \\ \hline \hline
Balance & >100 & 0.83 & >100 & 0.85 & 18.1 & 0.83 & 29.6 & 0.85 & 8.2  & 0.98 & 33.2 & 0.73  \\
Small   & >100 & 0.74 & >100 & 0.77 & 31.0 & 0.74 & 21.6 & 0.77 & 34.3 & 0.76 & 41.5 & 0.48 \\
Large   & >100 & 0.83 & >100 & 0.78 & 28.2 & 0.83 & 28.3 & 0.78 & 50.6 & 0.23 & 90.0 & 0.14 \\ 
\hline \hline
Average & >100 & 0.80 & >100 & 0.80 & 25.8 & 0.80 & 26.5 & 0.80 & 31.0 & 0.66 & 54.9 & 0.45 \\ 
\hline \hline
\multirow{3}{*}{Scenario}         & \multicolumn{4}{c|}{McPAT-Calib-Component}                               & \multicolumn{4}{c|}{McPAT-Calib-CompGroup}                       & \multicolumn{4}{c|}{PANDA}                                                                   \\ \cline{2-13} 
         & \multicolumn{2}{c|}{BOOM}     & \multicolumn{2}{c|}{XiangShan} & \multicolumn{2}{c|}{BOOM}       & \multicolumn{2}{c|}{XiangShan} & \multicolumn{2}{c|}{BOOM}     & \multicolumn{2}{c|}{XiangShan}  \\
 & MAPE & \multicolumn{1}{c|}{R} & MAPE            & R            & MAPE  & \multicolumn{1}{c|}{R}  & MAPE            & R            & MAPE & \multicolumn{1}{c|}{R} & MAPE            & R                        \\ \hline \hline
Balance & 6.2  & 0.98 & \textbf{14.0} & 0.97 & \textbf{6.2}  & \textbf{0.98} & 15.0 & \textbf{0.97} & 6.8  & 0.97 & 19.4 & 0.9 \\
Small   & 34.9 & 0.75 & 35.4 & 0.72 & 35.3 & 0.75 & 36.0 & 0.72 & \textbf{29.2} & \textbf{0.93} & \textbf{23.9} & \textbf{0.86} \\
Large   & 48.9 & 0.4  & 81.4 & 0.31 & 49.2 & 0.4  & 80.5 & 0.35 & \textbf{10.4} & \textbf{0.98} & \textbf{26.3} & \textbf{0.82} \\
\hline \hline
Average & 30.0 & 0.71 & 43.6 & 0.67 & 30.2 & 0.71 & 43.8 & 0.68 & \textbf{15.5} & \textbf{0.96} & \textbf{23.2} & \textbf{0.86} \\ 
\hline
\end{tabular}
}
    %   \vspace{-.08in}
        %\vspace{-1mm}
        \vspace{-.05in}
        \caption{Comparison between different architecture-level power models for total power prediction under different training scenarios. All MAPE values are reported as percentages. The best accuracies for each scenario are highlighted in bold. }
        \vspace{-.1in}
        \label{expresult}
\end{table*}

\begin{table*}[!t]
\centering
      %\vspace{-.35in}
    %  \renewcommand{\arraystretch}{1.05}
      \resizebox{1\textwidth}{!}{
\begin{tabular}{|c|cccc|cccc|cccc|}
\hline
\multirow{3}{*}{Component}         & \multicolumn{4}{c|}{McPAT}                               & \multicolumn{4}{c|}{McPAT-Plus}                       & \multicolumn{4}{c|}{McPAT-Calib}                                                                   \\ \cline{2-13} 
         & \multicolumn{2}{c|}{BOOM}       & \multicolumn{2}{c|}{XiangShan} & \multicolumn{2}{c|}{BOOM}     & \multicolumn{2}{c|}{XiangShan} & \multicolumn{2}{c|}{BOOM}     & \multicolumn{2}{c|}{XiangShan} \\
 & MAPE  & \multicolumn{1}{c|}{R}  & MAPE            & R            & MAPE & \multicolumn{1}{c|}{R} & MAPE            & R            & MAPE & \multicolumn{1}{c|}{R} & MAPE            & R            \\ \hline \hline
BP          & 67.5 & 0.37 & 55.4 & 0.78 & 96.6 & 0.37 & 90.0 & 0.78 & - & - & - & - \\
ICache      & 39.5 & 0.50 & 83.7 & 0.48 & 91.1 & 0.50 & 96.3 & 0.48 & - & - & - & - \\
IFU         & >100 & 0.35 & >100 & 0.78 & 41.2 & 0.35 & 66.6 & 0.78 & - & - & - & - \\
RNU         & >100 & 0.90 & >100 & 0.86 & >100 & \textbf{0.90} & >100 & 0.86 & - & - & - & - \\
ROB         & >100 & 0.65 & >100 & 0.94 & 44.1 & \textbf{0.65} & 67.0 & \textbf{0.94} & - & - & - & - \\
ISU         & >100 & 0.87 & 81.5 & 0.91 & 32.2 & 0.87 & 59.2 & 0.91 & - & - & - & - \\
Regfile     & >100 & 0.73 & \textbf{49.6} & 0.81 & 43.3 & 0.73 & 70.9 & 0.81 & - & - & - & - \\
FU Pool     & >100 & 0.51 & 67.9 & 0.53 & 45.0 & 0.51 & 92.8 & 0.53 & - & - & - & - \\
LSU         & >100 & 0.17 & >100 & 0.84 & >100 & 0.17 & >100 & 0.84 & - & - & - & - \\
DCache      & >100 & 0.71 & >100 & 0.90 & 46.8 & 0.71 & 57.1 & 0.90 & - & - & - & - \\
Other Logic & >100 & 0.70 & >100 & <0   & >100 & \textbf{0.70} & >100 & <0   & - & - & - & - \\   \hline \hline
\multirow{3}{*}{Component}         & \multicolumn{4}{c|}{McPAT-Calib-Component}                               & \multicolumn{4}{c|}{McPAT-Calib-CompGroup}                       & \multicolumn{4}{c|}{PANDA}                                                                   \\ \cline{2-13} 
         & \multicolumn{2}{c|}{BOOM}       & \multicolumn{2}{c|}{XiangShan} & \multicolumn{2}{c|}{BOOM}     & \multicolumn{2}{c|}{XiangShan} & \multicolumn{2}{c|}{BOOM}     & \multicolumn{2}{c|}{XiangShan} \\
 & MAPE  & \multicolumn{1}{c|}{R}  & MAPE            & R            & MAPE & \multicolumn{1}{c|}{R} & MAPE            & R            & MAPE & \multicolumn{1}{c|}{R} & MAPE            & R            \\ \hline \hline
BP          & \textbf{1.1}  & 1.00 & \textbf{12.4} & 0.90  & 1.2  & \textbf{1.00} & 12.6 & \textbf{0.90} & 1.6  & 0.99 & 24.5 & 0.78\\
ICache      & 18.7 & 0.97 & \textbf{36.2} & \textbf{0.92}  & 18.2 & 0.97 & 36.3 & 0.92 & \textbf{2.2}  & \textbf{1.00} & 52.8 & 0.77\\
IFU         & 14.2 & 0.42 & 16.9 & 0.86  & \textbf{13.1} & \textbf{0.48} & 16.6 & 0.87 & 36.3 & <0   & \textbf{15.5} & \textbf{0.90}\\
RNU         & 45.0 & 0.68 & 15.0 & 0.92  & 56.8 & 0.64 & 15.3 & \textbf{0.94} & \textbf{43.1} & 0.79 & \textbf{14.7} & 0.86\\
ROB         & 34.3 & 0.63 & 19.4 & 0.90  & \textbf{34.1} & 0.62 & \textbf{17.6} & 0.90 & 52.0 & 0.56 & 19.7 & 0.93\\
ISU         & 24.6 & 0.91 & \textbf{38.1} & 0.81  & 24.9 & \textbf{0.91} & 51.2 & 0.78 & \textbf{21.9} & 0.84 & 43.8 & \textbf{0.93}\\
Regfile     & \textbf{22.0} & 0.84 & 60.6 & 0.72  & 22.2 & \textbf{0.88} & 60.2 & 0.72 & 37.5 & 0.68 & 83.3 & \textbf{0.94}\\
FU Pool     & 10.5 & 0.94 & 7.3  & 0.97  & \textbf{10.0} & \textbf{0.94} & \textbf{7.2}  & \textbf{0.97} & 10.5 & 0.94 & 7.3  & 0.97\\
LSU         & >100 & <0   & 13.0 & 0.86  & >100 & <0   & 12.7 & 0.88 & \textbf{97.8} & \textbf{0.17} & \textbf{11.0} & \textbf{0.88}\\
DCache      & 23.7 & 0.87 & 23.0 & 0.92  & 21.6 & 0.89 & 22.1 & 0.92 & \textbf{16.9} & \textbf{0.96} & \textbf{16.1} & \textbf{0.95}\\
Other Logic & 28.6 & 0.44 & >100 & 0.32  & \textbf{28.1} & 0.44 & \textbf{>100} & \textbf{0.87} & 44.1 & 0.34 & >100 & 0.54 \\ \hline
\end{tabular}
}
    %   \vspace{-.08in}
        %\vspace{-1mm}
        \vspace{-.05in}
        \caption{Comparison between different architecture-level power models for per-component power prediction under the \emph{Balance} training scenario. All MAPE values are reported as percentages. McPAT-Calib is excluded because it does not provide per-component power information. The best accuracies for each scenario are highlighted in bold. }
        \vspace{-.2in}
        \label{expresultcomp}
\end{table*}

\subsection{Power Prediction Accuracy}

\subsubsection{Total Power Prediction}

Table~\ref{expresult} shows the accuracy comparison for total power prediction between our selected six models under different training scenarios. It shows that even the enhanced version of McPAT, McPAT-Plus, can not achieve a high accuracy, with MAPE over 15\% and correlation coefficient $R$ lower than 0.85 on all evaluations. In comparison, the ML-based architecture-level power model, McPAT-Calib, can achieve a high accuracy in the \emph{Balance} training scenarios on BOOM. The enhanced versions of McPAT-Calib, including McPAT-Calib-Component and McPAT-Calib-CompGroup, and the advanced model PANDA can further improve the accuracy. However, in the training scenarios \emph{Small} and \emph{Large}, where the testing data falls out of the training data distribution, the accuracy of the existing ML-based architecture-level power model drops dramatically. 
It shows that our dataset can provide a comprehensive evaluation for ML-based architecture-level power models, supporting different training scenarios that can evaluate the generalization ability of models. A comprehensive evaluation demonstrates that the generalization of ML models still needs to be improved in future research.

% It shows that the reliability of ML-based models still needs to be improved in future research. 

% Fig.~\ref{visboom} and \ref{visxs} visualize the prediction of different models under the \emph{Balance} training scenario, where each point represents a sample and points in the same color are from the same configuration. The visualization gives a clearer comparison between different models.  

\subsubsection{Per-Component Power Prediction}

Table~\ref{expresultcomp} shows the per-component prediction accuracy of different power models. McPAT-Calib can not provide valid values because it directly predicts the final total power and does not provide per-component power information. It demonstrates that, besides the total power prediction, the ML-based architecture-level power model can also achieve higher accuracy compared with the traditional models for most of the components from BOOM and XiangShan. 

However, it also demonstrates that ML-based architecture-level power models may also have a negative effect on some components. For example, for the RNU of BOOM CPU, the correlation coefficient $R$ of McPAT-Calib-CompGroup drops to 0.64 compared with analytical models that can achieve 0.90. 
It shows that our dataset can enable fine-grained evaluation for ML-based architecture-level power models and provide detailed information about the model accuracy, which shows the limitation of the existing ML-based architecture-level power models, driving potential research to improve these cases.

% It shows the limitation of the existing ML-based architecture-level power models, driving potential research to improve these cases.

\begin{table*}[!t]
\centering
      \resizebox{0.63\textwidth}{!}{
\begin{tabular}{|c|cccc|}
\hline
\multirow{2}{*}{Model} & \multicolumn{2}{c|}{BOOM} & \multicolumn{2}{c|}{XiangShan} \\ 
% \cline{2-5}
& MAPE & \multicolumn{1}{c|}{R} & MAPE & R \\
\hline
\hline
McPAT & 771 & 0.83 & 427 & 0.86 \\
McPAT-Plus & 18.3 & 0.83 & 22.5 & 0.84 \\
McPAT-Calib & \textbf{5.6} & 0.96 & \textbf{10.0} & \textbf{0.95} \\
McPAT-Calib-Comp & 6.2 & 0.95 & 11.5 & 0.94 \\
McPAT-Calib-CompGroup & 6.1 & \textbf{0.96} & 11.3 & 0.94 \\
PANDA & 7.2 & 0.95 & 11.6 & 0.92 \\
\hline
\end{tabular}
}
        \vspace{-.05in}
        \caption{Comparison between different architecture-level power models for cross-workload prediction. All MAPE values are reported as percentages. The best accuracies are highlighted in bold.}
        \vspace{-.2in}
        \label{crossworkload}
\end{table*}

\subsubsection{Cross-Workload Power Prediction}

Table~\ref{crossworkload} provides our experimental results that split the training and testing scenarios based on workloads, where we adopt 8-fold cross-validation for training-testing splitting, i.e., 7 workloads for training and 1 workload for testing. The experimental results show that the ML-based power models can also demonstrate advantages over the traditional analytical model in the cross-workload scenario. It indicates that ML-based power models have great potential.
\section{Conclusion}

% In this paper, we present ArchPower, the first open-source dataset for ML-based architecture-level power models. 
% ArchPower consists of data collected on 15 configurations of the BOOM CPU and 10 configurations of the XiangShan CPU, with each configuration running 8 representative real-world workloads. With ArchPower, ML-based architecture-level power models can be easily evaluated under different training scenarios. 
% We hope ArchPower can contribute to the further development of the ML-based architecture-level processor power model. 

In this paper, we present ArchPower, the first open-source dataset for ML-based architecture-level power models. 
ArchPower includes 200 data samples collected from 25 CPU configurations and 8 workloads. We consider the clock-gating and integrate realistic SRAM macros for power label collection. 
ArchPower allows anyone to easily replicate and further improve existing architecture-level power models. We expect ArchPower to reduce the hardware barrier and enable more brilliant AI solutions in hardware design and optimizations. 

\section*{Acknowledgement}
% \vspace{-.02in}
This work is supported by National Natural Science Foundation of China (NSFC) 62304192, Hong Kong Research Grants Council (RGC) YCRG Grant C6003-24Y, ECS Grant 26208723, and ACCESS – AI Chip Center for Emerging Smart Systems, supported by the InnoHK initiative of Innovation and Technology Commission of the Hong Kong Special Administrative Region Government.
% \vspace{-.05in}

\bibliographystyle{plain}
\bibliography{references_1}
%\bibliography{references_1}

% \newpage
\section*{NeurIPS Paper Checklist}

\begin{enumerate}

\item {\bf Claims}
    \item[] Question: Do the main claims made in the abstract and introduction accurately reflect the paper's contributions and scope?
    \item[] Answer: \answerYes{} % Replace by \answerYes{}, \answerNo{}, or \answerNA{}.
    \item[] Justification: We provide the first dataset for ML-based architecture-level power modeling.

\item {\bf Limitations}
    \item[] Question: Does the paper discuss the limitations of the work performed by the authors?
    \item[] Answer: \answerYes{} % Replace by \answerYes{}, \answerNo{}, or \answerNA{}.
    \item[] Justification: See Appendix~\ref{limitation}.

\item {\bf Theory assumptions and proofs}
    \item[] Question: For each theoretical result, does the paper provide the full set of assumptions and a complete (and correct) proof?
    \item[] Answer: \answerNA{} % Replace by \answerYes{}, \answerNo{}, or \answerNA{}.
    \item[] Justification: This work focuses on the introduction of our provided dataset.

    \item {\bf Experimental result reproducibility}
    \item[] Question: Does the paper fully disclose all the information needed to reproduce the main experimental results of the paper to the extent that it affects the main claims and/or conclusions of the paper (regardless of whether the code and data are provided or not)?
    \item[] Answer: \answerYes{} % Replace by \answerYes{}, \answerNo{}, or \answerNA{}.
    \item[] Justification: Information needed to reproduce the main experimental results can be found in https://github.com/hkust-zhiyao/ArchPower.

\item {\bf Open access to data and code}
    \item[] Question: Does the paper provide open access to the data and code, with sufficient instructions to faithfully reproduce the main experimental results, as described in supplemental material?
    \item[] Answer: \answerYes{} % Replace by \answerYes{}, \answerNo{}, or \answerNA{}.
    \item[] Justification: Data and code can be found in https://github.com/hkust-zhiyao/ArchPower.

\item {\bf Experimental setting/details}
    \item[] Question: Does the paper specify all the training and test details (e.g., data splits, hyperparameters, how they were chosen, type of optimizer, etc.) necessary to understand the results?
    \item[] Answer: \answerYes{} % Replace by \answerYes{}, \answerNo{}, or \answerNA{}.
    \item[] Justification: See Section~\ref{sec:setup}.

\item {\bf Experiment statistical significance}
    \item[] Question: Does the paper report error bars suitably and correctly defined or other appropriate information about the statistical significance of the experiments?
    \item[] Answer: \answerNo{} % Replace by \answerYes{}, \answerNo{}, or \answerNA{}.
    \item[] Justification: We evaluate models under different training scenarios explicitly and report the results for each training scenario. Besides, the reported result is the average across multiple runnings.

\item {\bf Experiments compute resources}
    \item[] Question: For each experiment, does the paper provide sufficient information on the computer resources (type of compute workers, memory, time of execution) needed to reproduce the experiments?
    \item[] Answer: \answerYes{} % Replace by \answerYes{}, \answerNo{}, or \answerNA{}.
    \item[] Justification: See Appendix~\ref{compres}.
    
\item {\bf Code of ethics}
    \item[] Question: Does the research conducted in the paper conform, in every respect, with the NeurIPS Code of Ethics \url{https://neurips.cc/public/EthicsGuidelines}?
    \item[] Answer: \answerYes{} % Replace by \answerYes{}, \answerNo{}, or \answerNA{}.
    \item[] Justification: Our research conforms with the NeurIPS Code of Ethics.

\item {\bf Broader impacts}
    \item[] Question: Does the paper discuss both potential positive societal impacts and negative societal impacts of the work performed?
    \item[] Answer: \answerNA{} % Replace by \answerYes{}, \answerNo{}, or \answerNA{}.
    \item[] Justification: The application targeted in our research is purely technical and therefore has no societal impact.
    
\item {\bf Safeguards}
    \item[] Question: Does the paper describe safeguards that have been put in place for responsible release of data or models that have a high risk for misuse (e.g., pretrained language models, image generators, or scraped datasets)?
    \item[] Answer: \answerNA{} % Replace by \answerYes{}, \answerNo{}, or \answerNA{}.
    \item[] Justification: Our paper poses no such risks.

\item {\bf Licenses for existing assets}
    \item[] Question: Are the creators or original owners of assets (e.g., code, data, models), used in the paper, properly credited and are the license and terms of use explicitly mentioned and properly respected?
    \item[] Answer: \answerYes{} % Replace by \answerYes{}, \answerNo{}, or \answerNA{}.
    \item[] Justification: See Appendix~\ref{license}.

\item {\bf New assets}
    \item[] Question: Are new assets introduced in the paper well documented and is the documentation provided alongside the assets?
    \item[] Answer: \answerYes{} % Replace by \answerYes{}, \answerNo{}, or \answerNA{}.
    \item[] Justification: See https://github.com/hkust-zhiyao/ArchPower.

\item {\bf Crowdsourcing and research with human subjects}
    \item[] Question: For crowdsourcing experiments and research with human subjects, does the paper include the full text of instructions given to participants and screenshots, if applicable, as well as details about compensation (if any)? 
    \item[] Answer: \answerNA{} % Replace by \answerYes{}, \answerNo{}, or \answerNA{}.
    \item[] Justification: Our paper does not involve crowdsourcing nor research with human subjects.

\item {\bf Institutional review board (IRB) approvals or equivalent for research with human subjects}
    \item[] Question: Does the paper describe potential risks incurred by study participants, whether such risks were disclosed to the subjects, and whether Institutional Review Board (IRB) approvals (or an equivalent approval/review based on the requirements of your country or institution) were obtained?
    \item[] Answer: \answerNA{} % Replace by \answerYes{}, \answerNo{}, or \answerNA{}.
    \item[] Justification: Our paper does not involve crowdsourcing nor research with human subjects.

\item {\bf Declaration of LLM usage}
    \item[] Question: Does the paper describe the usage of LLMs if it is an important, original, or non-standard component of the core methods in this research? Note that if the LLM is used only for writing, editing, or formatting purposes and does not impact the core methodology, scientific rigorousness, or originality of the research, declaration is not required.
    %this research? 
    \item[] Answer: \answerNA{} % Replace by \answerYes{}, \answerNo{}, or \answerNA{}.
    \item[] Justification: The core method development in this research does not involve LLMs as any important, original, or non-standard components.

\end{enumerate}

\newpage
\appendix

\section{More on Dataset and Evaluation Setting}

\subsection{Adopted CPU Configurations}

In our dataset, we adopt 25 CPU configurations in total, including 15 configurations of BOOM CPU and 10 configurations of XiangShan CPU. 
Table~\ref{boomconfigtable} and \ref{xsconfigtable} list all 25 CPU configurations adopted in our dataset.

\begin{table*}[!h]
\centering
      %\vspace{-.35in}
    %  \renewcommand{\arraystretch}{1.05}
      \resizebox{1\textwidth}{!}{
        \begin{tabular}{ |c||c c c c c c c c c c c c c c c| } 
\hline
Hardware Parameter  & B1 & B2 & B3 & B4 & B5 & B6 & B7 & B8 & B9 & B10 & B11 & B12 & B13 & B14 & B15 \\
\hline
\hline
FetchWidth & 4 & 4 & 4 & 4 & 4 & 8 & 8 & 8 & 8 & 8 & 8 & 8 & 8 & 8 & 8 \\
\hline
DecodeWidth & 1 & 1 & 1 & 2 & 2 & 2 & 3 & 3 & 3 & 4 & 4 & 4 & 5 & 5 & 5 \\
\hline
FetchBufferEntry & 5 & 8 & 16 & 8 & 16 & 24 & 18 & 24 & 30 & 24 & 32 & 40 & 30 & 35 & 40 \\
\hline
RobEntry & 16 & 32 & 48 & 64 & 64 & 80 & 81 & 96 & 114 & 112 & 128 & 136 & 125 & 130 & 140 \\
\hline
IntPhyRegister & 36 & 53 & 68 & 64 & 80 & 88 & 88 & 110 & 112 & 108 & 128 & 136 & 108 & 128 & 140 \\
\hline
FpPhyRegister & 36 & 48 & 56 & 56 & 64 & 72 & 88 & 96 & 112 & 108 & 128 & 136 & 108 & 128 & 140 \\
\hline
LDQ/STQEntry & 4 & 8 & 16 & 12 & 16 & 20 & 16 & 24 & 32 & 24 & 32 & 36 & 24 & 32 & 36 \\
\hline
BranchCount & 6 & 8 & 10 & 10 & 12 & 14 & 14 & 16 & 16 & 18 & 20 & 20 & 18 & 20 & 20 \\
\hline
Mem/FpIssueWidth & 1 & 1 & 1 & 1 & 1 & 1 & 1 & 1 & 2 & 1 & 2 & 2 & 2 & 2 & 2 \\
\hline
IntIssueWidth & 1 & 1 & 1 & 1 & 2 & 2 & 2 & 3 & 3 & 4 & 4 & 4 & 5 & 5 & 5 \\
\hline
DCache/ICacheWay & 2 & 4 & 8 & 4 & 4 & 8 & 8 & 8 & 8 & 8 & 8 & 8 & 8 & 8 & 8 \\
\hline
DTLBEntry & 8 & 8 & 16 & 8 & 8 & 16 & 16 & 16 & 32 & 32 & 32 & 32 & 32 & 32 & 32 \\
\hline
MSHREntry & 2 & 2 & 4 & 2 & 2 & 4 & 4 & 4 & 4 & 4 & 4 & 8 & 8 & 8 & 8 \\
\hline
ICacheFetchBytes & 2 & 2 & 2 & 2 & 2 & 4 & 4 & 4 & 4 & 4 & 4 & 4 & 4 & 4 & 4 \\
         \hline
        \end{tabular}
        }
    %   \vspace{-.08in}
        %\vspace{-1mm}
        \caption{The BOOM configurations adopted in our dataset, named B1-B15. The scales of these configurations are from small to large.}
        %\vspace{-5mm}
        \label{boomconfigtable}
\end{table*}

%\subsubsection{RISC-V XiangShan CPU}

\begin{table*}[!h]
\centering
      %\vspace{-.35in}
    %  \renewcommand{\arraystretch}{1.05}
      \resizebox{0.7\textwidth}{!}{
        \begin{tabular}{ |c||c c c c c c c c c c| } 
\hline
Hardware Parameter   & X1 & X2 & X3 & X4 & X5 & X6 & X7 & X8 & X9 & X10\\
\hline
\hline
FetchWidth &       4 & 4 & 4 & 4 & 4 & 8 & 8 & 8 & 8 & 8\\
\hline
DecodeWidth &      2 & 2 & 2 & 3 & 3 & 3 & 4 & 4 & 4 & 5\\
\hline
FetchBufferEntry &     8 & 16 & 24 & 16 & 24 & 24 & 24 & 32 & 32 & 24\\
\hline
RobEntry &      16 & 32 & 48 & 64 & 64 & 80 & 81 & 96 & 114 & 112\\
\hline
IntPhyRegister &      36 & 53 & 68 & 64 & 80 & 88 & 88 & 110 & 112 & 108\\
\hline
FpPhyRegister &      36 & 53 & 68 & 64 & 80 & 88 & 88 & 110 & 112 & 108\\
\hline
LDQ/STQEntry &      16 & 20 & 24 & 20 & 24 & 28 & 24 & 32 & 40 & 32\\
\hline
BranchCount &      7 & 7 & 7 & 7 & 7 & 7 & 7 & 7 & 7 & 7\\
\hline
Mem/FpIssueWidth &      2 & 2 & 2 & 2 & 2 & 2 & 2 & 2 & 2 & 2\\
\hline
IntIssueWidth &      2 & 2 & 2 & 2 & 4 & 4 & 4 & 6 & 6 & 6\\
\hline
DCache/ICacheWay &      4 & 4 & 8 & 4 & 4 & 8 & 8 & 8 & 8 & 8\\
\hline
DTLBEntry &      8 & 8 & 16 & 8 & 8 & 16 & 16 & 16 & 32 & 32\\
\hline
MSHREntry &      2 & 2 & 4 & 2 & 2 & 4 & 4 & 4 & 4 & 4\\
\hline
ICacheFetchBytes &      2 & 2 & 2 & 2 & 2 & 2 & 2 & 2 & 2 & 2\\
         \hline
        \end{tabular}
        }
    %   \vspace{-.08in}
        %\vspace{-1mm}
        \caption{The XiangShan configurations adopted in our dataset, named X1-X10. The scales of these configurations are from small to large.}
        %\vspace{-5mm}
        \label{xsconfigtable}
\end{table*}

\subsection{Evaluation Metrics}
We adopt the mean absolute percentage error (MAPE) %, denoted as Eq.(\ref{eq:mape}), 
and the correlation coefficient $R$, %denoted as Eq.(\ref{eq:r}), 
between label $Y_i$ and prediction $\hat{Y}_i$ to evaluate the power modeling accuracy of the ML-based architecture-level power model, as shown in Eq.(\ref{eq:metrics}). 
\begin{equation}
    {MAPE} = \frac{1}{n} \sum_{i=1}^{n} \left| \frac{Y_i - \hat{Y}_i}{Y_i} \right| \times 100\%
    ,\,\,\,\,
    R = \frac{\sum_{i=1}^{n} (\hat{Y}_i - \bar{\hat{Y}})(Y_i - \bar{Y})}{\sqrt{\sum_{i=1}^{n} (\hat{Y}_i - \bar{\hat{Y}})^2 \sum_{i=1}^{n} (Y_i - \bar{Y})^2}}
    \label{eq:metrics}
\end{equation}
% \begin{equation}
%     R = \frac{\sum_{i=1}^{n} (\hat{Y}_i - \bar{\hat{Y}})(Y_i - \bar{Y})}{\sqrt{\sum_{i=1}^{n} (\hat{Y}_i - \bar{\hat{Y}})^2 \sum_{i=1}^{n} (Y_i - \bar{Y})^2}}
%     \label{eq:r}
% \end{equation}

\subsection{Compute Resources}
\label{compres}

We perform our experiments on a server with Intel\textsuperscript{\textregistered} Xeon\textsuperscript{\textregistered} Gold 6438Y+ processor. The model evaluation is fast and efficient, taking less than one minute for each model. The memory requirement is within 10MB. Reproducing all of our results takes less than ten minutes.

\subsection{Licenses}
\label{license}

Chipyard framework and BOOM CPU are under BSD-3-Clause. OpenXiangShan framework and XiangShan CPU are under Mulan PSL v2. Riscv-tests is under BSD-3-Clause.

\section{Limitations and Future Work}
\label{limitation}

While ArchPower provides the first dataset for the ML-based architecture-level power models, there are still some limitations that can be improved in future work: 1) Due to the difficulty of RTL code collection for the CPU, the diversity of architectures and the number of configurations are limited in size. Now there are only two CPU architectures with 25 configurations in our dataset. 2) The real-world single-thread workloads provided in riscv-tests are limited. Therefore, now we only include 8 workloads in our dataset for each CPU configuration. 

For future work, we will have follow-up updates to ArchPower to address the two limitations above: 1) We will continue to collect new CPU architectures and provide more configurations to improve the scale of our dataset. 2) We will collect or write more real-world workloads to improve the workload diversity in our dataset.

\section{Result Visualization}
% Fig.~\ref{visboomA} and \ref{visxsA} visualize the prediction of different models on BOOM and XiangShan under the \emph{Balance} training scenario. 

This section visualizes the prediction of different models on BOOM and XiangShan under different training scenarios. Each point represents a sample, and points in the same color are from the same configuration. The visualization gives a clearer comparison between different models.  

\newpage

\begin{figure*}[!h]
\centering
%\vspace{-.2in}

\hspace{-5mm}
\subfigure[McPAT]{
    \centering
    \includegraphics[height=0.25\textwidth]{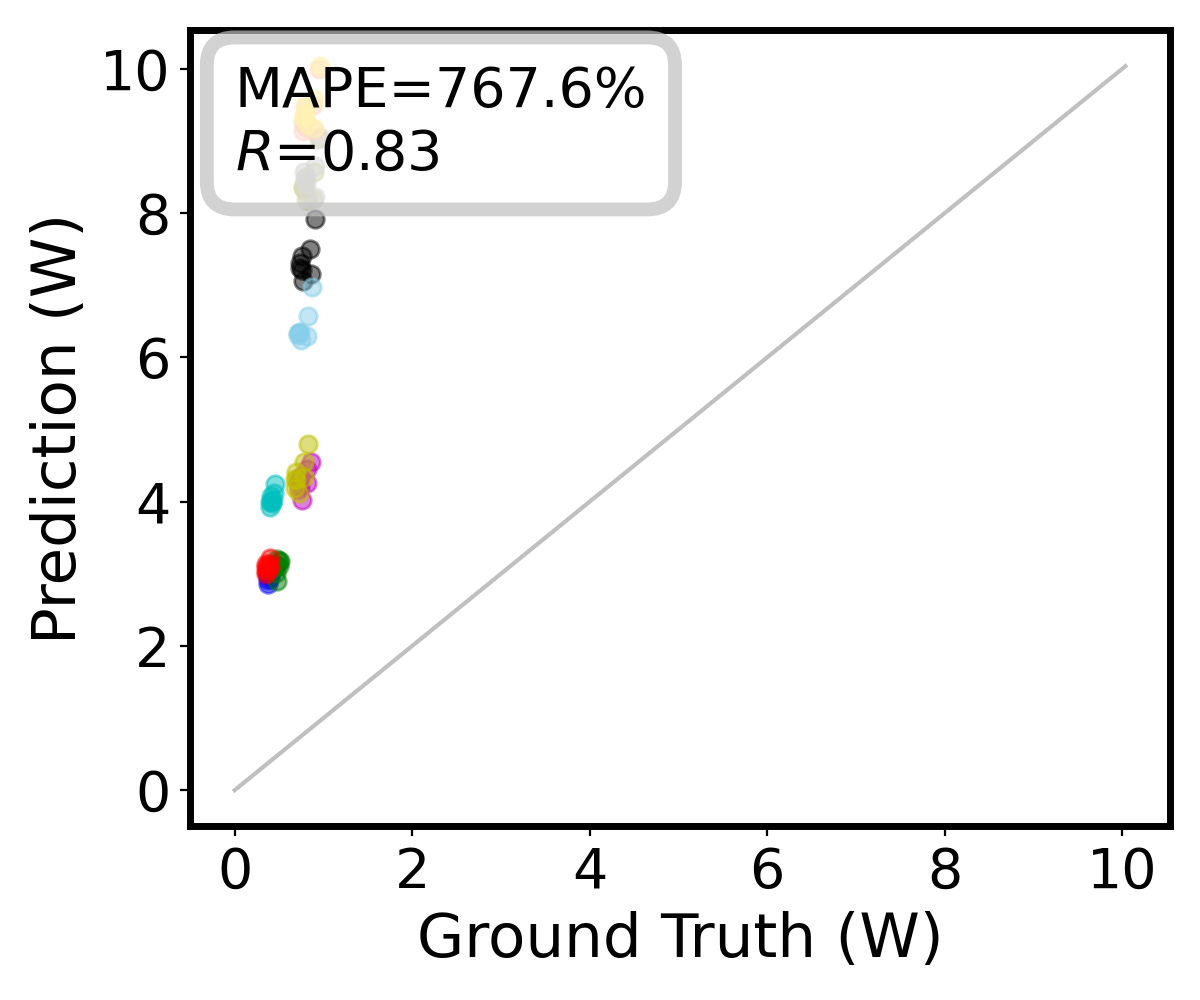}
    %\label{McPAT}
}
\hspace{-3mm}
\subfigure[McPAT-Plus]{
    \centering
    \includegraphics[height=0.25\textwidth]{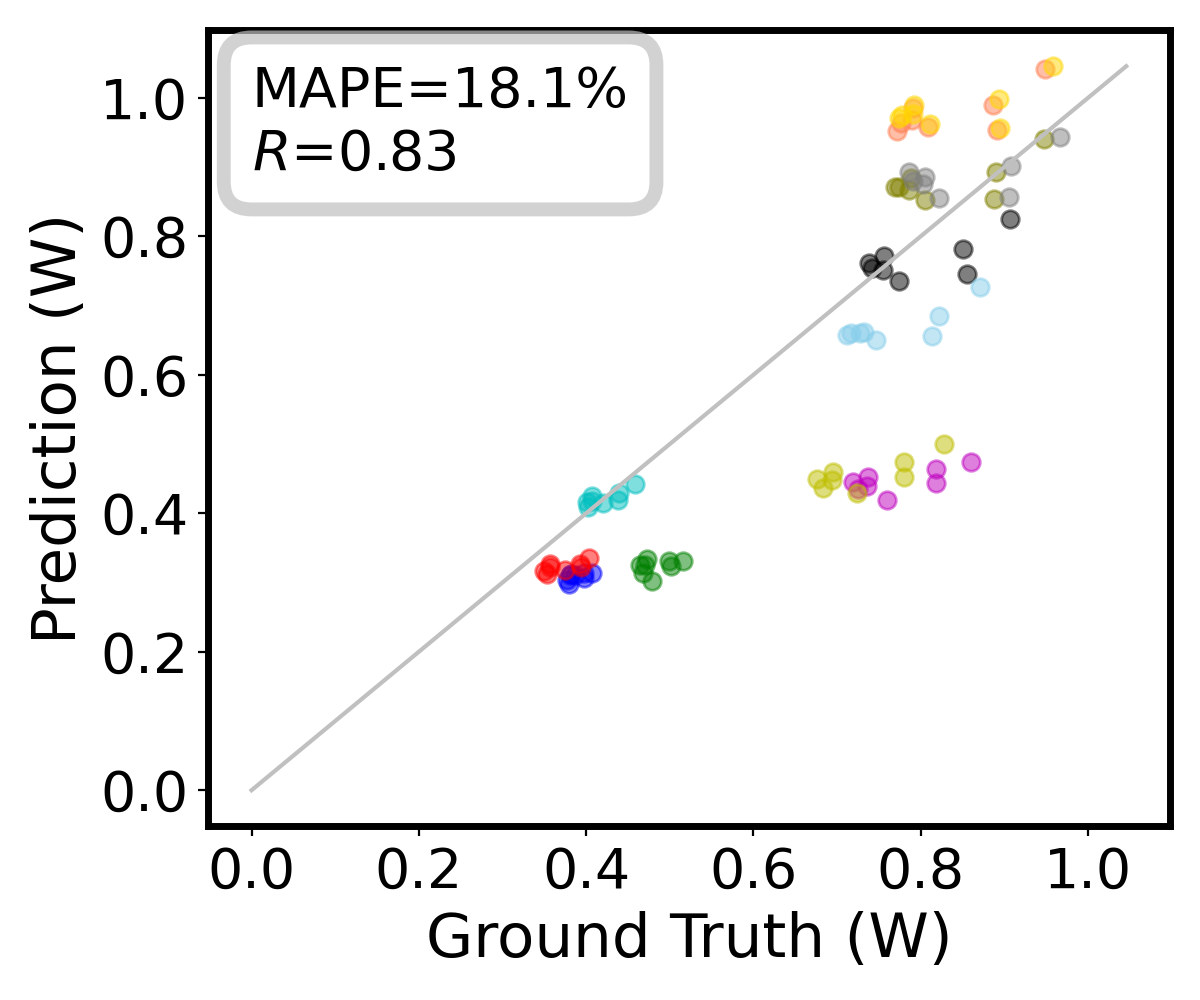}
    %\label{McPAT-Plus}
}
\hspace{-3mm}
\subfigure[McPAT-Calib]{
    \centering
    \includegraphics[height=0.25\textwidth]{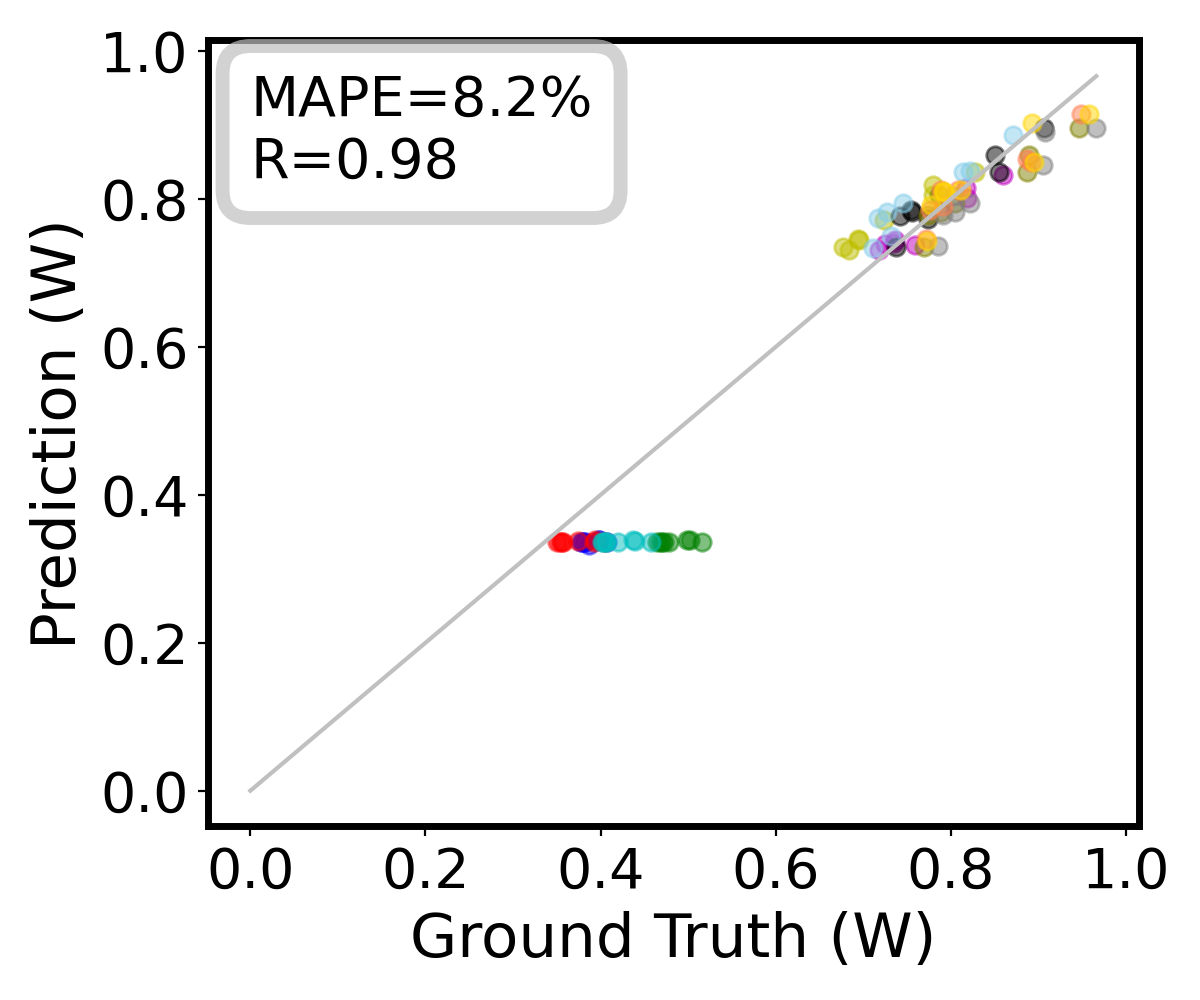}
    %\label{archbp}
}

\hspace{-5mm}
\subfigure[McPAT-Calib-Component]{
    \centering
    \includegraphics[height=0.25\textwidth]{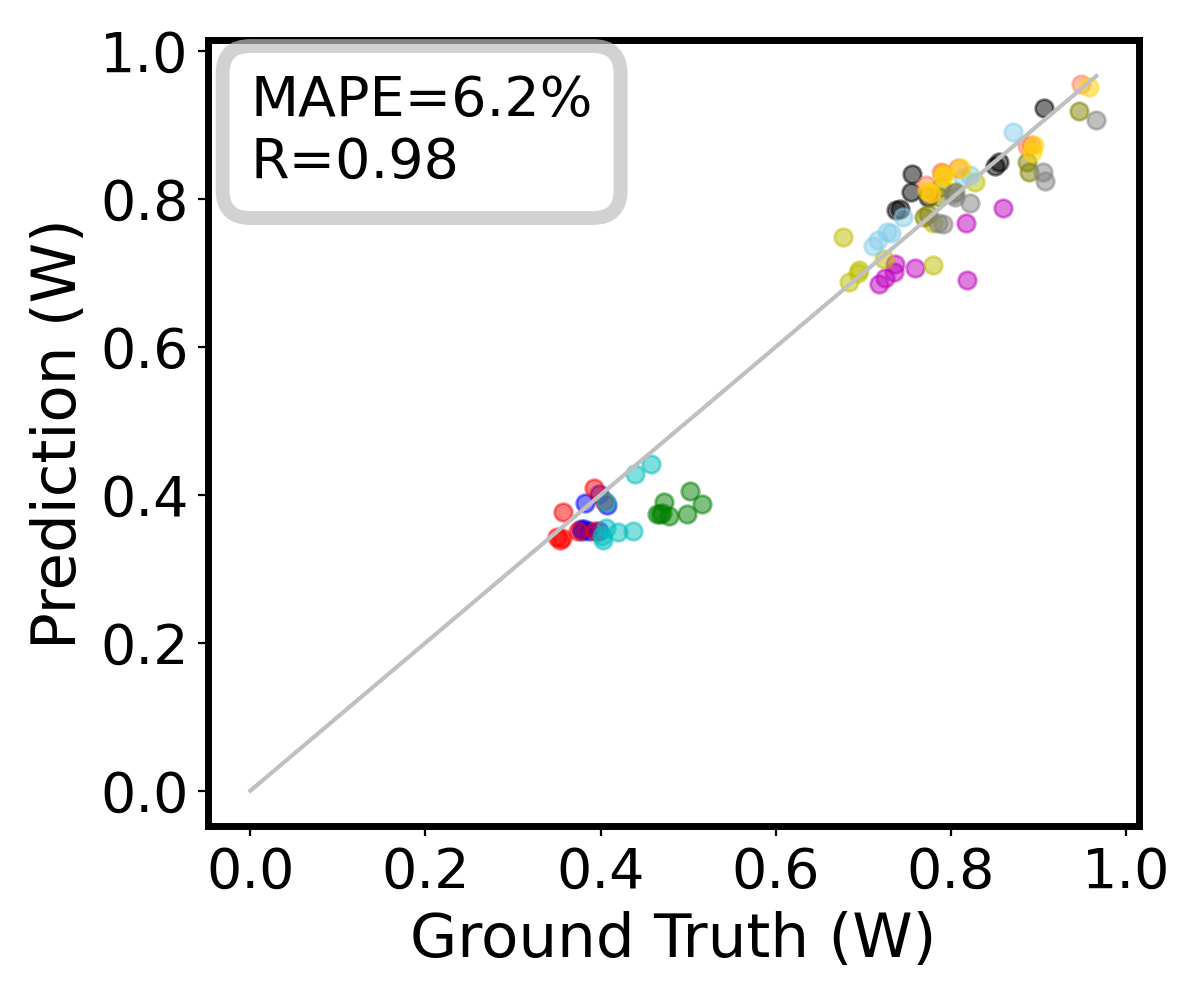}
    %\label{McPAT-Calib-Component}
}
\hspace{-3mm}
\subfigure[McPAT-Calib-CompGroup]{
    \centering
    \includegraphics[height=0.25\textwidth]{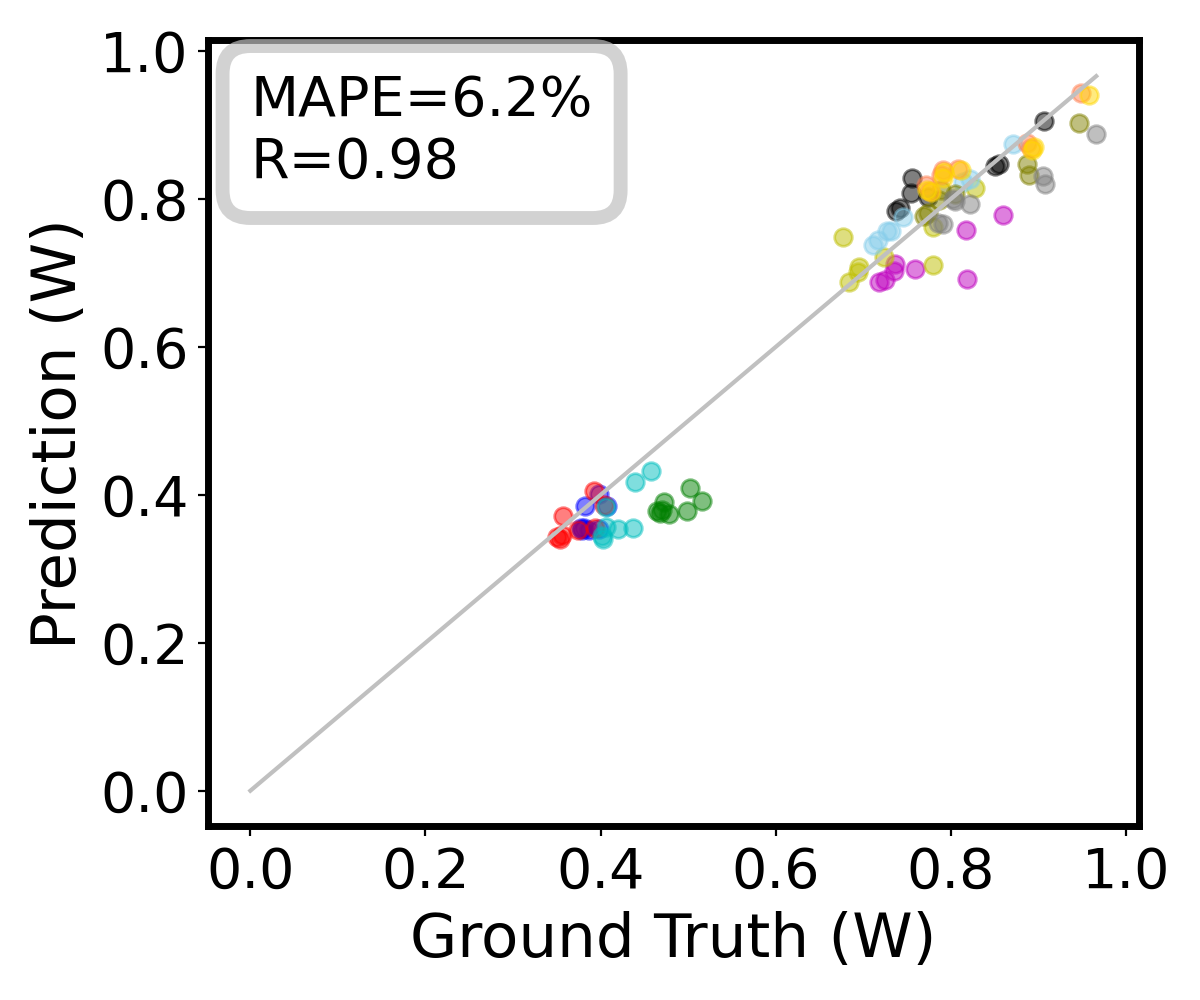}
    %\label{McPAT-Calib-Component}
}
\hspace{-3mm}
\subfigure[PANDA]{
    \centering
    \includegraphics[height=0.25\textwidth]{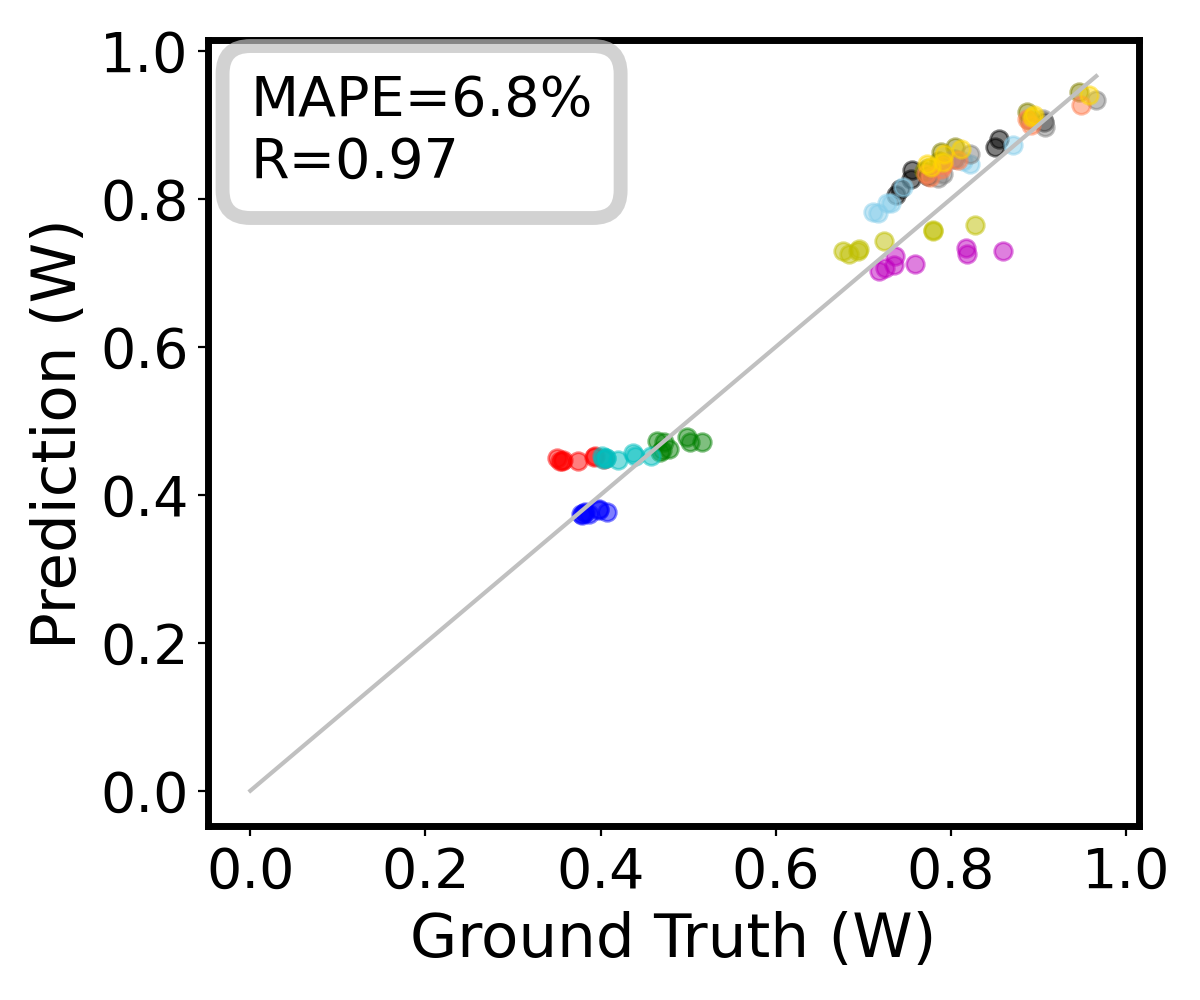}
    %\label{archbp}
}

%\vspace{-.1in}
\caption{Predictions with different models on BOOM CPU under \emph{Balance} training scenario.}
%\vspace{-.2in}
\label{visboomA}
\end{figure*}

\begin{figure*}[!h]
\centering

\hspace{-5mm}
\subfigure[McPAT]{
    \centering
    \includegraphics[height=0.25\textwidth]{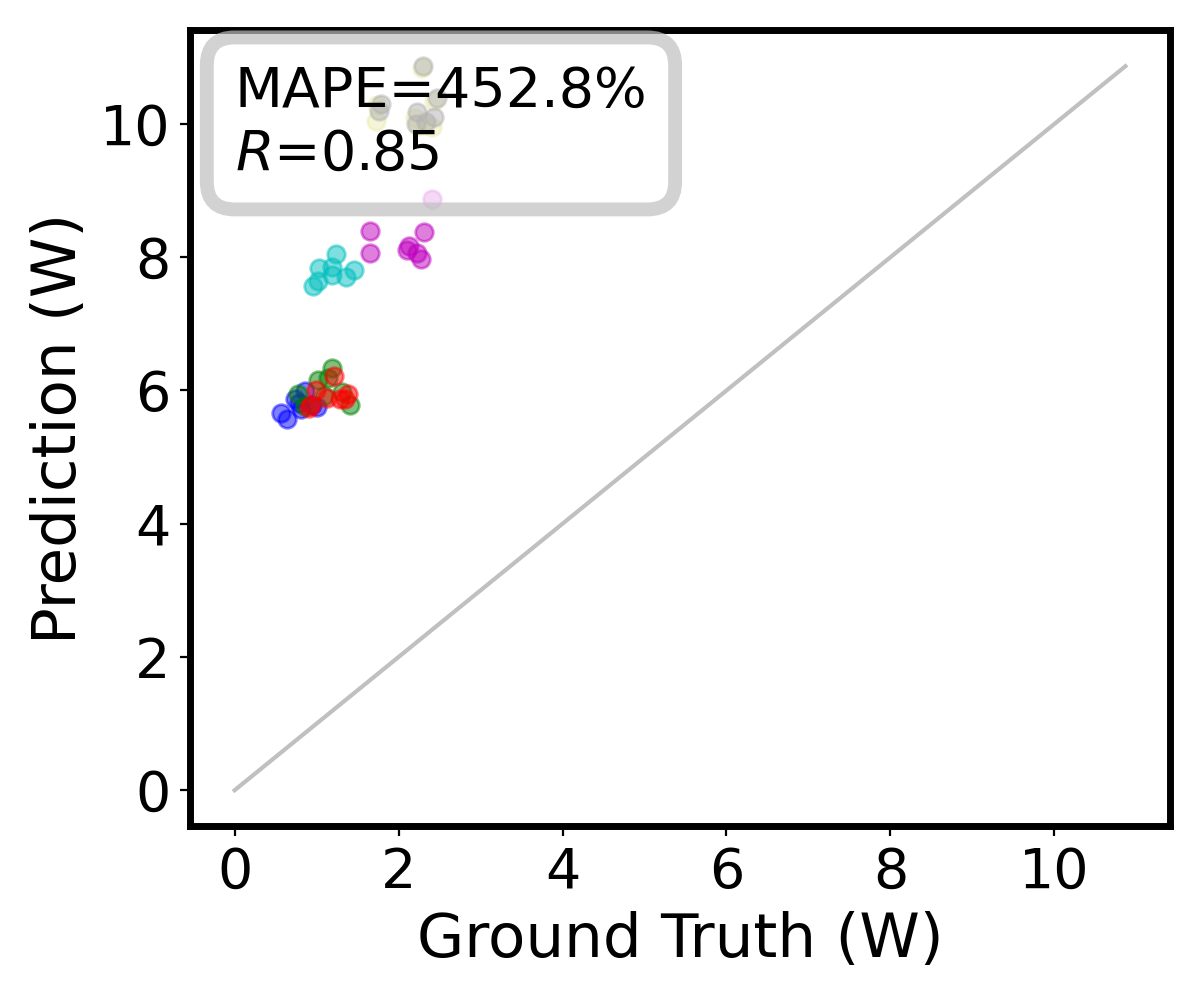}
    %\label{McPAT}
}
\hspace{-3mm}
\subfigure[McPAT-Plus]{
    \centering
    \includegraphics[height=0.25\textwidth]{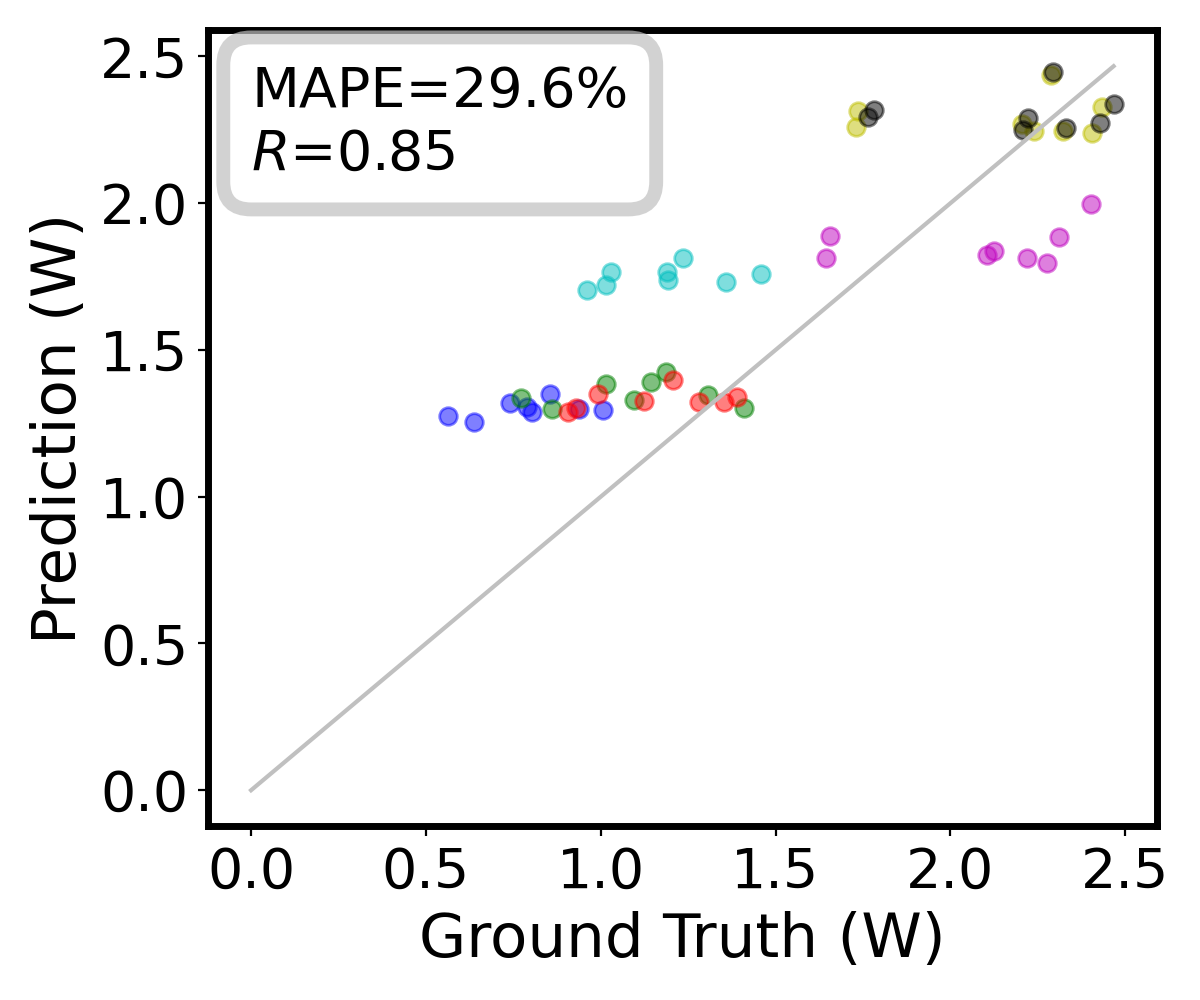}
    %\label{McPAT-Plus}
}
\hspace{-3mm}
\subfigure[McPAT-Calib]{
    \centering
    \includegraphics[height=0.25\textwidth]{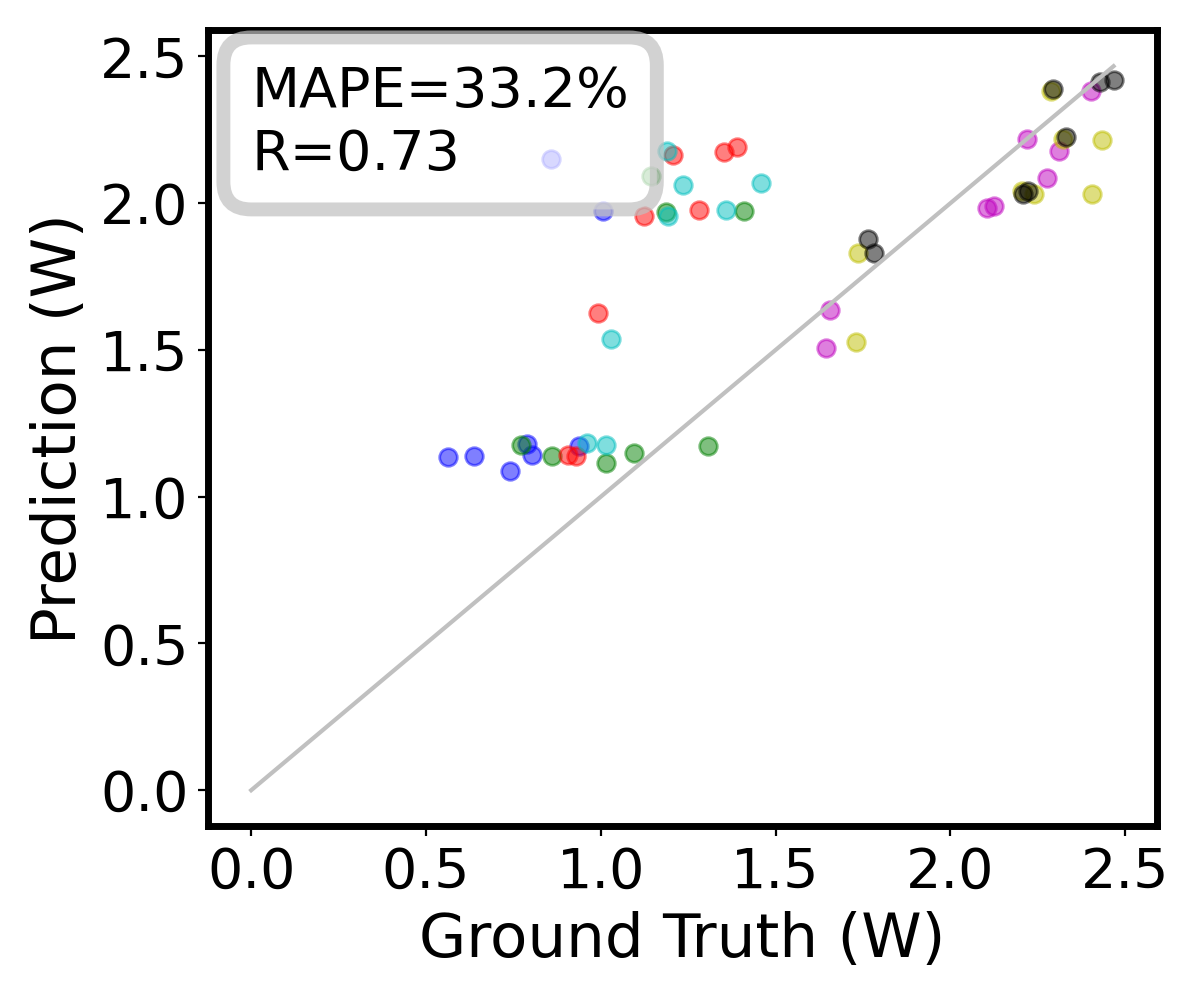}
    %\label{archbp}
}

\hspace{-5mm}
\subfigure[McPAT-Calib-Component]{
    \centering
    \includegraphics[height=0.25\textwidth]{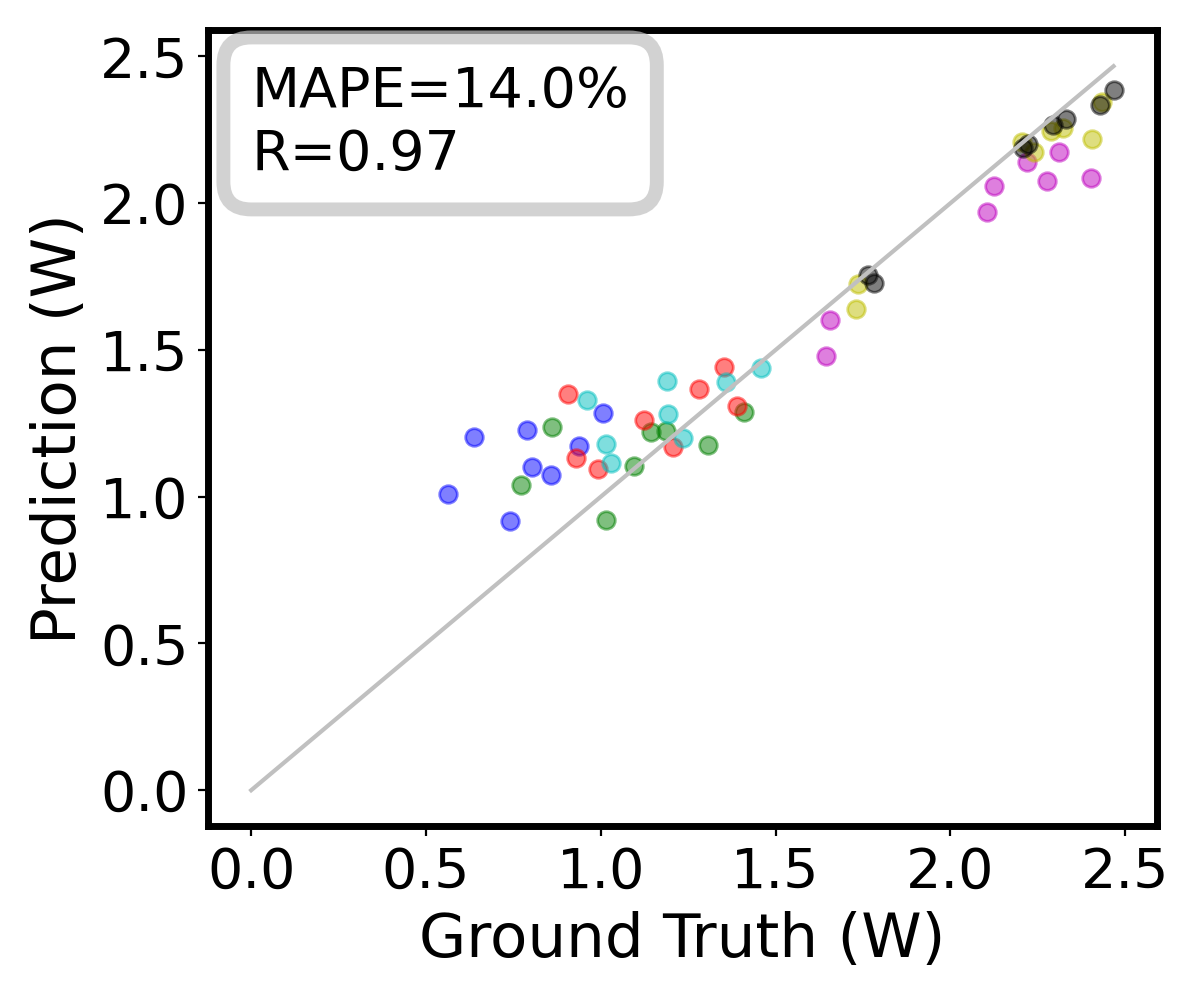}
    %\label{McPAT-Calib-Component}
}
\hspace{-3mm}
\subfigure[McPAT-Calib-CompGroup]{
    \centering
    \includegraphics[height=0.25\textwidth]{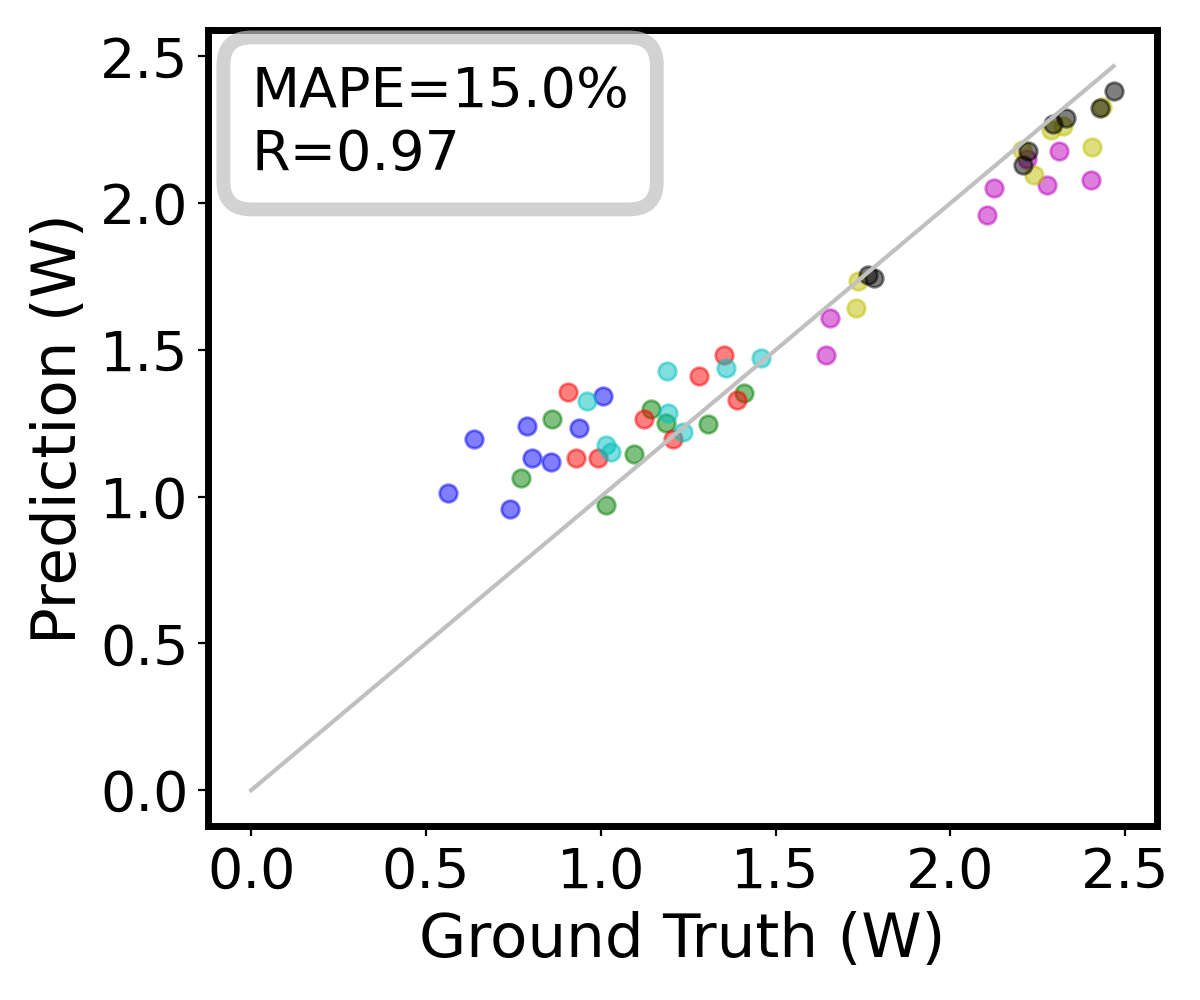}
    %\label{McPAT-Calib-Component}
}
\hspace{-3mm}
\subfigure[PANDA]{
    \centering
    \includegraphics[height=0.25\textwidth]{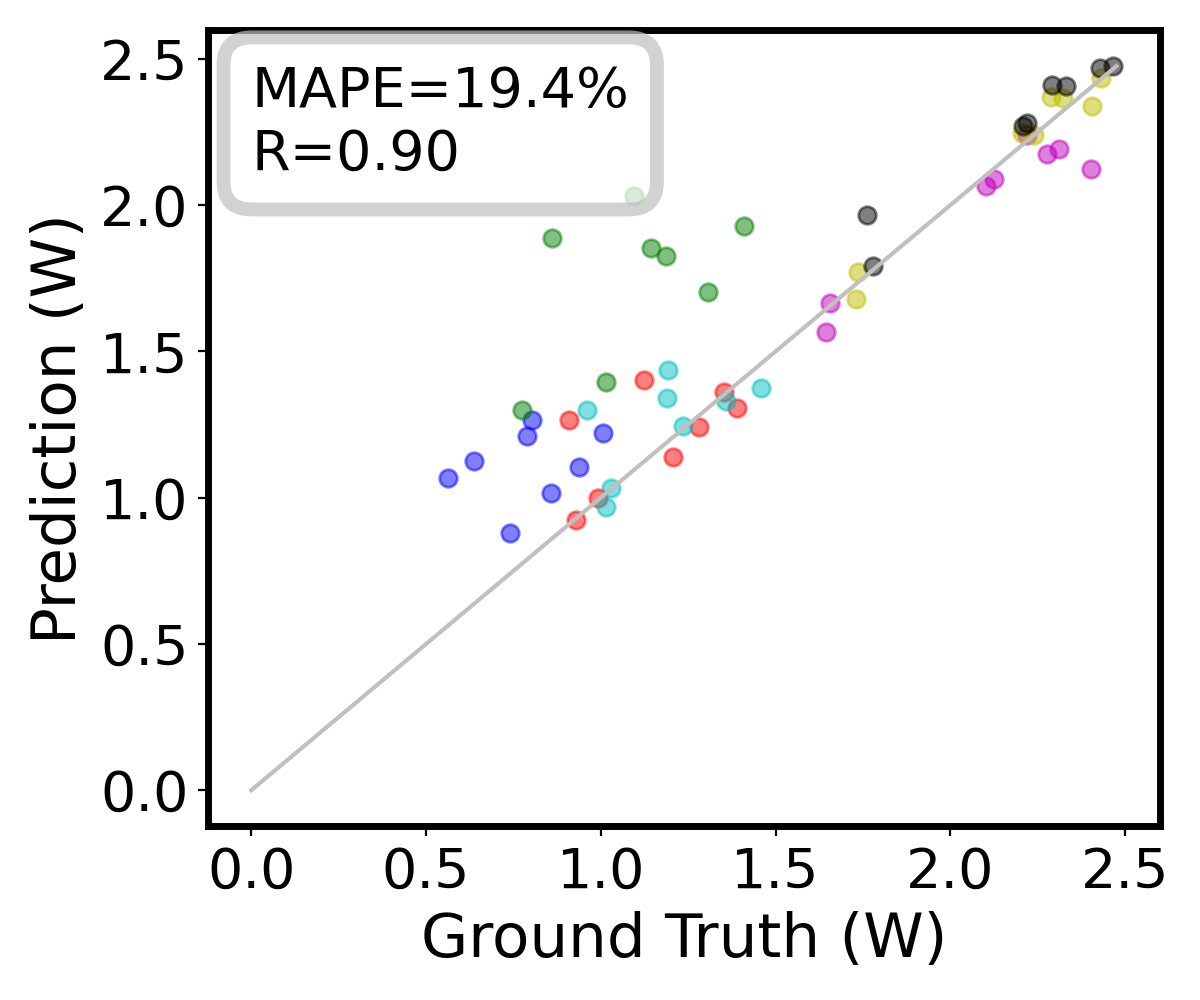}
    %\label{archbp}
}

%\vspace{-.1in}
\caption{Predictions with different models on XiangShan CPU under \emph{Balance} training scenario.}
%\vspace{-.2in}
\label{visxsA}
\end{figure*}

Fig.~\ref{visboomA} and \ref{visxsA} visualize the prediction of different models on BOOM and XiangShan under the \emph{Balance} training scenario.

\newpage

\begin{figure*}[!h]
\centering

\hspace{-5mm}
\subfigure[McPAT]{
    \centering
    \includegraphics[height=0.25\textwidth]{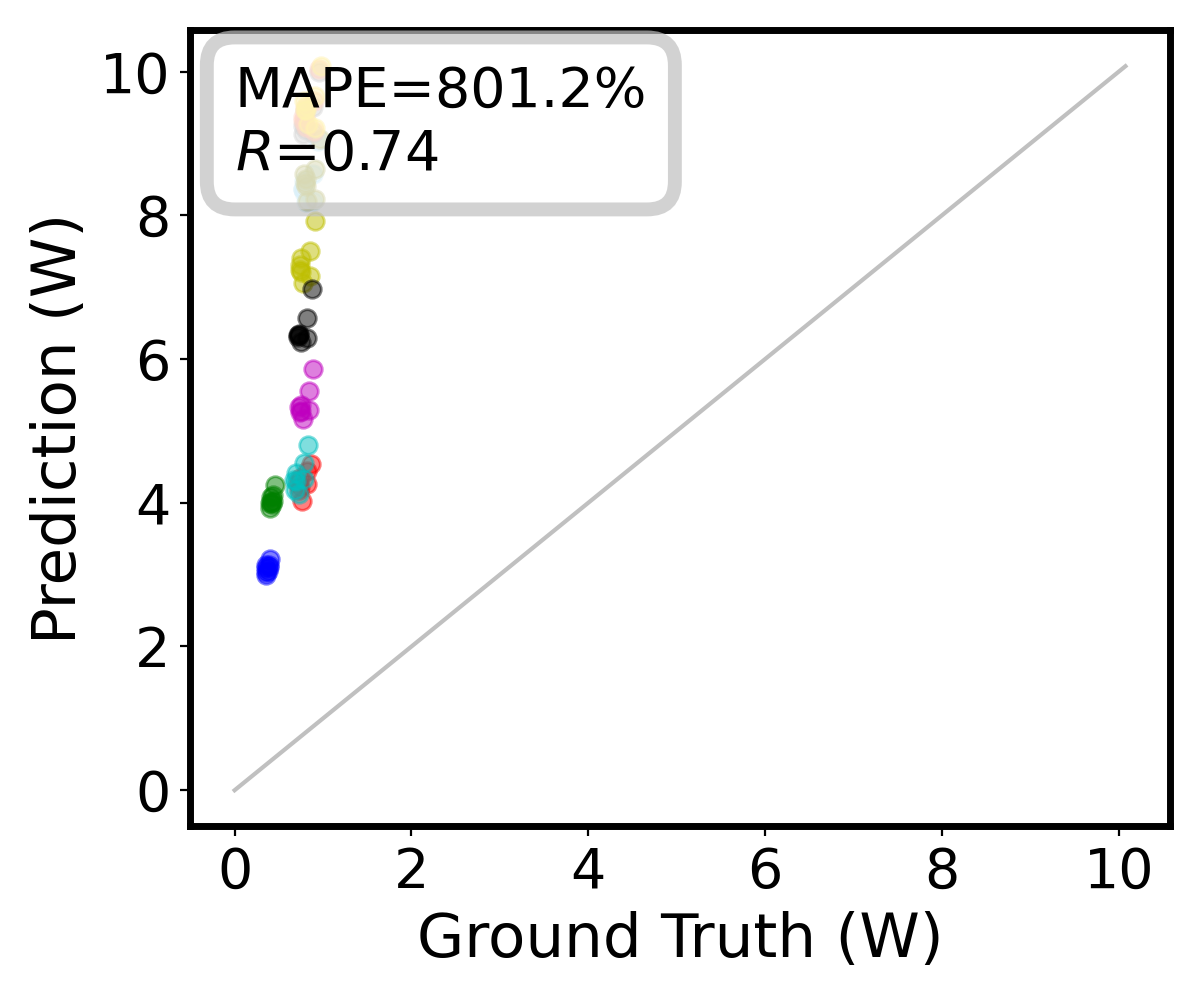}
    %\label{McPAT}
}
\hspace{-3mm}
\subfigure[McPAT-Plus]{
    \centering
    \includegraphics[height=0.25\textwidth]{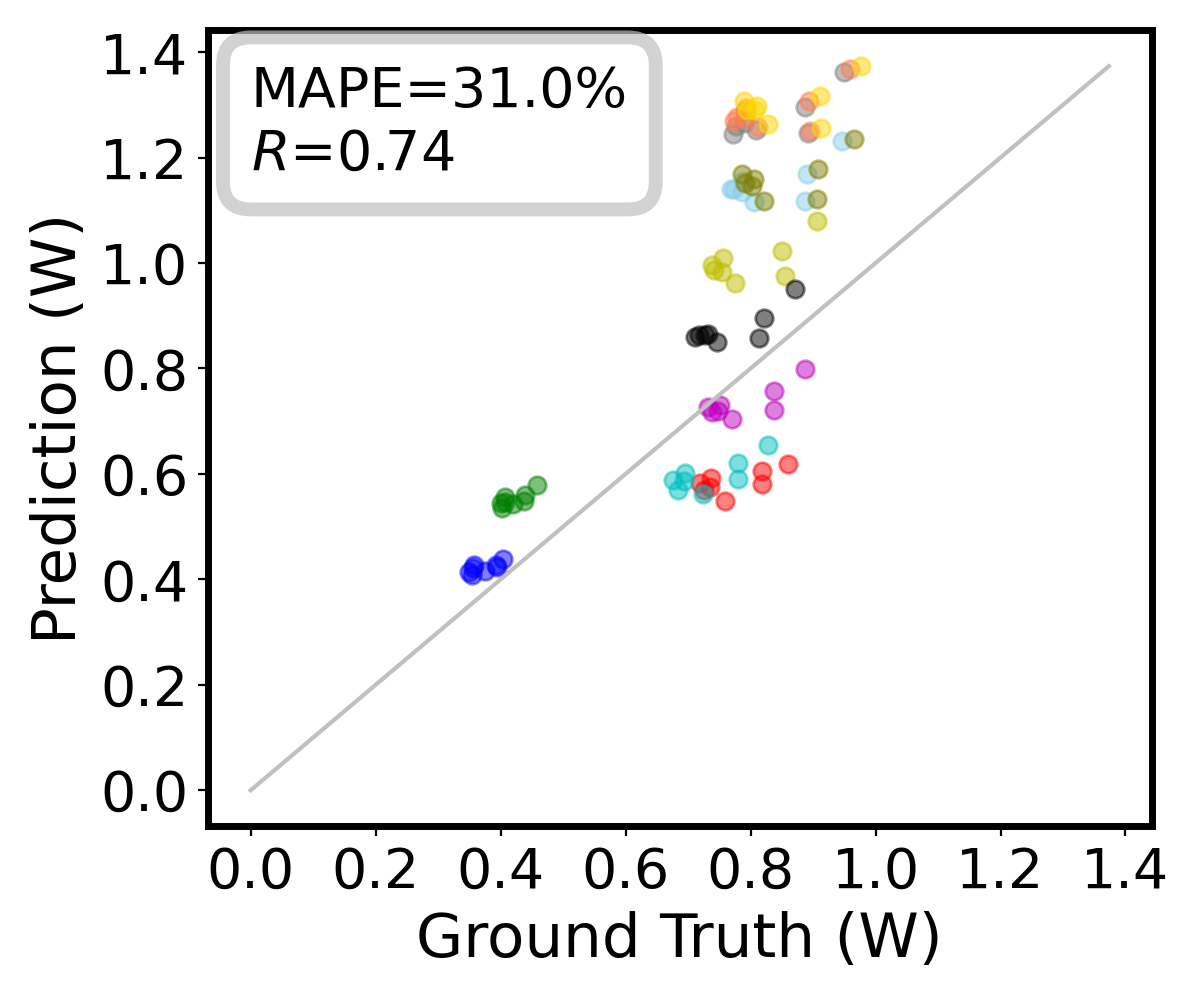}
    %\label{McPAT-Plus}
}
\hspace{-3mm}
\subfigure[McPAT-Calib]{
    \centering
    \includegraphics[height=0.25\textwidth]{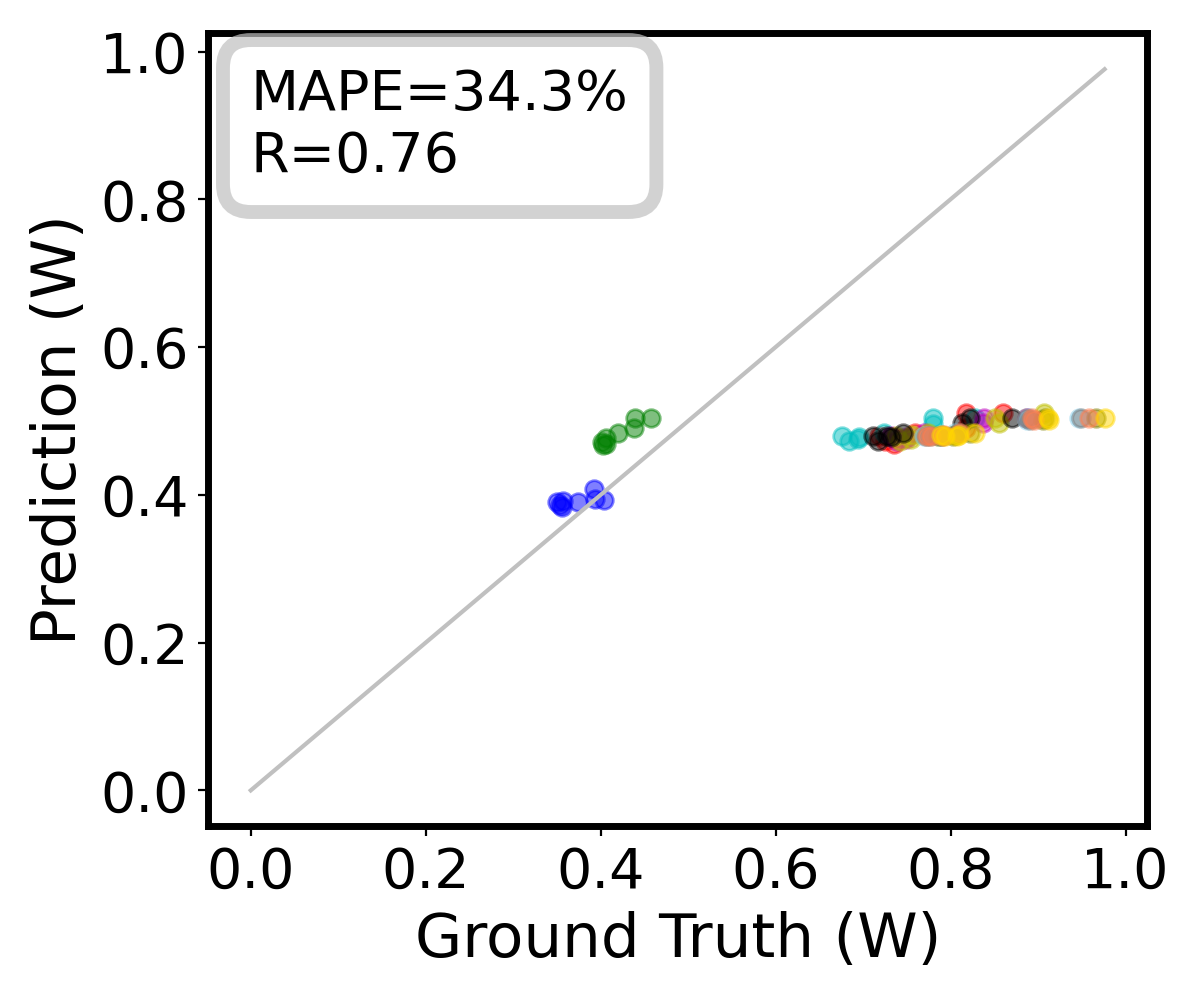}
    %\label{archbp}
}

\hspace{-5mm}
\subfigure[McPAT-Calib-Component]{
    \centering
    \includegraphics[height=0.25\textwidth]{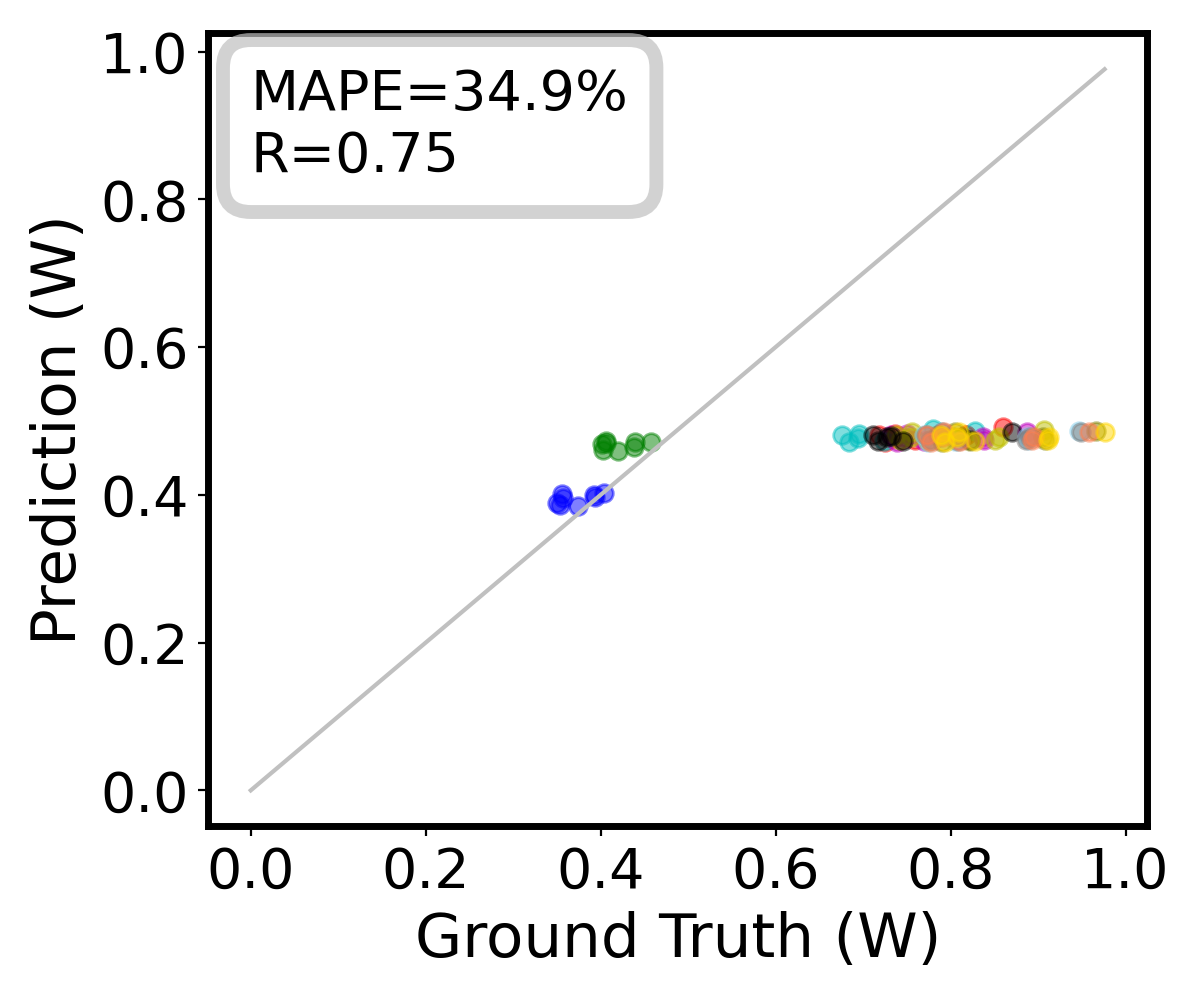}
    %\label{McPAT-Calib-Component}
}
\hspace{-3mm}
\subfigure[McPAT-Calib-CompGroup]{
    \centering
    \includegraphics[height=0.25\textwidth]{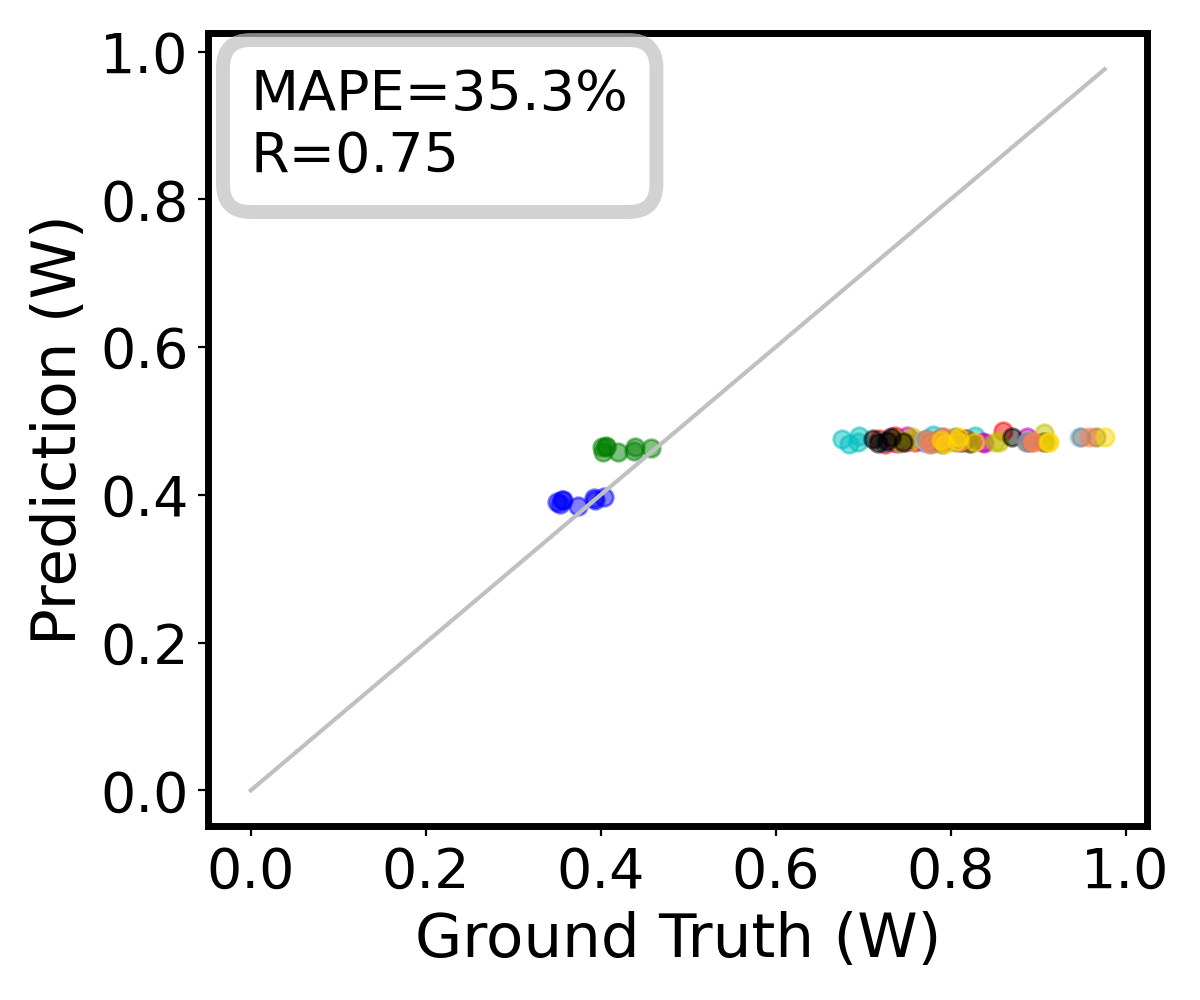}
    %\label{McPAT-Calib-Component}
}
\hspace{-3mm}
\subfigure[PANDA]{
    \centering
    \includegraphics[height=0.25\textwidth]{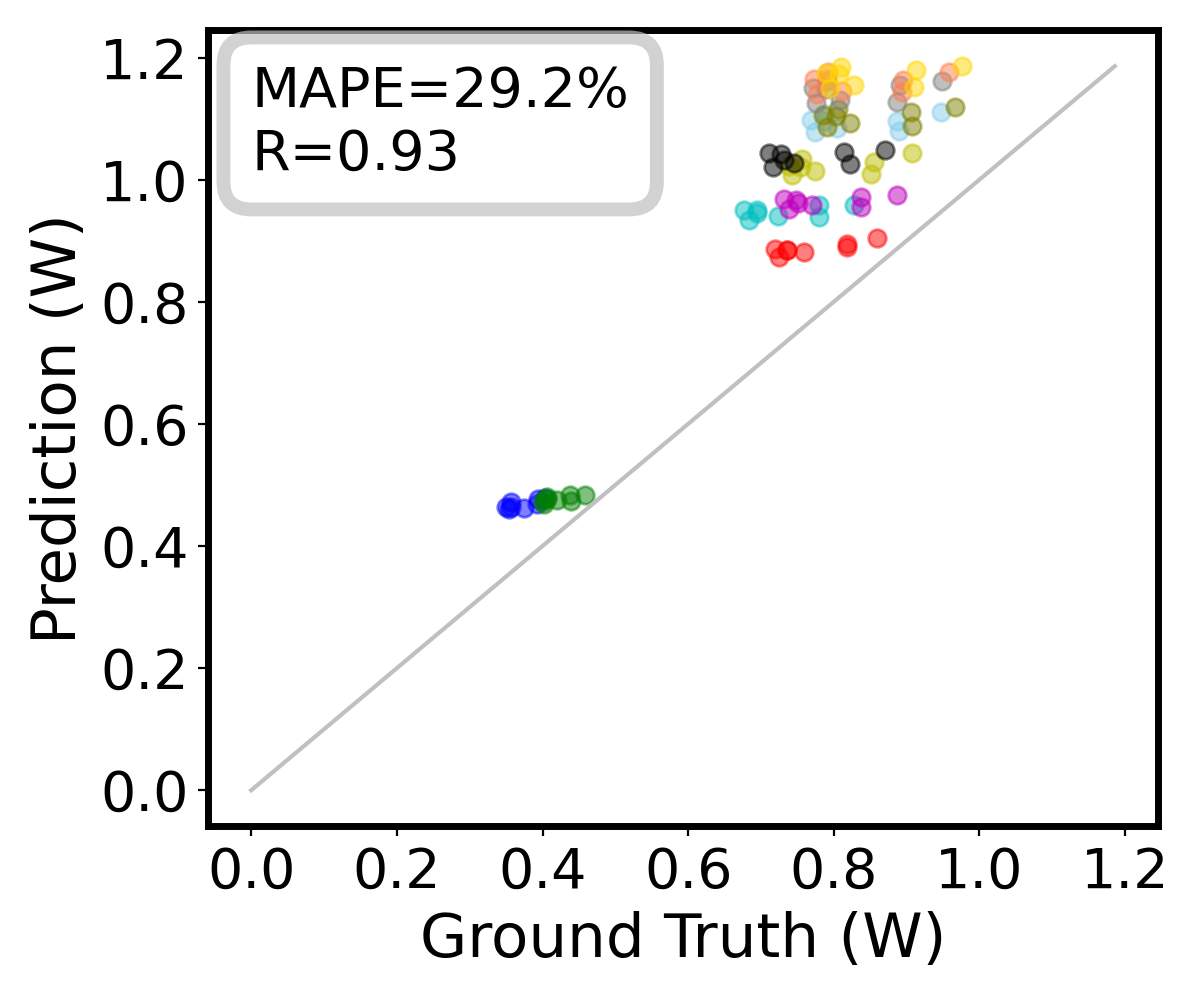}
    %\label{archbp}
}

%\vspace{-.1in}
\caption{Predictions with different models on BOOM CPU under \emph{Small} training scenario.}
%\vspace{-.2in}
\label{visboomB}
\end{figure*}

\begin{figure*}[!h]
\centering

\hspace{-5mm}
\subfigure[McPAT]{
    \centering
    \includegraphics[height=0.25\textwidth]{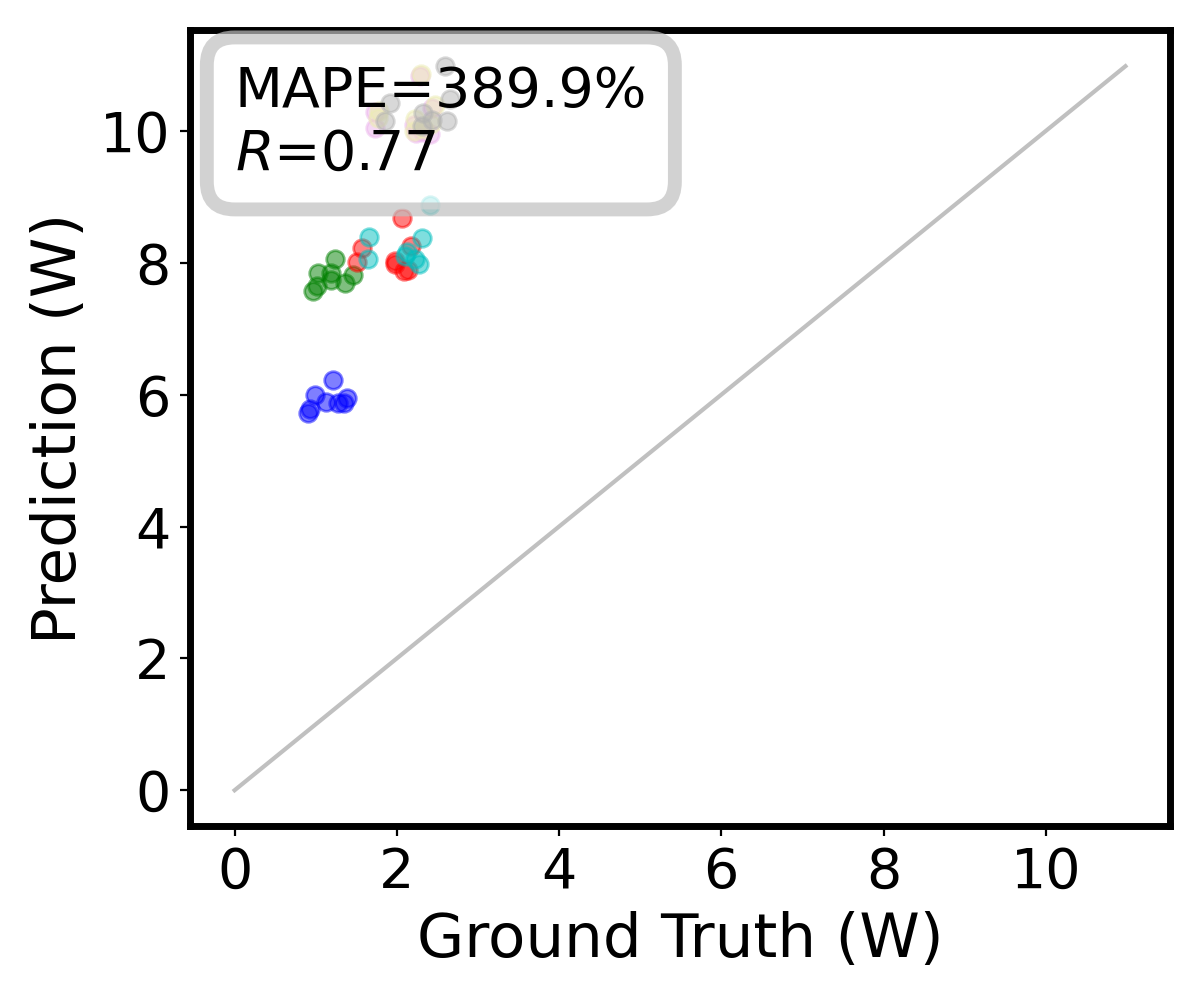}
    %\label{McPAT}
}
\hspace{-3mm}
\subfigure[McPAT-Plus]{
    \centering
    \includegraphics[height=0.25\textwidth]{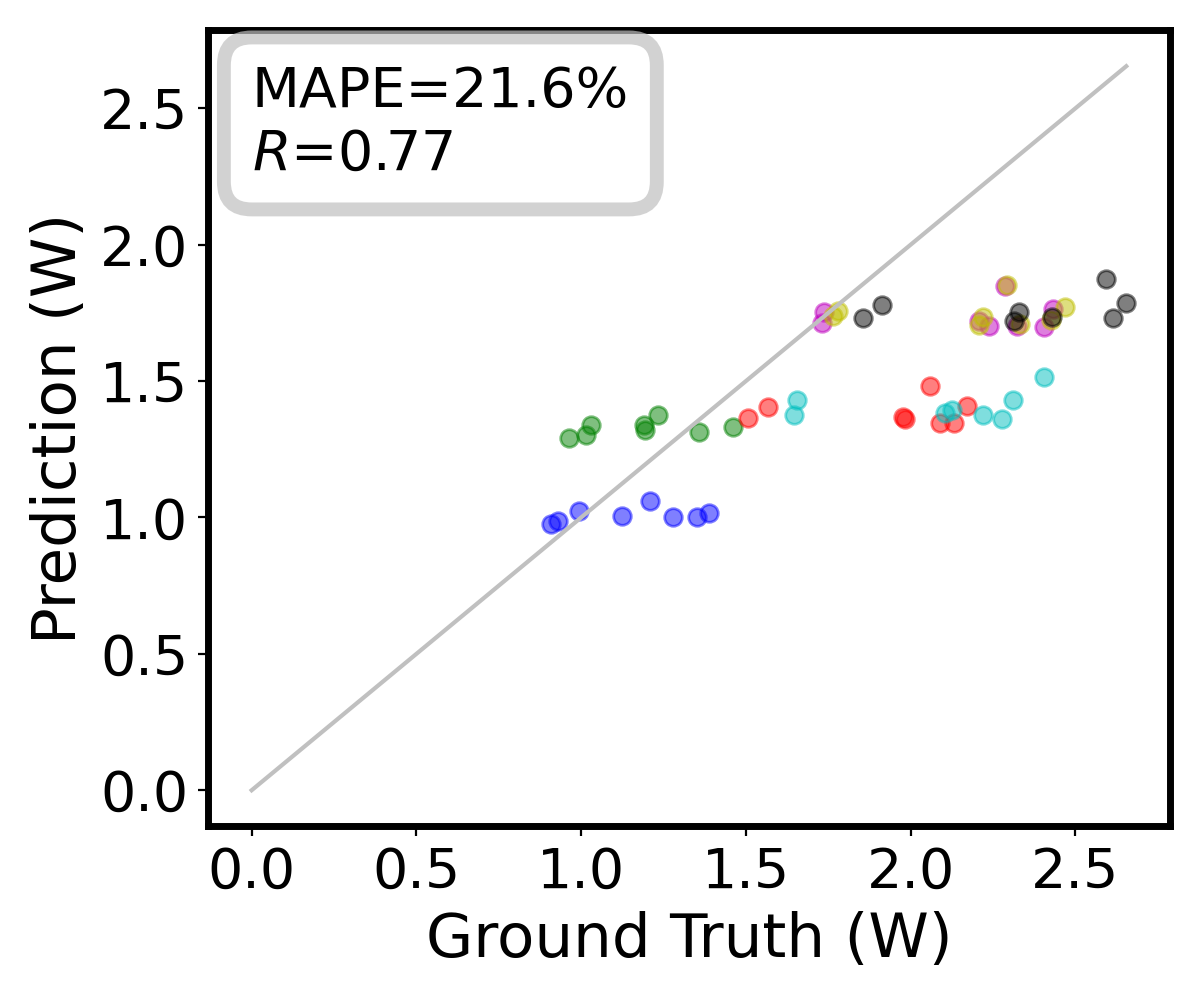}
    %\label{McPAT-Plus}
}
\hspace{-3mm}
\subfigure[McPAT-Calib]{
    \centering
    \includegraphics[height=0.25\textwidth]{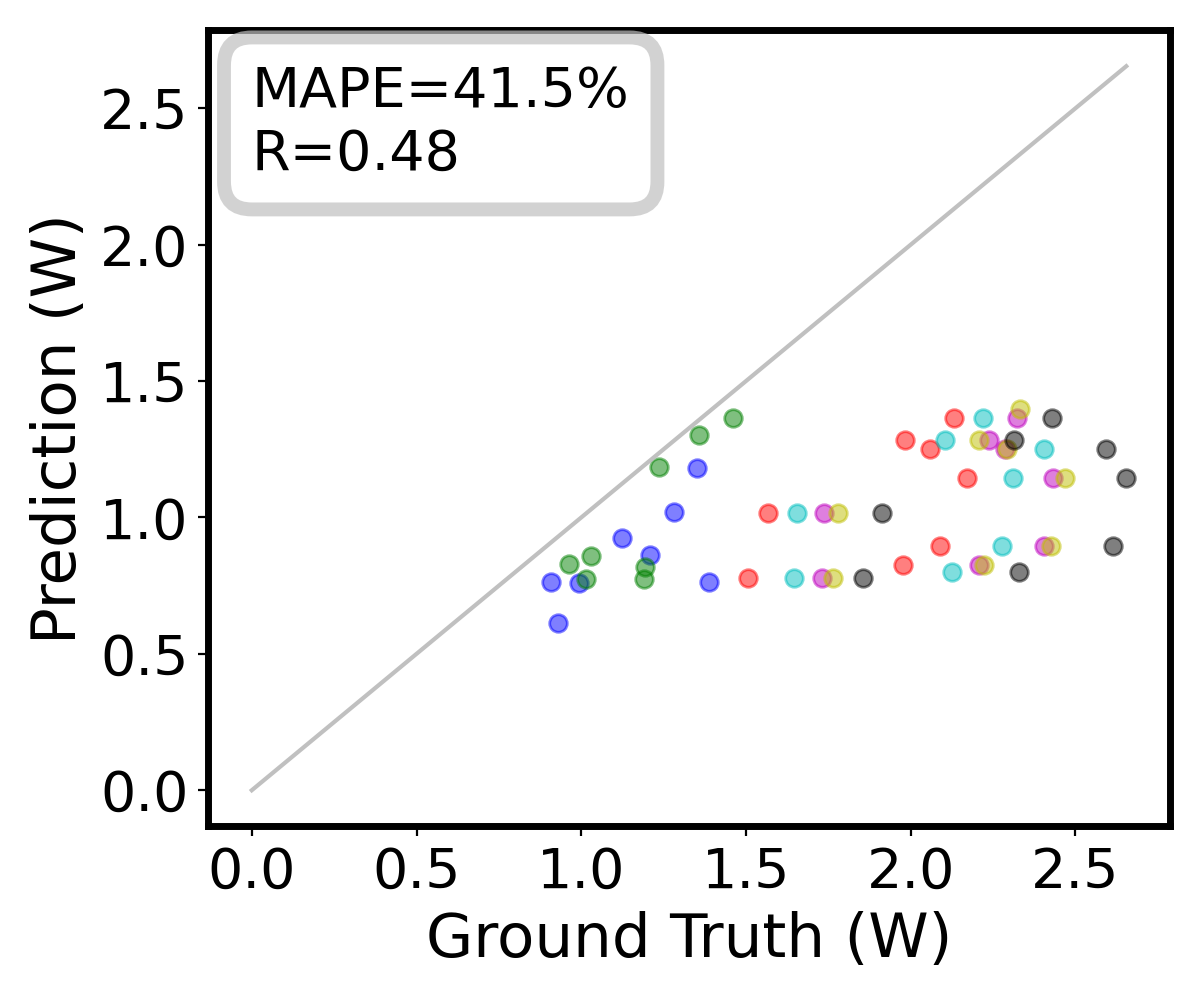}
    %\label{archbp}
}

\hspace{-5mm}
\subfigure[McPAT-Calib-Component]{
    \centering
    \includegraphics[height=0.25\textwidth]{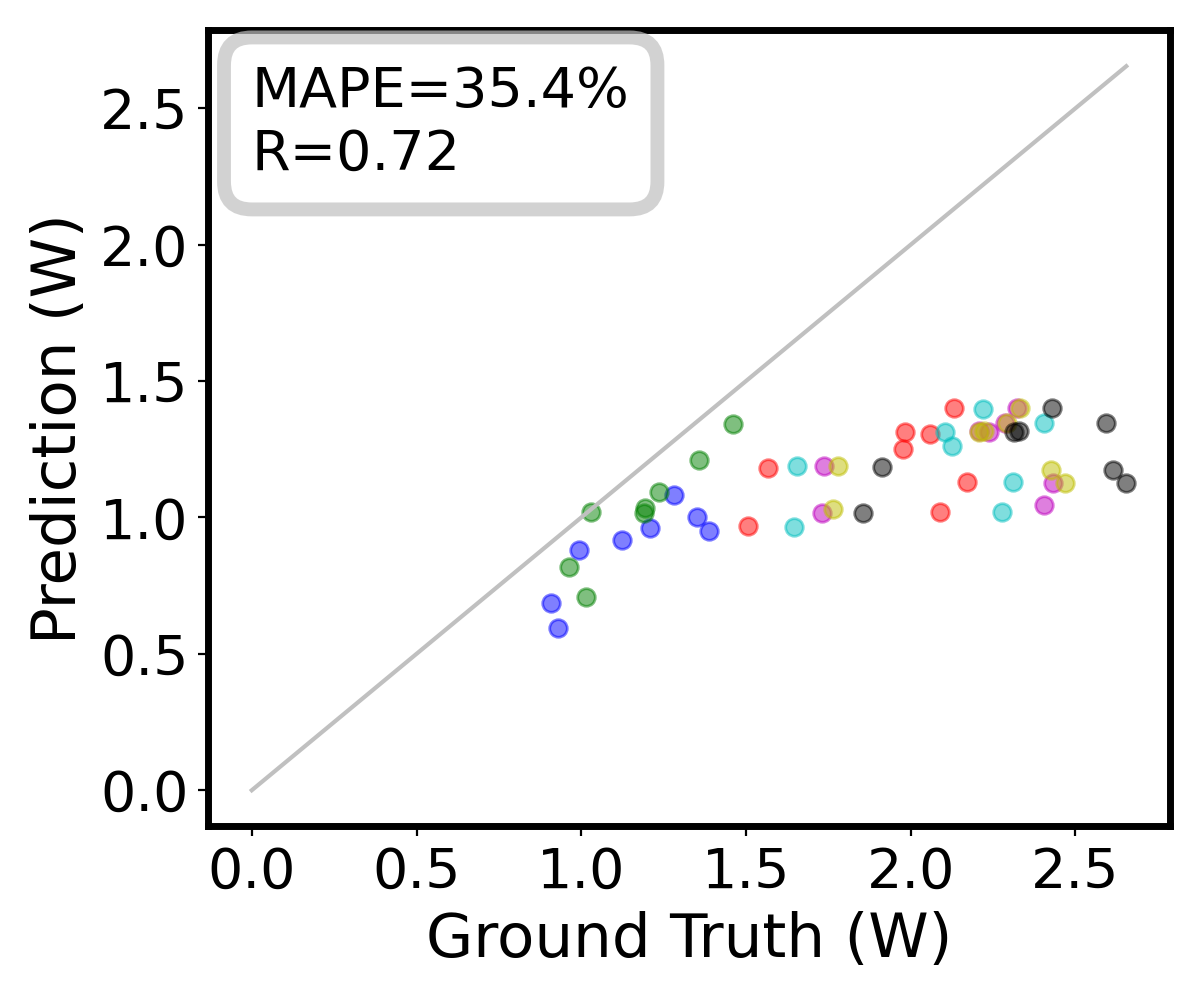}
    %\label{McPAT-Calib-Component}
}
\hspace{-3mm}
\subfigure[McPAT-Calib-CompGroup]{
    \centering
    \includegraphics[height=0.25\textwidth]{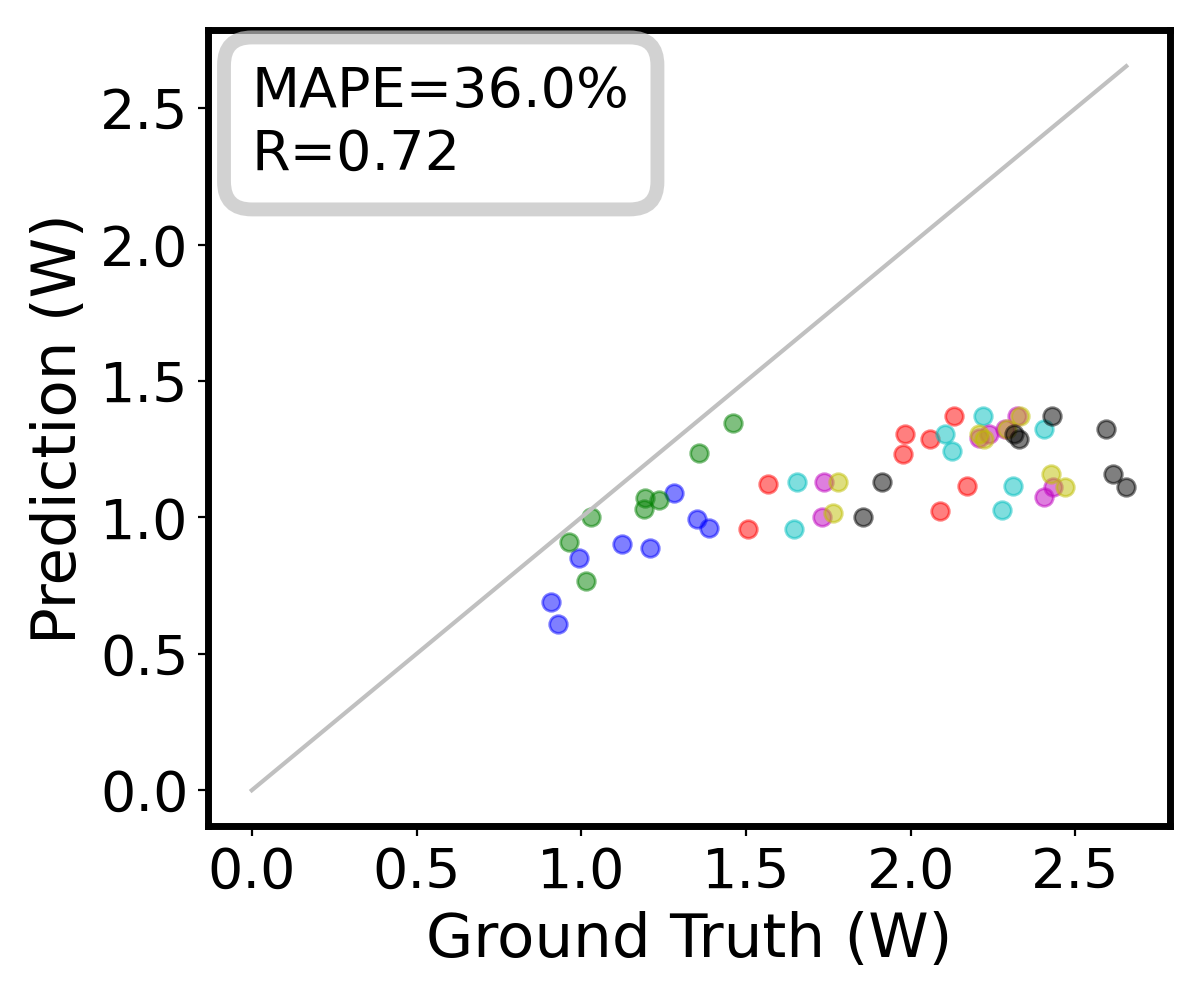}
    %\label{McPAT-Calib-Component}
}
\hspace{-3mm}
\subfigure[PANDA]{
    \centering
    \includegraphics[height=0.25\textwidth]{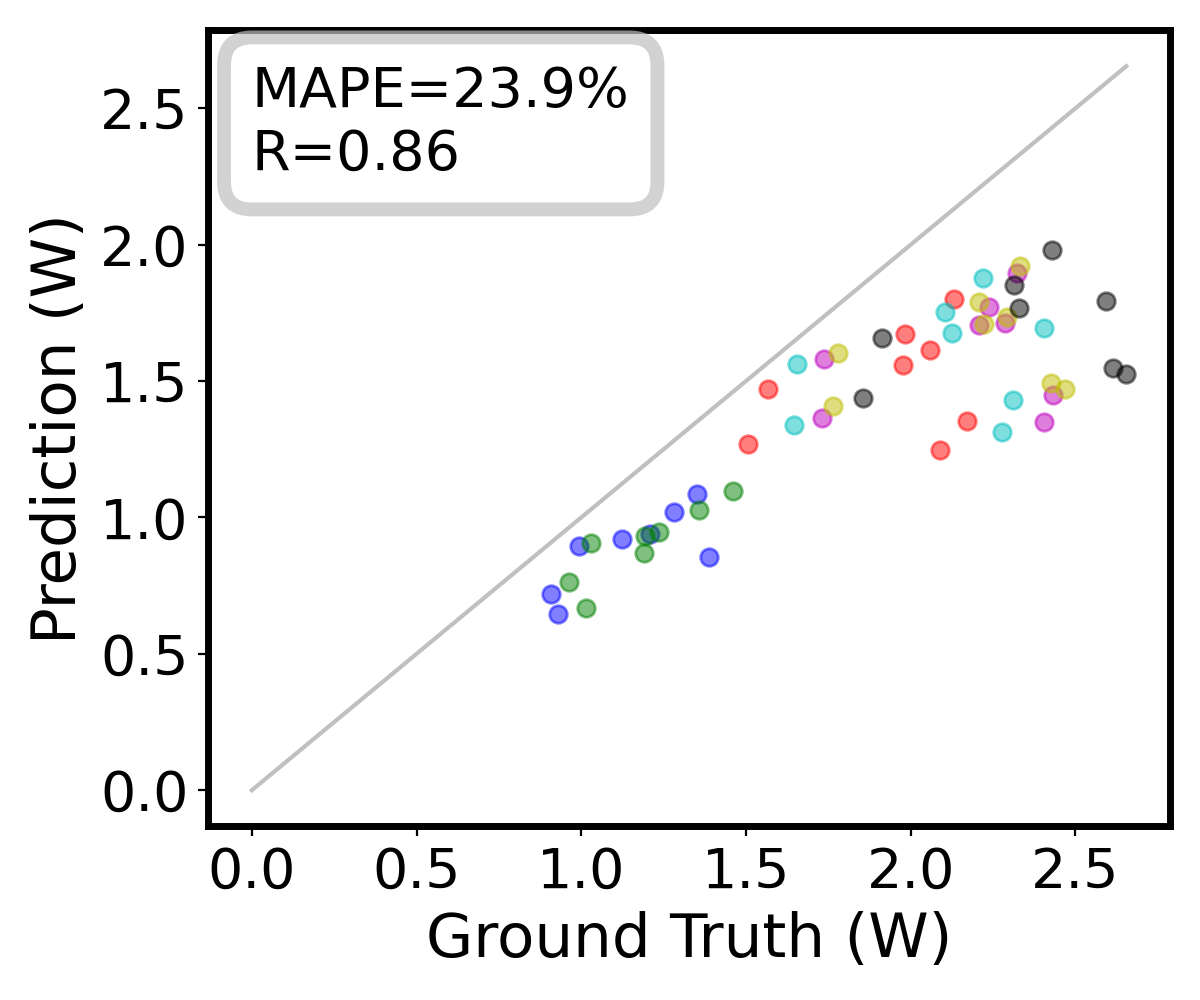}
    %\label{archbp}
}

%\vspace{-.1in}
\caption{Predictions with different models on XiangShan CPU under \emph{Small} training scenario.}
%\vspace{-.2in}
\label{visxsB}
\end{figure*}

Fig.~\ref{visboomB} and \ref{visxsB} visualize the prediction of different models on BOOM and XiangShan under the \emph{Small} training scenario. 

\newpage

\begin{figure*}[!h]
\centering

\hspace{-5mm}
\subfigure[McPAT]{
    \centering
    \includegraphics[height=0.25\textwidth]{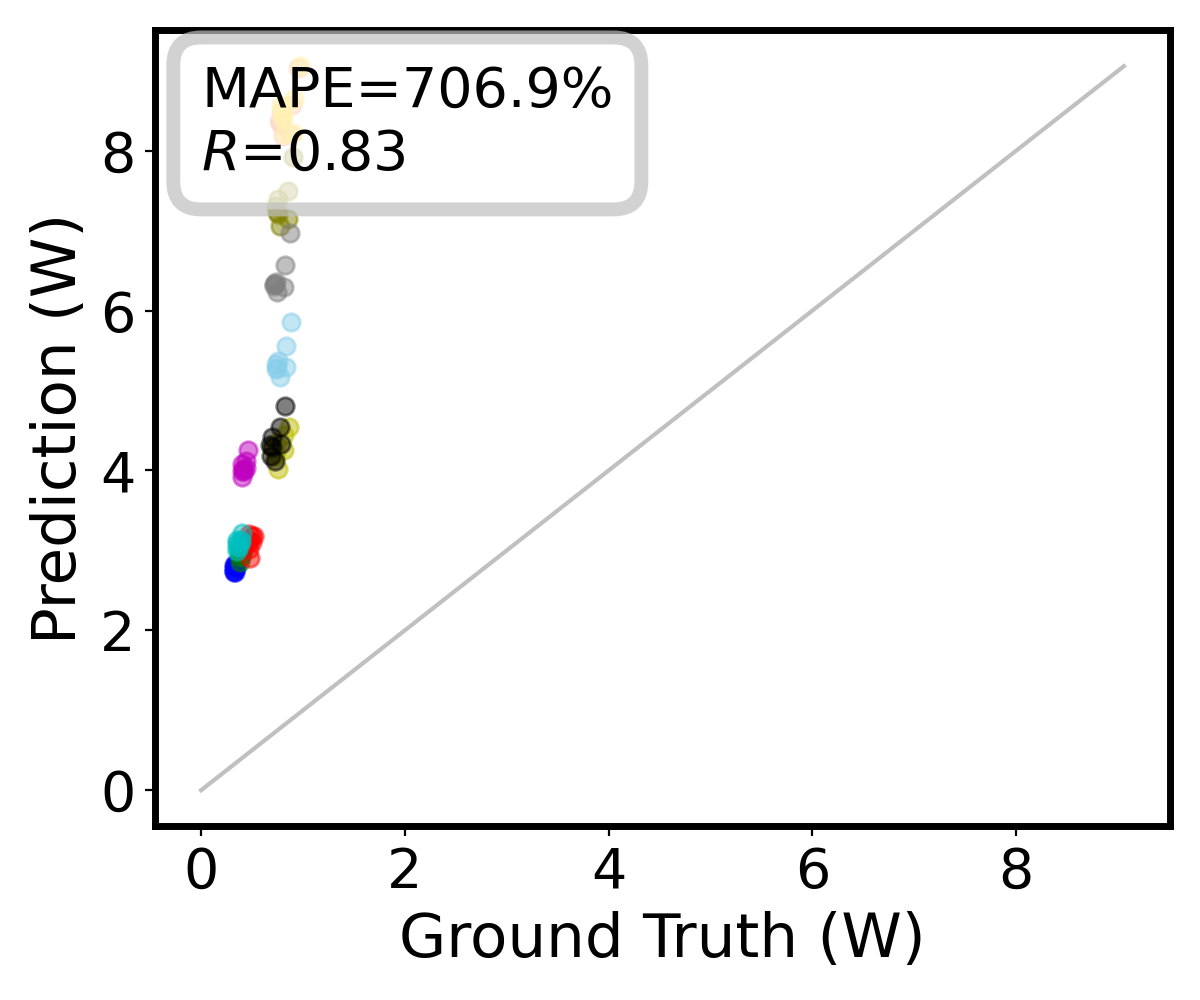}
    %\label{McPAT}
}
\hspace{-3mm}
\subfigure[McPAT-Plus]{
    \centering
    \includegraphics[height=0.25\textwidth]{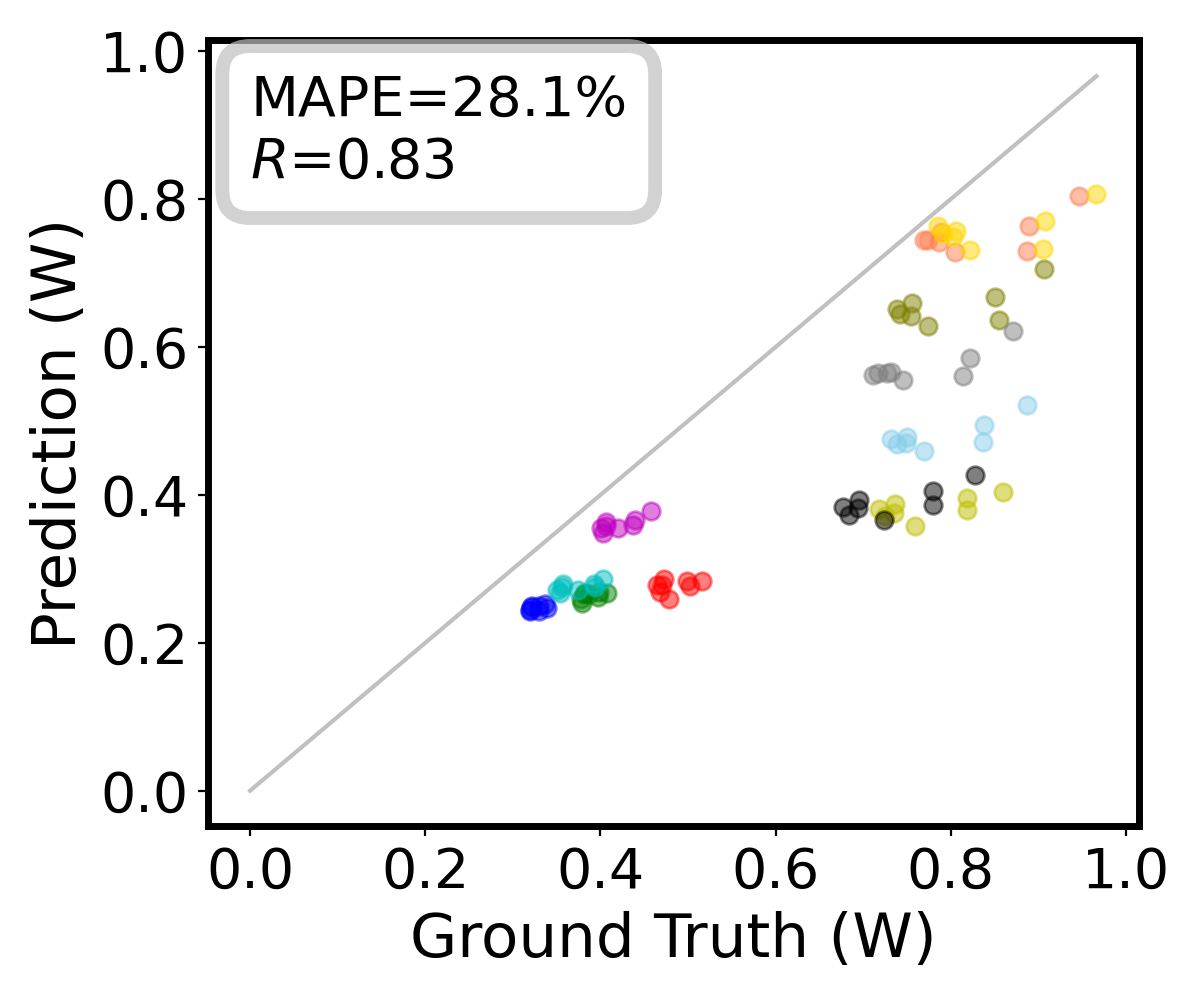}
    %\label{McPAT-Plus}
}
\hspace{-3mm}
\subfigure[McPAT-Calib]{
    \centering
    \includegraphics[height=0.25\textwidth]{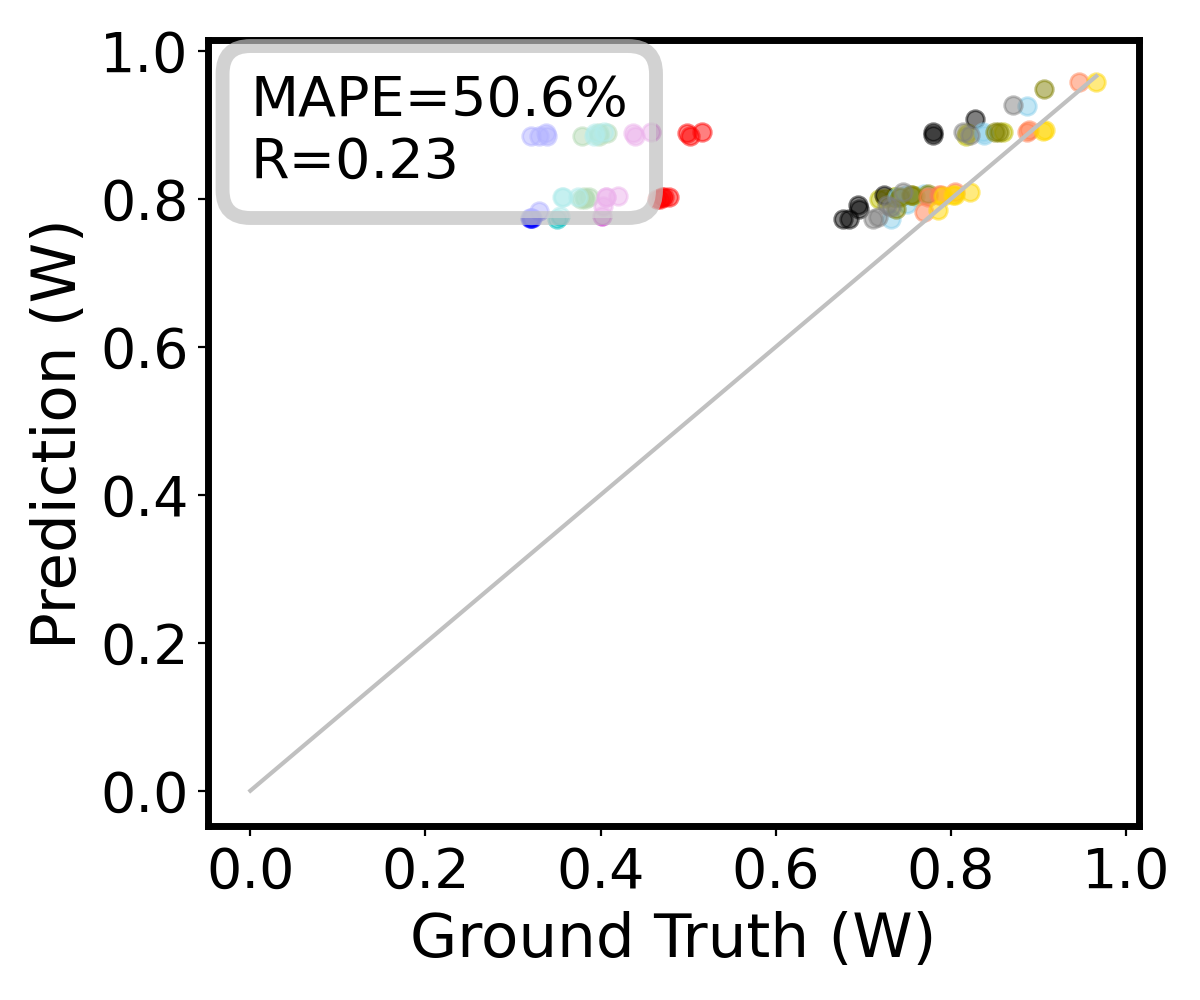}
    %\label{archbp}
}

\hspace{-5mm}
\subfigure[McPAT-Calib-Component]{
    \centering
    \includegraphics[height=0.25\textwidth]{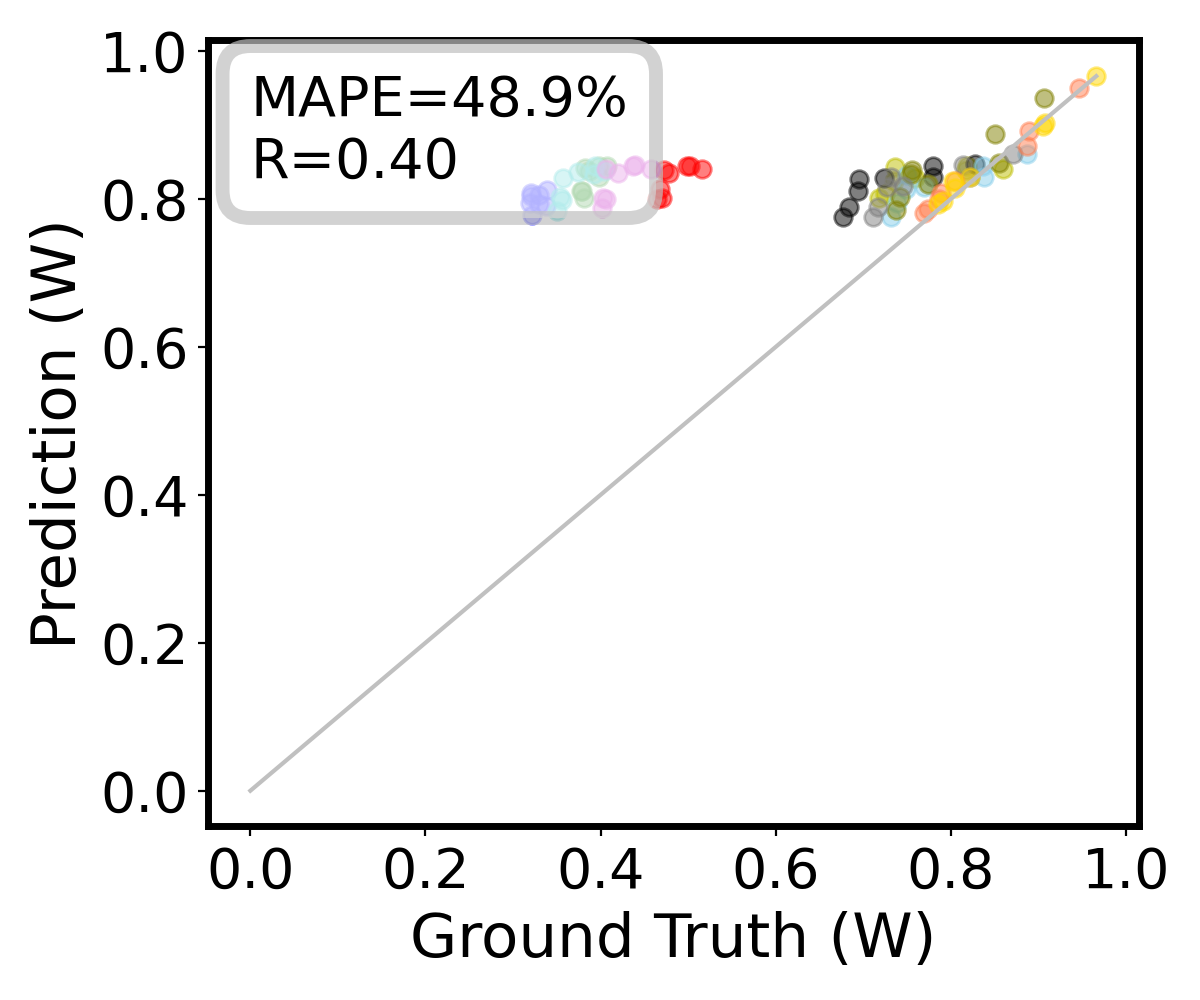}
    %\label{McPAT-Calib-Component}
}
\hspace{-3mm}
\subfigure[McPAT-Calib-CompGroup]{
    \centering
    \includegraphics[height=0.25\textwidth]{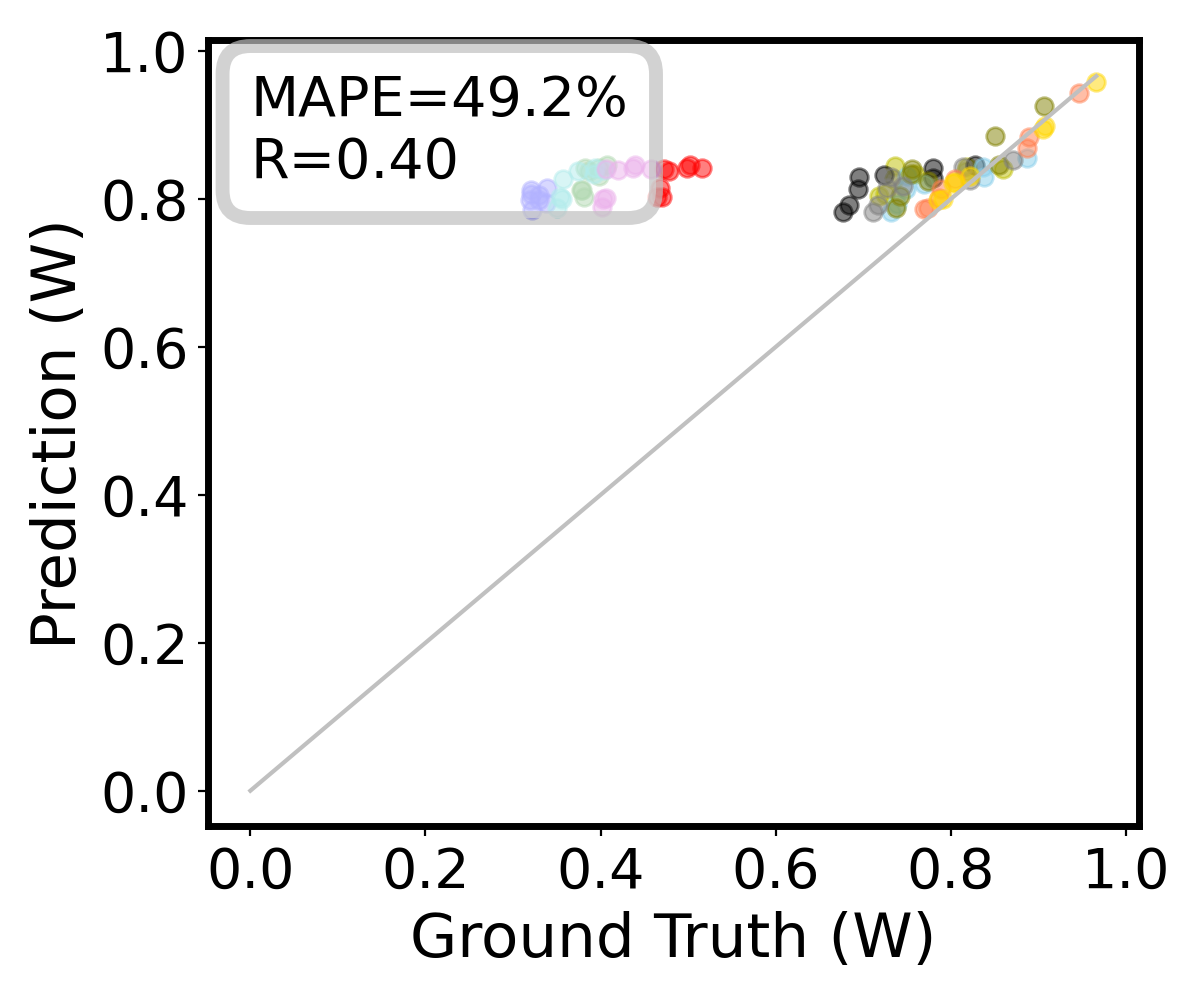}
    %\label{McPAT-Calib-Component}
}
\hspace{-3mm}
\subfigure[PANDA]{
    \centering
    \includegraphics[height=0.25\textwidth]{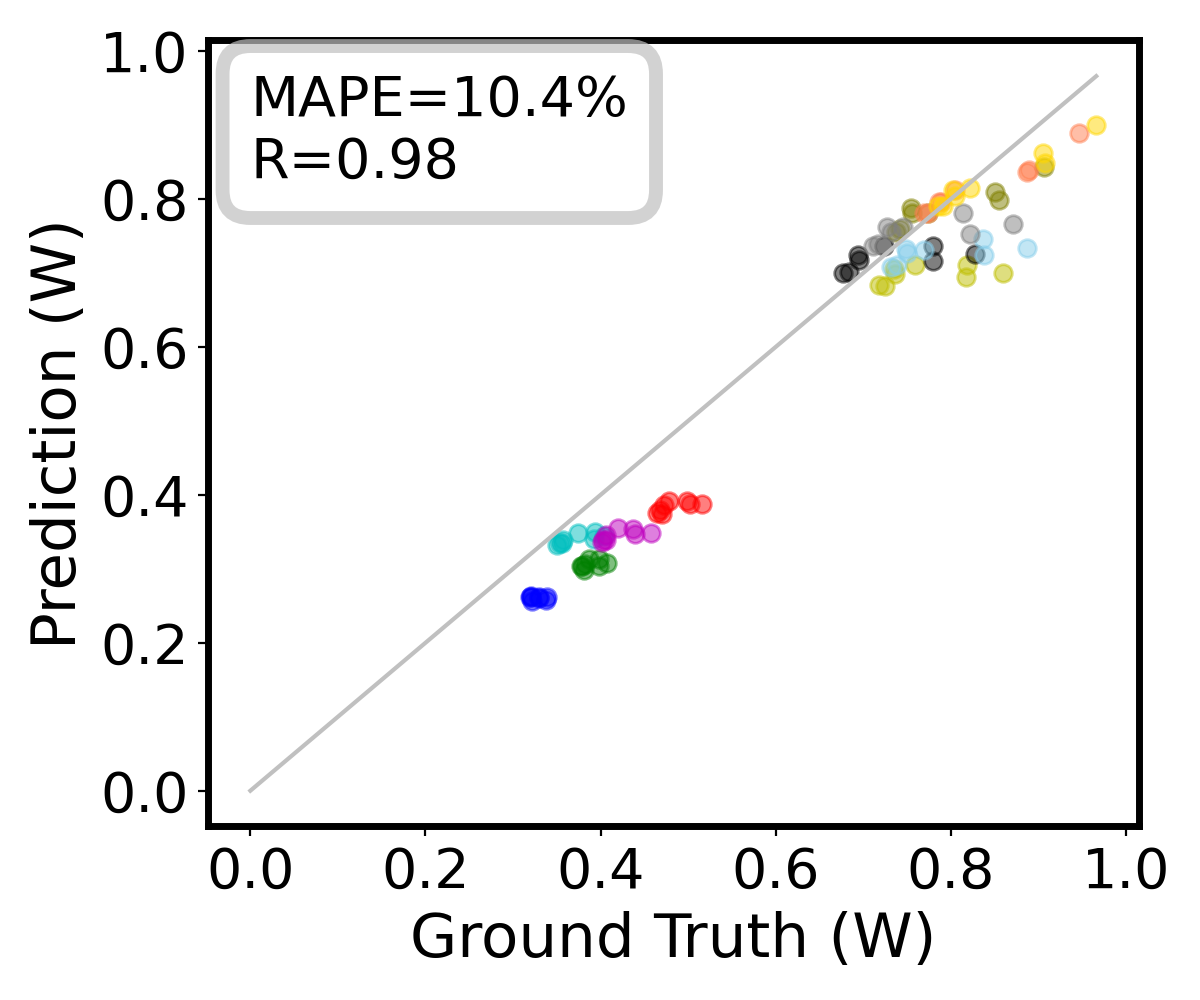}
    %\label{archbp}
}

%\vspace{-.1in}
\caption{Predictions with different models on BOOM CPU under \emph{Large} training scenario.}
%\vspace{-.2in}
\label{visboomC}
\end{figure*}

\begin{figure*}[!h]
\centering

\hspace{-5mm}
\subfigure[McPAT]{
    \centering
    \includegraphics[height=0.25\textwidth]{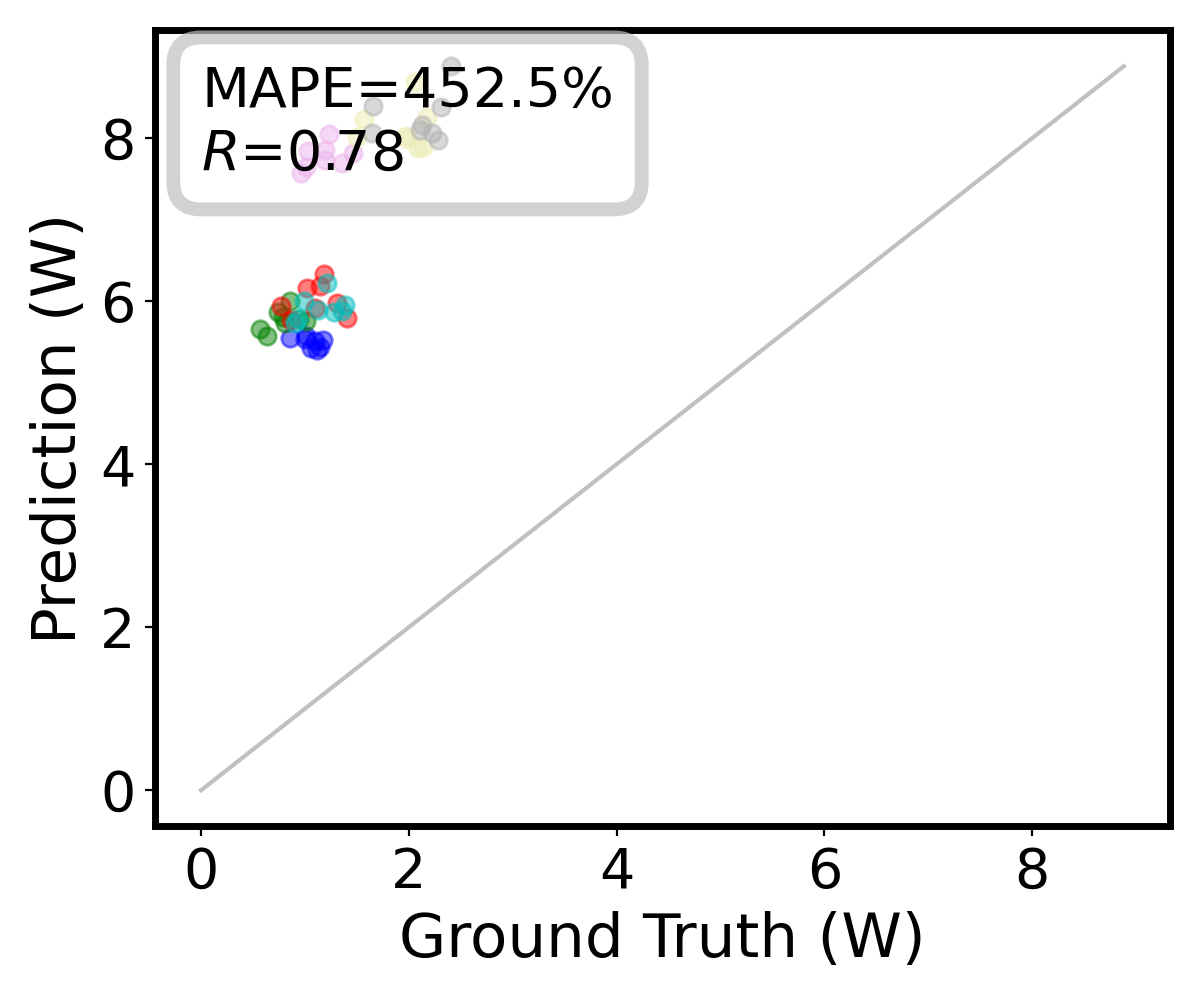}
    %\label{McPAT}
}
\hspace{-3mm}
\subfigure[McPAT-Plus]{
    \centering
    \includegraphics[height=0.25\textwidth]{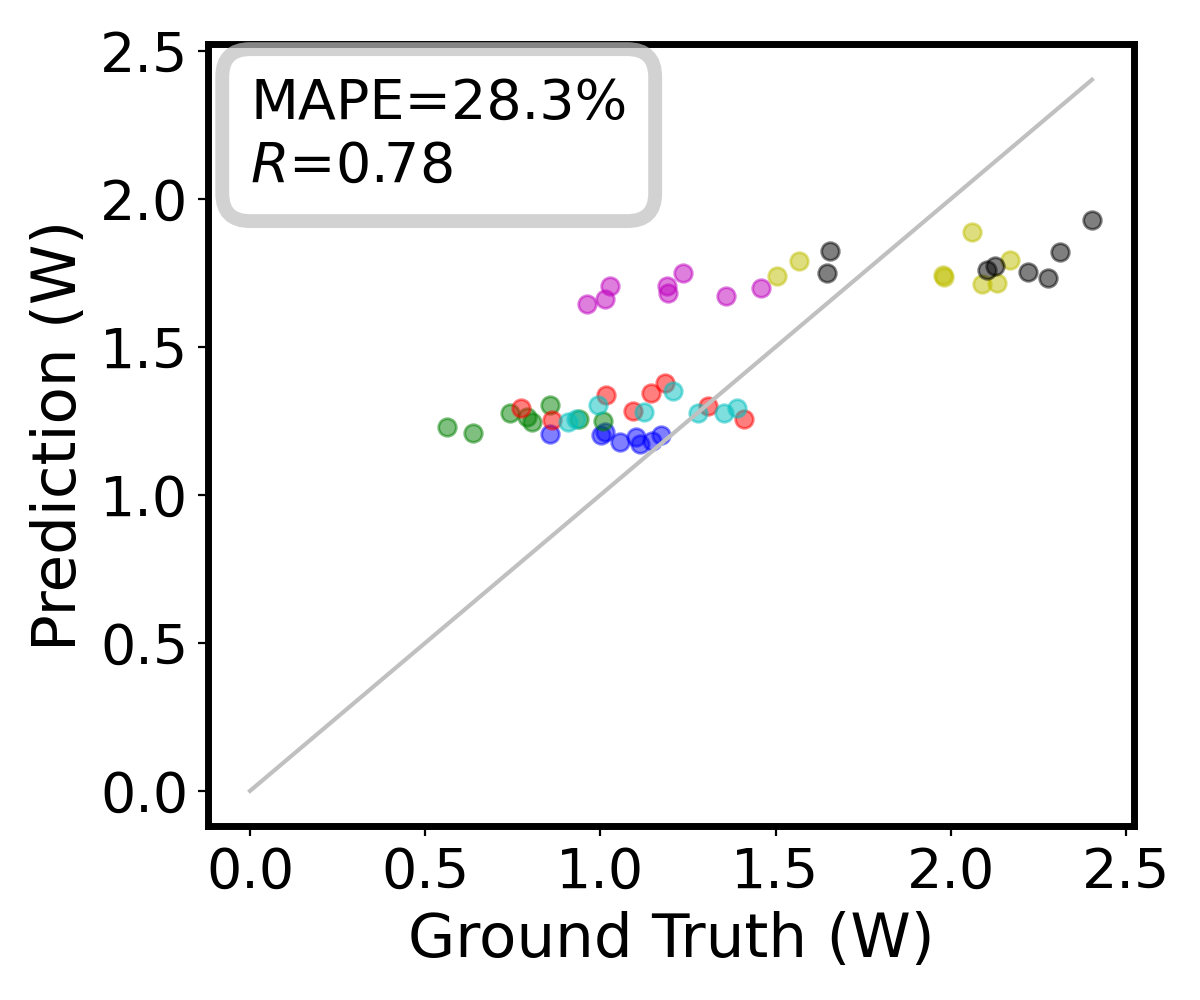}
    %\label{McPAT-Plus}
}
\hspace{-3mm}
\subfigure[McPAT-Calib]{
    \centering
    \includegraphics[height=0.25\textwidth]{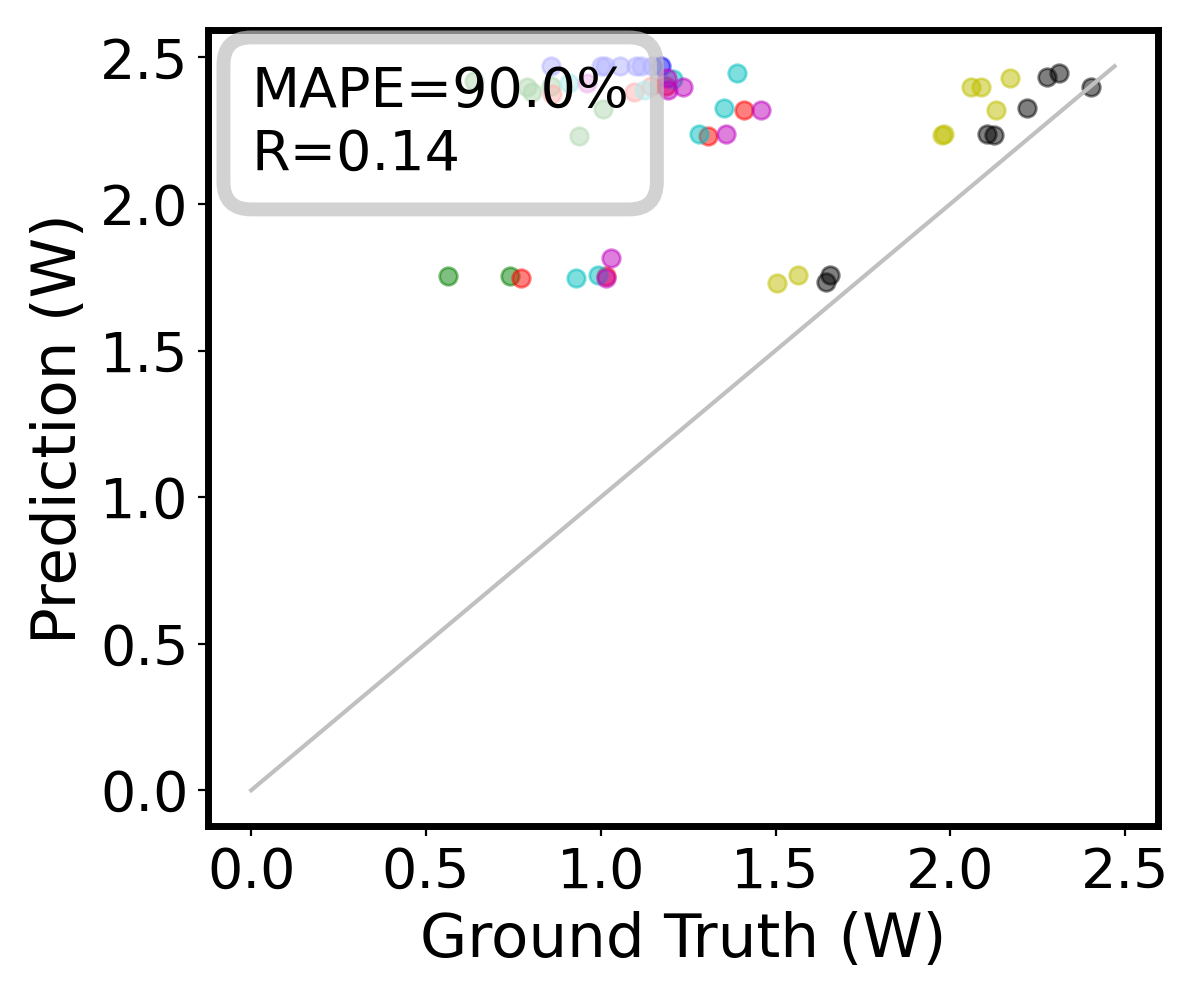}
    %\label{archbp}
}

\hspace{-5mm}
\subfigure[McPAT-Calib-Component]{
    \centering
    \includegraphics[height=0.25\textwidth]{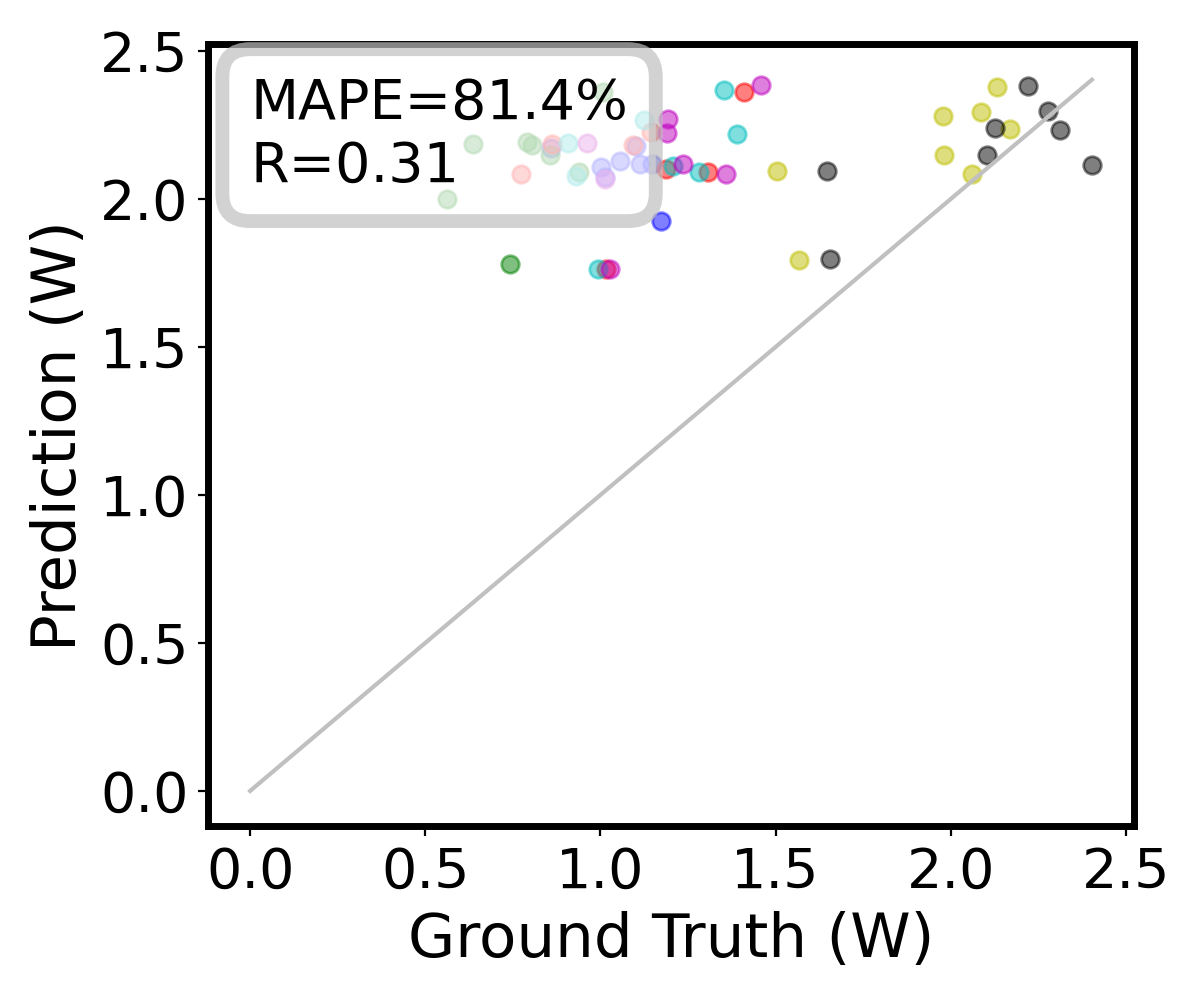}
    %\label{McPAT-Calib-Component}
}
\hspace{-3mm}
\subfigure[McPAT-Calib-CompGroup]{
    \centering
    \includegraphics[height=0.25\textwidth]{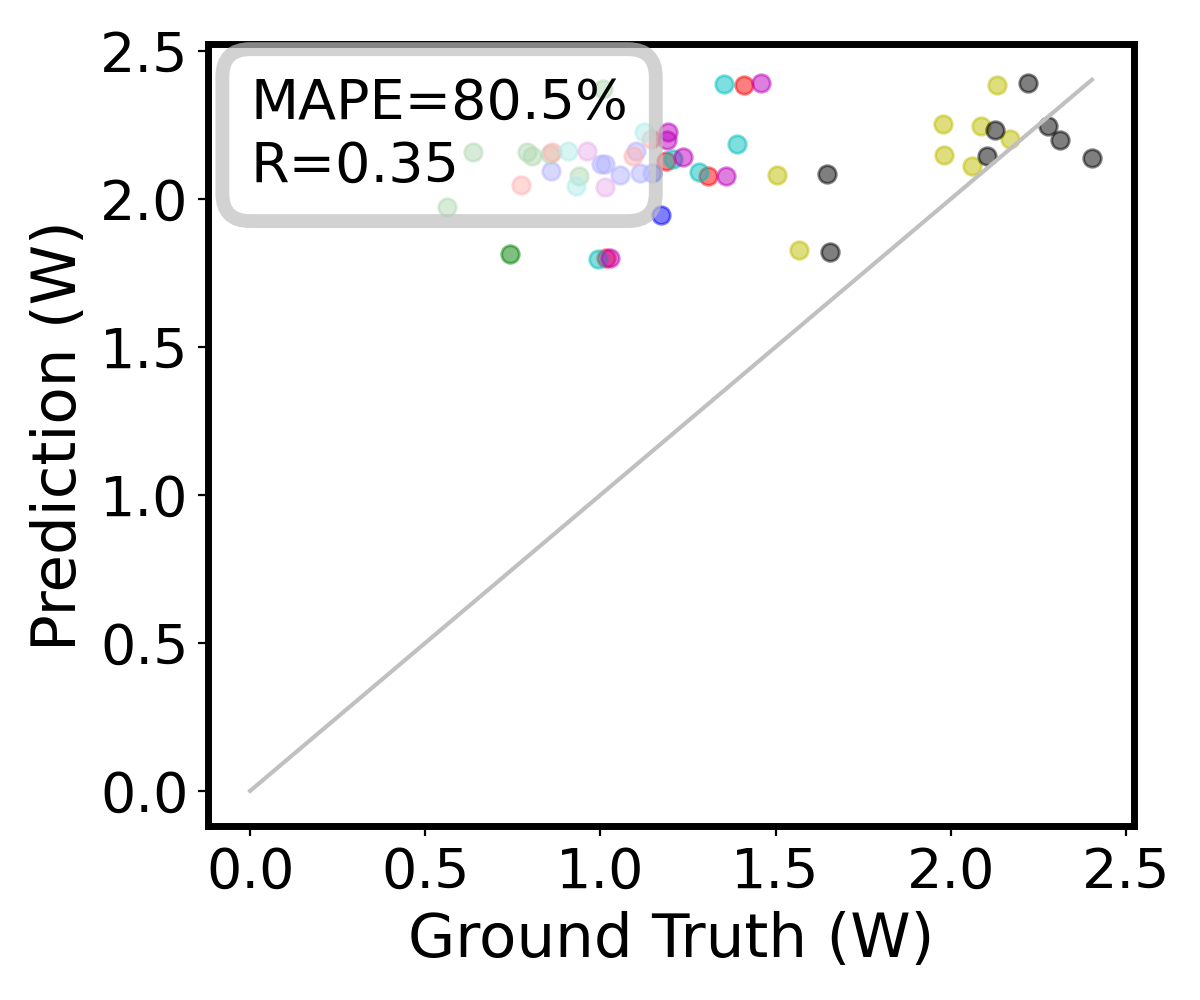}
    %\label{McPAT-Calib-Component}
}
\hspace{-3mm}
\subfigure[PANDA]{
    \centering
    \includegraphics[height=0.25\textwidth]{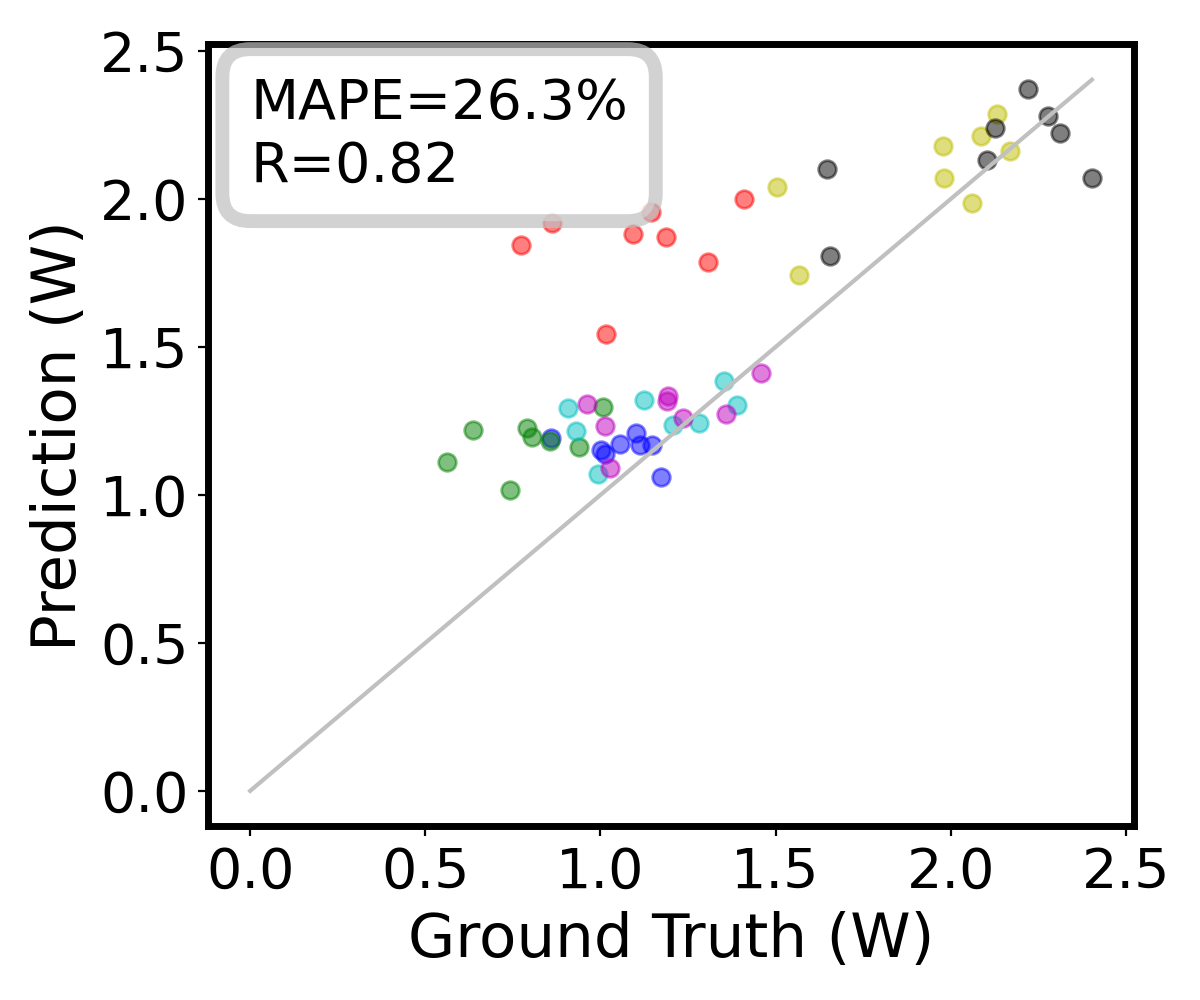}
    %\label{archbp}
}

%\vspace{-.1in}
\caption{Predictions with different models on XiangShan CPU under \emph{Large} training scenario.}
%\vspace{-.2in}
\label{visxsC}
\end{figure*}

Fig.~\ref{visboomC} and \ref{visxsC} visualize the prediction of different models on BOOM and XiangShan under the \emph{Large} training scenario. 

% \newpage

% \subsection{Compute Resources}
% \label{compres}

% We perform our experiments on a 64-core server. The model evaluation is fast and efficient, taking less than one minute for each model. The memory requirement is within 10MB. Reproducing all of our results takes less than ten minutes.

% \subsection{Licenses}
% \label{compres}

% Chipyard framework and BOOM CPU are under BSD-3-Clause. OpenXiangShan framework and XiangShan CPU are under Mulan PSL v2. Riscv-tests is under BSD-3-Clause.

\end{document}